%% file: proposal.tex
\documentclass[12pt]{book}
\usepackage{times}
\usepackage{epsfig}
\usepackage{rotating}
\usepackage{subfigure}
\begin{document}
\newcommand{\vC}{Cherenkov}
\newcommand{\PM}{photomultiplier tube }
\newcommand{\PMs}{photomultiplier tubes }
\newcommand{\tom}{$\theta_{OM}$}
\newcommand{\f}[2]{\frac{#1}{#2}}
\newcommand{\bal}{\begin{eqnarray}}
\newcommand{\eal}{\end{eqnarray}}
\newcommand{\bec}{\begin{displaymath}}
\newcommand{\eec}{\end{displaymath}}
\newcommand{\becnum}{\begin{equation}}
\newcommand{\eecnum}{\end{equation}}
\newcommand{\numu}{$\mbox{$\mathrm{\nu_\mu}$}$ }
\newcommand{\nue}{$\nu_e$ }
\newcommand{\anumu}{$\nu_\mu$ }
\newcommand{\potasiu}{$^{40}$K }
\newcommand{\hms}{hm3s\_00\_100\_8m\_48floors }
\newcommand{\hmh}{hm3h\_8m\_1s }
\newcommand{\hmd}{hm2d\_4m\_1s }
\newcommand{\spec}{oscspec\_1\_48\_6 }
\newcommand{\mixt}{hmixte\_4m\_1s }
\newcommand{\mstring}{osc\_48fl\_16st\_60m }
\newcommand{\putpicture}[3]{%
  \begin{center}
    \resizebox{#1mm}{#2mm}{\includegraphics{#3}}      
  \end{center}}
\newcommand{\mypicture}[3]{%
    \resizebox{#1mm}{#2mm}{\includegraphics{#3}} }    
\input{front_page}

\newpage
\chapter{Introduction} 
\label{chap:introduction}
\input{chap1}

\newpage 
\chapter{Scientific programme} 
\label{chap:scientific_programme}
\input{chap2}
\newpage
\chapter{Detection principles} 
\label{chap:detection_principles}
\input{chap3}
\newpage
\chapter{R \& D programme} 
\label{chap:prototypes}
\input{chap4}

\newpage
\chapter{Detector design} 
\label{chap:detector_design}
\input{chap5}

\newpage 
\chapter{Detector performance} 
\label{chap:detector_performance}
\input{chap6}

\newpage
\chapter{Complementary techniques} 
\label{chap:complementary_techniques}
\input{chap7}
\newpage
\chapter{Cost and schedule} 
\label{chap:cost}
\input{chap8}

\newpage
\chapter{Conclusion} 
\label{chap:conclusion}
\input{chap9}

\newpage
\addcontentsline{toc}{chapter}{References}

\input{references}


\end{document}

%% file: front_page.tex
\begin{titlepage}
\vglue 3.5cm
\centerline{\LARGE \bf A Deep Sea Telescope for }
\centerline{\LARGE \bf High Energy Neutrinos}
\vglue 3cm
\centerline{\Large \bf The ANTARES Collaboration}
\vglue 4cm
\centerline{\large 31 May, 1999}
\vglue 4cm
\hglue 4.5cm 
\parbox[c]{4cm}{ 
CPPM-P-1999-02\\
DAPNIA 99-01\\
IFIC/99-42\\
SHEF-HEP/99-06}
\newpage
\thispagestyle{empty}
\vspace*{\stretch{1}}
\centering{
\fbox{\parbox[c]{6cm}{
	This document may be retrieved from the
	Antares web site:\\
	http://antares.in2p3.fr/antares/\\
}}
}
\vspace*{\stretch{1}}
\newpage
\thispagestyle{empty}
\vspace*{\stretch{1}}
\centerline {\large \bf Abstract}
\vglue 1cm

The ANTARES Collaboration proposes to construct a large area water
Cherenkov detector in the deep Mediterranean Sea, optimised for the
detection of muons from high-energy astrophysical neutrinos.  This paper
presents the scientific motivation for building such a device, along with
a review of the technical issues involved in its design and construction.
\vglue 5mm

The observation of high energy neutrinos will open a new window on the
universe.  The primary aim of the experiment is to use neutrinos as a tool
to study particle acceleration mechanisms in energetic astrophysical
objects such as active galactic nuclei and gamma-ray bursts, which may
also shed light on the origin of ultra-high-energy cosmic rays.  At
somewhat lower energies, non-baryonic dark matter (WIMPs) may be detected
through the neutrinos produced when gravitationally captured WIMPs
annihilate in the cores of the Earth and the Sun, and neutrino
oscillations can be measured by studying distortions in the energy
spectrum of upward-going atmospheric neutrinos.
\vglue 5mm

The characteristics of the proposed site are an important consideration in
detector design.  The paper presents measurements of water transparency,
counting rates from bioluminescence and potassium 40, bio-fouling of the
optical modules housing the detectorÕs photomultipliers, current speeds
and site topography.  These tests have shown that the proposed site
provides a good-quality environment for the detector, and have also
demonstrated the feasibility of the deployment technique. 
\vglue 5mm

The present proposal concerns the construction and deployment of a
detector with surface area 0.1~km$^2$.  The conceptual design for such a
detector is discussed, and the physics performance evaluated for
astrophysical sources and for neutrino oscillations.  An overview of costs
and schedules is presented.
\vglue 5mm

It is concluded that a 0.1~km$^2$ detector is technically feasible at
realistic cost, and offers an exciting and varied physics and astrophysics
programme.  Such a detector will also provide practical experience which
will be invaluable in the design and operation of future detectors on the
astrophysically desirable 1~km$^2$ scale.

\vspace*{\stretch{1}}

\newpage
~
\thispagestyle{empty}

\end{titlepage}
\thispagestyle{empty}
\newpage
\input{authors}
\newpage
~
\thispagestyle{empty}
\setcounter{page}{0}
\tableofcontents

%% file: authors.tex
\pagestyle{empty}
\newpage
\small
%
%
%
%
%
\pretolerance=10000
{\raggedbottom
\centerline{\large\bf The ANTARES Collaboration}
\footnotesize
\vspace{0.5cm}
\centerline{\bf Particle Physics Institutes}
\vspace{0.5cm}
\begin{sloppypar}
\noindent
E.~Aslanides,
J-J.~Aubert, 
S.~Basa, 
F.~Bernard, 
V.~Bertin,
M.~Billault,
P-E.~Blanc,
J.~Brunner,
A.~Calzas,
F.~Cassol,
J.~Carr,
C.~Carloganu,
J-J.~Destelle,
P-Y.~Duval,
F.~Hubaut,
E.~Kajfasz,
M.~Jaquet,
D.~Laugier,
A.~Le Van Suu,
P.L.~Liotard,
L.~Martin,
F.~Montanet,
S.~Navas,
C.~Olivetto,
P.~Payre,
A.~Pohl,
R.~Potheau,
M.~Raymond,
M.~Talby,
C.~Tao,
E.~Vigeolas.
\vglue 0.1cm
\indent
\parbox{13cm}{\sl
Centre de Physique des Particules de Marseille (CPPM),
(CNRS/IN2P3 - Universit\'e de la M\'editerran\'ee Aix-Marseille II),
163 Avenue de Luminy, Case 907, 13288 Marseille Cedex 09, France}
\end{sloppypar}
\vspace{10mm}

\begin{sloppypar}
\noindent
S.~Anvar,
R.~Azoulay,
R. W.~Bland,
F.~Blondeau,
N.~de Botton,
N.~Bottu,
P-H.~Carton,
P.~Deck,
F. E.~Desages,
G.~Dispau,
F.~Feinstein,
P.~Goret,
L.~Gosset,
J-F.~Gournay,
J.~R.~Hubbard,
M.~Karolak,
A.~Kouchner,
D.~Lachartre,
H.~Lafoux,
P.~Lamare,\linebreak
J-C.~Languillat,
J-P.~Laugier,
H.~Le Provost,
S.~Loucatos,
P.~Magnier,
B.~Mazeau,
P.~Mols,
L.~Moscoso,
N.~Palanque-Delabrouille,
P.~Perrin,
J.~Poinsignon,
Y.~Queinec,
Y.~Sacquin,
J-P.~Schuller,
T.~Stolarczyk,
A.~Tabary,
Y.~Tayalati,
P.~Vernin,
D.~Vignaud,
D.~Vilanova.
\vglue 0.1cm
\indent
\parbox{13cm}{\sl
DAPNIA/DSM, CEA/Saclay,
91191 Gif sur Yvette Cedex, France}
\end{sloppypar}
\vspace{10mm}
\begin{sloppypar}
\noindent
Y.~Benhammou,
F.~Drouhin,
D.~Huss,
A.~Pallares,
T.~Tzvetanov.
\vglue 0.1cm
\indent
\parbox{13cm}{\sl
Groupe de Recherches en Physique des Hautes Energies
(GRPHE),
(Universit\'e de Haute Alsace), 61 Rue Albert Camus, 68093
Mulhouse Cedex, France}
\end{sloppypar}
\vspace{10mm}

\begin{sloppypar}
\noindent
M.~Danilov,
R.~Kagan,
A.~Rostovstev.
\vglue 0.1cm
\indent
\parbox{13cm}{\sl
Institute for Theoretical and Experimental Physics (ITEP),
B. Cheremushkinskaya 25,
117259 Moscow, Russia}
\end{sloppypar}
\vspace{10mm}

\begin{sloppypar}
\noindent
E.~Carmona,
R.~Cases,
J. J.~Hern\'andez,
J.~Z\'u\~niga.
\vglue 0.1cm
\indent
\parbox{13cm}{\sl
Instituto de F\'{\i}sica Corpuscular,
CSIC - Universitat de Val\`encia,
46100 Burjassot,
Valencia, Spain}
\end{sloppypar}
\vspace{10mm}

\begin{sloppypar}
\noindent
C.~Racca,
A.~Zghiche.
\vglue 0.1cm
\indent
\parbox{13cm}{\sl
Institut de Recherches Subatomiques (IReS),
(CNRS/IN2P3 - Universit\'e Louis Pasteur), BP 28,
67037 Strasbourg Cedex 2, France}
\end{sloppypar}
\vspace{10mm}
\pagebreak
\begin{sloppypar}
\noindent
R.~van Dantzig,
J.~Engelen,
A.~Heijboer,
M.~de Jong,
E.~Kok,
P.~Kooijman,
G.J.~Nooren,
J.~Oberski,
P.~de Witt Huberts,
E.~de Wolf.
\vglue 0.1cm
\indent
\parbox{13cm}{\sl
NIKHEF and University of Amsterdam,
Kruislaan 409,
PO Box 41882,
1009 BD Amsterdam, Netherlands}
\end{sloppypar}
\vspace{10mm}

\begin{sloppypar}
\noindent
D.~Evans,
G.~Mahout,
I.~Kenyon,
P.~Jovanovic,
P.~Newman,
T.~McMahon.
\vglue 0.1cm
\indent
\parbox{13cm}{\sl
University of Birmingham, School of Physics and Astronomy,
Edgbaston, Birmingham B15 2TT,
United Kingdom}
\end{sloppypar}
\vspace{10mm}

\begin{sloppypar}
\noindent
B.~Brooks,
S.~Cooper,
J.~Fopma,
N.~Jelley,
W.~Schuster,
S.~Tilav,
D.~Wark.
\vglue 0.1cm
\indent
\parbox{13cm}{\sl
University of Oxford, Department of Physics,
Nuclear and Astrophysics Laboratory,
Keble Road,
Oxford OX1 3RH, United Kingdom}
\end{sloppypar}
\vspace{10mm}

\begin{sloppypar}
\noindent
S.~Cartwright,
V.~Kudryavtsev,
J.~McMillan,
N.~Spooner,
L.~Thompson.
\vglue 0.1cm
\indent
\parbox{13cm}{\sl
University of Sheffield,
Department of Physics and Astronomy,
Sheffield, S3 7RH, United Kingdom}
\end{sloppypar}
\vspace{10mm}

\begin{sloppypar}
\noindent
R.~Triay. 
\vglue 0.1cm
\indent
\parbox{13cm}{\sl
Centre de Physique Th\'eorique (CPT),
(CNRS),
163 Avenue de Luminy, Case 907, 13288 Marseille Cedex 09, France}
\end{sloppypar}
\vspace{10mm}

\vspace{5mm}
\centerline{\bf Astronomy Institutes}
\vspace{5mm}

\begin{sloppypar}
\noindent
A.~Mazure.
\vglue 0.1cm
\indent
\parbox{13cm}{\sl
Laboratoire d'Astronomie Spatiale,
Institut Gassendi pour la Recherche \linebreak Astronomique en Provence (IGRAP),
(CNRS/INSU - Universit\'e de Provence Aix-Marseille I),
Les Trois Lucs,
Traverse du Siphon,
13012 Marseille Cedex, France}
\end{sloppypar}
\vspace{10mm}

\begin{sloppypar}
\noindent
P.~Amram,
J.~Boulesteix,
M.~Marcelin.
\vglue 0.1cm
\indent
\parbox{13cm}{\sl
Observatoire de Marseille,
Institut Gassendi pour la Recherche Astronomique en Provence (IGRAP),
(CNRS/INSU - Universit\'e de Provence Aix-Marseille I),
2 Place Le Verrier,
13248 Marseille Cedex 4, France}
\end{sloppypar}
\vspace{2mm}

\vspace{10mm}

\pagebreak
\vspace{5mm}
\centerline{\bf Sea Science Institutes}
\vspace{5mm}
\begin{sloppypar}
\noindent
F.~Blanc,
G.~Coustillier,
J-L.~Fuda,
C.~Millot.
\vglue 0.1cm
\indent
\parbox{13cm}{\sl
Centre d'Oc\'eanologie de Marseille,
(CNRS/INSU - Universit\'e de la M\'editerran\'ee),
Station Marine d'Endoume-Luminy,
Rue de la Batterie des Lions, 13007 Marseille, France}
\end{sloppypar}
\vspace{10mm}

\begin{sloppypar}
\noindent
C.~Comp\`ere,
J.F.~Drogou,
D.~Festy,
G.~Herrouin,
Y.~Le Guen,
L.~Lemoine,
A.~Massol,
F.~Maz\'eas,
J.P.~Morel,
J.F.~Rolin,
P.~Valdy.
\vglue 0.1cm
\indent
\parbox{13cm}{\sl
IFREMER, Centre de Toulon/La Seyne sur Mer, Port Br\'egaillon,
Chemin Jean-Marie Fritz,
83500 La Seyne sur Mer, France}
\indent
\parbox{13cm}{\sl
IFREMER, Centre de Brest,
29280 Plouzan\'e, France}
\end{sloppypar}
}
\normalsize
\newpage
\pagestyle{plain}

%% file: chap1.tex
This document presents
the scientific motivation for building 
a high energy neutrino undersea detector, with an
effective area of 0.1 km$^2$, 
along with a review of the technical issues involved in its
design and construction.  The planned apparatus consists of an
array of photomultipliers, arranged in a lattice near the sea bed at a depth 
of 2400~m, to detect the Cherenkov light from muons produced by
neutrino interactions in the seawater and rock beneath.
Since 1996, the ANTARES collaboration
has conducted an extensive R\&D programme in close collaboration 
with experts in marine technology.  This programme has addressed 
most of the critical technical aspects relevant to the 
construction and operation of a neutrino telescope, providing 
the information and experience necessary
to be confident that the planned array is feasible at a reasonable 
cost.  The neutrino detection capabilities of the design have been
evaluated in detailed simulation studies and its potential science
mission explored.

\section{Why neutrino astronomy?}
Most of our current knowledge of the Universe comes from the
observation of photons.  Photons have many advantages as cosmic
information carriers: they are copiously produced, they are stable and
electrically neutral, they are easy to detect over a wide energy
range, and their spectrum carries detailed information about the
chemical and physical properties of the source.  Their disadvantage is
that the hot, dense regions which form the central engines of stars,
active galactic nuclei and other astrophysical energy sources are
completely opaque to photons, and therefore we cannot investigate the
properties of these regions by direct observation, but only by
indirect inference.  For example, the photons we observe from the Sun
come from its  photosphere, far removed from the
hydrogen-fusing core.
Moreover, high energy photons interact with photons of the infrared radiation
background and with the cosmic microwave background to create 
electron-positron pairs; this is the Greisen-Zatsepin-Kuz'min effect
(GZK)~\cite{rf:gzk}. This effect suppresses any possibility of surveying the
sky over distances greater than 100\,Mpc with high energy 
(\(>\)10 TeV) gamma rays.

In order to observe the inner workings of the astrophysical objects and
to obtain a description of the Universe over a larger range of 
energies, we
need a probe which is electrically neutral, so that its trajectory
will not be affected by magnetic fields, stable so that it will reach
us from distant sources, and weakly interacting so that it will
penetrate regions which are opaque to photons.  The only candidate
currently known to exist is the neutrino.

Some astrophysical sources are known to emit neutrinos: hydrogen
fusion produces electron neutrinos as by-products, and solar neutrino
astronomy has a 30~years long history; the conversion of iron
nuclei to neutrons when a neutron star is formed in the heart of a
supernova produces a burst of neutrinos (augmented by the thermal
production of neutrino-antineutrino pairs), and one such burst was
observed by Kamiokande and IMB for Supernova 1987A; cosmology predicts
a low-energy relic neutrino background similar to the low-energy relic
photons of the Cosmic Microwave Background, but these would have an
effective temperature of around 1.9 K and are very
difficult to observe.

Astrophysical sources of high-energy neutrinos have not been observed
directly, but their existence can be inferred from the properties of
cosmic rays.  Primary cosmic rays are protons, with some
admixture of heavier nuclei; the energy spectrum is a power law which
extends to extremely high energies, values exceeding $10^{20}$ eV having
been observed in recent years.  Protons themselves have limited use as
astrophysical information carriers because they are charged, and
therefore subject to deflection by cosmic magnetic fields: only the
very highest-energy cosmic rays are likely to retain any memory of the
source direction.  The exact source of the high-energy cosmic rays is
thus unknown, although supernova remnants and active galactic nuclei
have been proposed.  Whatever the source, it is clear that
accelerating protons to such high energies is likely to generate a
large associated flux of photo-produced pions, which decay to
yield gamma rays and neutrinos.  These will remember the source
direction, and so the existence of a general flux of very high energy
cosmic-ray protons implies the existence of sources of
high-energy neutrinos.

Neutrino astronomy thus offers the possibility of observing
sources which correspond to the central engines of the most energetic
astrophysical phenomena.  As discussed below, it also provides long
baselines for neutrino oscillation studies, and can explore useful
regions of supersymmetric parameter space in the context of dark
matter.  The drawback, of course, is that the weak interactions of
neutrinos imply that a very massive detector with extremely good
background rejection is required to observe a measurable flux.

\section{The view from a neutrino telescope}
The ANTARES scientific programme is described in detail in chapter 2,
but a brief overview is presented here.  It is
convenient to divide the programme into three broad subject areas:
particle physics (neutrino oscillations), particle astrophysics
(searches for neutralino dark matter) and astronomy.

Within the minimal standard model, neutrinos are strictly massless,
but the need to incorporate non-zero masses has long been
anticipated.  Neutrino astronomy was instrumental in establishing
non-zero neutrino masses as an important topic: the
lower-than-predicted flux of electron neutrinos from the Sun (the
Solar Neutrino Problem) is now generally believed to be explained by
neutrino oscillations either {\it in vacuo} or more probably within
the Sun itself, enhanced by the high electron density in the solar
core (the MSW effect).

Neutrino oscillation solutions to the solar neutrino problem involve
the conversion of electron neutrinos into some other flavour.  The
energies involved are not well matched to the detection capabilities
of ANTARES.  However, recent results from Super-Kamiokande appear to
show a similar flux reduction
occurring for {\it muon} neutrinos generated by cosmic-ray
interactions at the top of the Earth's atmosphere.  The effect is
interpreted as evidence for the oscillation of $\nu_\mu$ into either
$\nu_\tau$ or a so-called ``sterile'' neutrino, with a large mixing
angle and a squared mass difference of $\sim 10^{-3}-10^{-2}$ eV$^2$.
This is a result of major importance which urgently requires
confirmation by an experiment with independent systematics.  As
discussed below, the proposed ANTARES configuration should be capable
of exploring the region of parameter space favoured by the
Super-Kamiokande data.

In recent years it has become generally accepted by astrophysicists
that most of the matter in the universe is non-luminous ``dark
matter''.  The clearest
evidence for this is the observed flatness of the rotation curves of disk galaxies,
which imply a dynamical mass far in excess of that accounted for by
the constituent stars and gas.  Constraints
from the observed abundances of light elements indicate that much of
the dark matter in the cosmos must be non-baryonic.  No presently known
particle has the required properties, but a good theoretical candidate
is the stable neutral particle expected in most versions of
supersymmetry theory.  

The detection and identification of a relic
cosmological population of supersymmetric particles would be of
immense importance to both cosmology and particle theory.  Neutrino
telescopes are not directly sensitive to a weakly interacting 
massive particle (WIMP).  However, supersymmetric WIMPs will
accumulate in the cores of the Sun and the Earth or in the
centre of the Galaxy
through gravitational
capture.  The resulting high space
density leads to  annihilation reactions, which
will yield high-energy neutrinos through the decays of the 
gauge bosons and heavy particles produced.  The proposed detector would be
sensitive to these neutrinos over a useful range of WIMP masses.
Compared to ongoing direct detection experiments, 
neutrino telescopes are generally more suitable for higher
masses, although resonances in the Earth's capture
cross-section enhance the signal strongly at certain lower
masses, particularly around 56~GeV.
But even a confirmation of a
prior direct detection would provide useful information about the
couplings of the WIMP, and thus help to constrain the parameters of
the theory.

As discussed above, the two confirmed astrophysical
neutrino sources are:
the Sun and supernovae as exemplified by SN 1987A.  Both of
these produce low energy neutrinos which could not be tracked by
ANTARES, although a nearby supernova could be detected through a
transient increase in the overall singles rate.  
However, there are a
number of candidate astrophysical sources of high-energy neutrinos
which would be detected by ANTARES.  Since pions produced by high-energy protons
are the likeliest source of
high-energy neutrinos, 
candidate astrophysical neutrino sources are the proton
accelerators, which might explain the High Energy Cosmic
Rays spectrum.
High energy $\gamma$-rays may be associated with high energy protons and their
subsequent decays from $\pi^0$, but they are also produced by  
synchrotron radiation of fast electrons in the presence of magnetic fields.
The observation of neutrino sources would unambiguously discriminate 
between the two acceleration mechanisms.

Candidate sources can be identified both
within the Galaxy---accreting binaries containing neutron stars or
black holes, supernovae and young supernova remnants---and elsewhere,
most notably active galactic nuclei (AGN) and gamma-ray bursters.
This represents a rich spectrum of possible sources, both steady and
transient, covering a wide range of neutrino energies.

It is worth noting that the history of astronomical observation
suggests that the likeliest outcome of opening a new observational
window is the discovery of a completely unexpected class of sources.
Such discoveries are by definition difficult to anticipate, but one
possible pointer is that the very highest-energy cosmic rays, those
above $10^{20}$ eV, remain difficult to explain in present models.  This
puzzle could be solved by the observation of their associated neutrinos.
 
\section{Present and future neutrino telescopes}

The proposed ANTARES detector is one of a number of present and
proposed neutrino telescope projects.  Of these, the Lake Baikal
detector, consisting of eight strings supporting a total of 192
optical modules, was the first to demonstrate the feasibility of the
technique, but is limited in depth and spatial extent by the nature of its
site and so cannot be extended to a full-scale neutrino telescope.
The AMANDA array at the South Pole uses ice rather than water as the
detector medium, which gives it lower backgrounds but poorer
directional accuracy.  An array with an effective area of around
$10^4$ m$^2$, AMANDA B, has been deployed at a depth of 1500--2000 m
and has published data on atmospheric muons; the installation of
AMANDA-II, which will have an effective area several times
larger, has
started during the 1997--98 season.  An eventual scale-up to a full
km-scale array (ICECUBE) is proposed.  Also at the proposal stage is
NESTOR, with a deep site in the eastern Mediterranean off
Pylos (Greece).

The present ANTARES proposal is in many respects complementary to AMANDA.  A
water-based detector is more flexible and has better directional
sensitivity, at the cost of higher background noise.  Together, the
two detectors cover the whole sky, with a substantial amount of overlap
for cross-checking.

The Mediterranean Sea represents an environment for a neutrino
telescope that is quite different from those of the operating arrays
at Lake Baikal (a freshwater lake which freezes over in winter) and
the South Pole.  Therefore, since its creation in 1996, ANTARES has
followed a first phase R\&D programme focused on three major
milestones:

\begin{itemize}
\item construction and deployment of test lines dedicated to measuring
environmental parameters such as optical background, biofouling and
water transparency;

\item development of prototype strings to acquire the necessary
expertise to deploy and operate an undersea detector up to the
kilometre scale;

\item development of software tools to explore the physics
capabilities of the detector.
\end{itemize}

To date, the majority of the deployments have been performed in the
Mediterranean Sea 30 km off the coast near Toulon (France), at a depth
of 2400~m.

This first phase has demonstrated that the
deployment and physics operation of such a detector is feasible.  The
present proposal advances the ANTARES programme to a second phase,
namely the construction of an array of 1000 optical modules to form a
high-energy neutrino detector with effective area 0.1~km$^2$.
Such an array would have the following achievable physics goals:

\begin{itemize}
\item the study of the high energy neutrino flux (in the TeV--PeV
range) with unprecedented angular resolution;

\item the measurement of atmospheric neutrino oscillations in the
region of parameter space favoured by Super-Kamiokande;

\item a search for supersymmetric dark matter covering a region of
model parameter space which is interesting for cosmology and
particle physics.
\end{itemize}

It would also provide practical experience and expertise which
will be invaluable for the anticipated third phase during which a
larger scale detector will be constructed, capable of conducting a search for
astrophysical sources.

%% file: chap2.tex

The science mission of a neutrino telescope such as ANTARES is vast, 
encompassing neutrino astronomy, neutrino oscillations and non-baryonic matter 
in the form of neutralinos, heavy metastable relic particles or topological defects.
This chapter gives a brief account of the physics of the main candidate neutrino
sources and the mechanisms by which they might produce a significant signal.

\section{Astrophysical sources} 
The principal mechanism for generating high energy neutrinos
is through decay cascades induced by high energy
protons.  Interactions of protons with matter or radiation produce
mesons whose leptonic decay modes will yield neutrinos.  A list of possible
high energy astrophysical neutrino sources therefore involves a list of 
candidate astrophysical sites for proton acceleration.

\subsection{X-ray binaries}

This class of binary stars, which are among the
brightest cosmic X-ray sources, consists
of compact objects, such as neutron stars or black holes, which accrete
matter from their normal companion stars.  The accretion process
leads to plasma waves in the strong magnetic field of the compact object,
which bring protons to high energies by stochastic acceleration.
Interactions of the accelerated particles with the accreting matter 
or
with the companion star would then produce a neutrino flux comparable to
that in high-energy particles with a spectral index close to 2.

\subsection{Supernova remnants and cosmic rays}

Explosions of massive stars (supernovae) produce an expanding shell of
material which is known from radio observations to accelerate high-energy
particles. In some cases, the residue of the supernova is a neutron star
which is detectable as a pulsar.  Protons inside supernova shells can be
accelerated by a first-order Fermi mechanism if (as seems likely) the 
shell is turbulent. If a pulsar is present there are additional acceleration
mechanisms: in the magnetosphere of the pulsar, or at the front of the
shock wave produced by the magneto-hydrodynamic wind in the
shell~\cite{Berez90}.  The interaction of these protons with the matter of
the shell gives rise to neutrinos and photons (from charged and neutral
pion decays respectively). An especially promising source is
the ion doped wind model based on ultra high energy ion acceleration by
the pulsar~\cite{Bednarek97}.  The ions drift across the magnetic field in
the surrounding supernova remnant filaments at a speed of order 10 times
the hydrodynamic expansion velocity and interact with the thermal gas to
produce neutrinos via $\pi^\pm$ production.  The minimum flux predicted is
substantially above background for the Crab remnant at $10^{16}$ eV, but
the theory of ion acceleration is too primitive to constrain the spectrum
of injected ions and hence the production rate of energetic neutrinos at
ANTARES energies.

It is thought that supernova remnants are the principal galactic source of
cosmic-ray protons, but, as yet, there is no direct confirmation of this
hypothesis. Recent observations~\cite{Esposito96} above $10^8$\,eV by the 
EGRET detector
have found $\gamma-$ray signals associated with at least 2 supernova
remnants (IC 443 and $\gamma$ Cygni). However, electromagnetic radiation
is an ambiguous signature, as it can come from either accelerated electrons
or protons. An observation of neutrinos would provide a clear indication
of proton acceleration with the direction identifying the source.

\subsection{Active galactic nuclei}

\newcommand{\gtsim}{\mbox{{\raisebox{-0.4ex}{$\stackrel{>}{{\scriptstyle\sim}}
$}}}}
\newcommand{\ltsim}{\mbox{{\raisebox{-0.4ex}{$\stackrel{<}{{\scriptstyle\sim}}
$}}}}

Active Galactic Nuclei (AGN) such as quasars are, averaged over time,
the most powerful known objects in the Universe.  Their observed total
luminosities are in the range $10^{35} - 10^{41}$ W~\cite{Gaisser95}.
In the generic model of AGN, these high luminosities arise from accretion of matter, at a
rate of at least a few solar masses per year, onto a super-massive black
hole, ranging from $10^6$ to $10^{10}$ solar masses. A minority of AGN also 
produces relativistic jets which transport synchrotron-emitting
electrons, most easily observable in the radio, over distances up to
1~Mpc.  Objects of this type whose jets are directed almost exactly
towards us appear as intense, compact and variable sources because their
emission is amplified by Doppler boosting.  Such objects are known as ``blazars''. 

An unexpected result, now well established by the EGRET
satellite~\cite{Egret94}, is that many blazars
 emit $\gamma$ rays with energies up to $\approx 10^{10}$\,eV.
Even more surprisingly, four members of this class have been detected 
by atmospheric Cherenkov telescopes as
highly variable emitters of $\gamma$ rays with energies 
exceeding $10^{12}$\,eV~\cite{TeV}.
The mechanism by which such high energy particles are
generated is a matter of active debate, as is the jet composition.

Electrons are thought to be accelerated by the first-order Fermi process,
and this is also efficient for protons.  In generic AGN models, which do not 
necessarily involve jets, protons may be accelerated by shock waves 
associated with the accretion flow into the black hole.  These protons may 
then interact with the matter of the accretion disk or with its ambient 
radiation field. 
Such models produce neutrinos with no associated high energy gammas, because
the photons do not escape from the core of the AGN.
  
In active galactic nuclei with jets, 
protons may also be accelerated in the inner regions of
jets.  These will then produce neutrinos
by interaction with ambient radiation either emitted by the accretion disk or generated as synchrotron radiation within the jet.  Similarly, Fermi
acceleration in the hot-spots of powerful (Fanaroff-Riley Class II) radio
sources may produce a population of high-energy protons~\cite{Biermann89}.
Such protons contribute to the high-energy $\gamma$-ray emission
via a proton-induced cascade~\cite{Mannheim98}, which will also generate
neutrinos.  
Both neutrino and photon flux will be 
increased by Doppler beaming in the case of blazars.

Recently limits have been set~\cite{WaxBah99} on the production of high energy
neutrinos in AGN jets.  These upper bounds, which are related to the 
measured flux of high energy cosmic rays, define fluxes of neutrinos 
significantly smaller than those of previous models.  However, it remains 
possible that high-energy protons are found in some, but not all, AGN jets.

Theoretical models of these sources are subject to large uncertainties,
so that observations are of paramount importance.

\subsection{Gamma ray bursts}

Gamma Ray Bursts (GRBs) are the most spectacularly
violent phenomena in the universe known to this date, and up until two 
years ago they were declared to be 
one of the outstanding mysteries of modern astrophysics.
However, early 1997 brought a major break though when
the BeppoSAX satellite \cite{SAX} located a burst precisely enough to permit
the identification of its optical counterpart. Since then, further advances
have been made, with 
the discovery of GRB afterglows, measurements of their redshift and recognition
of the magnitude of the energy release occurring in these second-long flashes. 

So far about a dozen GRB afterglows have been detected.
The distances measured place 
the burst sources at cosmological distances with redshifts in the 
range $z=0.8-3.4$ and indicate
an energy release of $10^{45\pm1}$ J in $\gamma$-rays alone 
(assuming isotropic emission). The location of GRBs in their host galaxies
somewhat correlates the population of GRB progenitors with the star
formation rates, and supports models which involve a black hole formation
through coalescence of a binary system of either a black hole-neutron star
 or a neutron star-neutron star.

Detailed studies of afterglows from X-rays to optical and
radio wavelengths
provided crucial constraints on physical parameters for theoretical models.
In the current standard model for $\gamma$-ray bursts and their afterglows, 
the {\it fireball-plus-blastwave} model~\cite{Piran98},
the initial event deposits a solar mass of energy into a region with 
a radius of about 100 km. The resulting fireball expands ultra-relativistically
with Lorentz factor $\gamma\geq 300$ into the surrounding medium. While
the forward shock sweeps material and heats it, the reverse shock collides
with the ejecta. Therefore, the afterglow is produced by synchrotron radiation
when external shocks decelerate.
However, the multi-peaked light curve of the gamma-ray burst itself is produced
by the collisions of several internal shocks which are catching up each other
 with different Lorentz factors within the inner engine. There is clear
evidence of these three distinct regions of the fireball in the light curve of
GRB990123~\cite{Galama99}.  

A hadronic component would naturally be expected in such
 extreme phenomena. Nearby GRBs could therefore
be the long sought after sources of the highest energy cosmic rays. 
A possible example of such a relatively local GRB is GRB980425, whose location was
observed to be coincident with the extremely bright supernova SN 1998bw.  If this
association is real, the GRB occurred at a distance of a few Mpc\cite{Baron} and
is a potential source of ultra-high-energy cosmic rays.
In such a case, the cosmic ray observations set a model independent
upper bound to the flux of high energy neutrinos produced by photo-meson
interactions of the high energy protons with the radiation field of 
the source~\cite{WaxBah99}. 
This upper bound implies a muon flux from 
diffuse neutrinos of $\sim 20$/km$^2$/year (in $4\pi$ sr)
 in the energy range above 100 TeV.

 The rate predicted in reference~\cite{WaxBah99} 
 is very low for ANTARES, which has a 
 detector area of 0.1~km$^2$
 and an angular acceptance of $2\pi$ sr. Nevertheless, a significantly 
 higher event rate can be obtained by lowering the energy threshold 
 to around 100 GeV.
  The background is greatly reduced by requiring a spatial and temporal coincidence
with an observed GRB, offering
a unique opportunity for
high energy neutrino detectors to observe neutrinos associated 
with individual bursts.

\subsection{Relic sources}

In recent years, the Fly's Eye atmospheric fluorescence detector and
the AGASA air shower array have convincingly detected~\cite{UHEdata}
cosmic rays with energies exceeding $\sim5\times10^{19}$~eV --- the
Greisen-Zatsepin-Kuz'min (GZK) cutoff set by interactions on the
2.7~K black body cosmic microwave background. These ultra high-energy
cosmic rays (UHECR) constitute a population distinct from those at
lower energies ($<5\times10^{18}$~eV), in having a flatter
spectrum. The
depth in the atmosphere at which the shower reaches its maximum suggests a
correlated change in the composition
from iron nuclei to protons between $10^{18}$ and
$10^{19}$~eV~\cite{UHEreview}. The lack of any detectable anisotropy
argues against a local origin in the Galactic disc where the presumed
sources of low energy cosmic rays reside. However, there are no
potential extra-galactic sources such as active galaxies near enough
(within $\sim50$~Mpc) to evade the GZK cutoff. Thus the origin of the
UHECR is a major puzzle for standard physics and astrophysics.

An exciting possibility is that UHECRs result from the decay of
massive particles, rather than being accelerated up from low
energies. The most popular models in this context are based on the
annihilation or collapse of topological defects (TDs) such as cosmic
strings or monopoles formed in the early universe~\cite{UHEtopdef}. When
TDs are destroyed, their energy is released as massive gauge and Higgs
bosons with masses of ${\cal O}(10^{25})$~eV if such defects have
formed at the GUT-symmetry breaking phase transition. The decays of
these bosons can generate cascades of high energy nucleons,
$\gamma$-rays and neutrinos. These models are constrained both by
considerations of the cosmological evolution of TDs~\cite{UHEtopevol} and
observational bounds on the extra-galactic $\gamma$-ray background
\cite{UHEtopgamma}. These require the mass of the decaying bosons to be
less than $\sim10^{21}$~eV, i.e. well below the GUT scale. Since only
GUT scale TDs have independent motivation, e.g. to provide `seeds' for
the formation of large-scale structure, the above constraint thus
disfavours TDs as the source of UHECRs.

A more recent suggestion is that UHECRs arise from the decays of
metastable relics with masses exceeding $\sim10^{21}$~eV which
constitute a fraction of the dark matter~\cite{UHEbkv,UHEbs}. Such particles
can be naturally produced with a cosmologically-interesting abundance
during re-heating following inflation~\cite{UHEpartprod}. A lifetime
exceeding the age of the universe is natural if they have only
gravitational interactions, e.g. if they are `cryptons' --- bound
states from the hidden sector of string theory~\cite{UHEben,UHEbs}. This
interpretation also naturally accounts for the required mass. A
detailed study of the fragmentation of such heavy particles has been
performed~\cite{UHEbs} in order to calculate the expected spectra of
nucleons, $\gamma$-rays and neutrinos. Such particles would, like all
`cold dark matter' particles, be strongly clustered in the Galactic
halo, i.e. within $\sim100$~kpc. Therefore, the extra-galactic contribution to
the cosmic ray flux would be negligible in comparison and the observed
flux fixes the ratio of the halo density to the lifetime. As an example,
if such particles comprise all of the halo dark matter then the
required lifetime is $\sim10^{20}$ yr, with a proportionally shorter
lifetime for a smaller contribution.

Most of the energy in the cascade ends up as neutrinos and the
predicted flux is then
$F(>E_\nu)\approx10^{8}(E_\nu/1~TeV)^{-1}$~km$^{-2}$y$^{-1}$sr$^{-1}$,
through normalisation to the observed UHECR flux. Furthermore, the
neutrinos should be well correlated in both time and arrival direction
with the UHECRs, given the relatively short propagation length in the
halo. A small departure from isotropy of ${\cal O}(20\%)$ should also
be observed, since our location is $\sim8$~kpc from the galactic
centre~\cite{UHEaniso}. This anisotropy will be less than that
for UHECRs, however, since there is no GZK cutoff to reduce the 
extra-galactic (isotropic) flux of neutrinos~\cite{UHEggs}.

The essential point is that {\em whatever} process creates the UHECRs,
it is exceedingly likely that there is a concomitant production of
very high energy neutrinos. Measurement of the neutrino flux will, at
the very least, provide important clues as to the origin of UHECRs and
may even provide dramatic evidence for new physics.

\section{Neutrino oscillations} 
The observation of atmospheric neutrinos with a neutrino telescope
provides a means to study neutrino
oscillations with a base-line length up to the order of the diameter of the
Earth.  The focus of investigation is the muon neutrino, but there is also a
very interesting possible signature for extremely high
energy ($>100$~TeV) tau neutrinos.

\subsection{Atmospheric $\nu_\mu$ oscillations}

Atmospheric neutrinos are emitted in the decay of hadrons produced by
the interactions of cosmic rays with atmospheric nuclei. The production
of electron neutrinos and of muon neutrinos is dominated by the processes
${\pi}^{\pm} \rightarrow {\mu}^{\pm} + {\nu}_{\mu}/{\overline{\nu}}_{\mu}$
followed by ${\mu}^{\pm}  \rightarrow e^{\pm} + {\overline{\nu}}_{\mu} /
{\nu}_{\mu} + {\nu}_e/{\overline{\nu}}_e$. In an infinite medium the ratio, 
$r$, of the
flux of ${\nu}_{\mu}$ and ${\overline{\nu}}_{\mu}$ to the flux of
${\nu}_{e}$ and ${\overline{\nu}}_{e}$ is expected to be two. Since 
the atmosphere is not an infinite medium,  this ratio increases with
increasing neutrino energy,
because not all high energy muons 
can decay before they are absorbed by the ground. Furthermore, the 
magnetic field of the Earth has some influence on low energy charged 
particles and this modifies the energy spectra below a few GeV.

These effects are taken into account for the calculations of the predicted
neutrino fluxes. The overall normalisation uncertainty is estimated to be
about 20\%, which is due to systematic theoretical uncertainties
in the energy spectra of the primary cosmic rays and to 
the uncertainties in their composition.
Generally, experimental results
are reported as $R = r_{\mathrm{DATA}}/r_{\mathrm{MC}}$ in order to 
cancel common systematic uncertainties thereby reducing the overall
uncertainty in R to about 5\%. 

Measurements published by underground 
experiments~\cite{kamioka,imb,superk,soudan}
show evidence of a 
deficit in the number of muon neutrinos with respect to electron neutrinos.
No anomaly was observed by the Fr\'ejus~\cite{frejus1, frejus2} and 
NUSEX~\cite{nusex} experiments.

Neutrino oscillations have been proposed as an explanation for the  
low value of the ratio $R$. With the hypothesis of two-neutrino mixing,
the oscillation probability is:

\begin{displaymath}
 P = \sin^2 2 \theta \sin^2 \left( 1.27 \frac{L}{E} 
\Delta m^2 \right)   
\end{displaymath}
where $\theta$ is the mixing angle, $L$ is the distance travelled by the 
neutrino (in km), $E$ is the neutrino 
energy (in GeV) and $\Delta m^2$ is the difference of 
the square of the masses (in eV$^2$). As the neutrinos are produced in
the atmosphere, the distance $L$ ranges between 15\,km, for vertically 
downward-going neutrinos, and almost 13\,000\,km, for vertically
upward-going neutrinos.

\begin{figure}[h] 
\begin{center}
\mbox{\epsfig{file=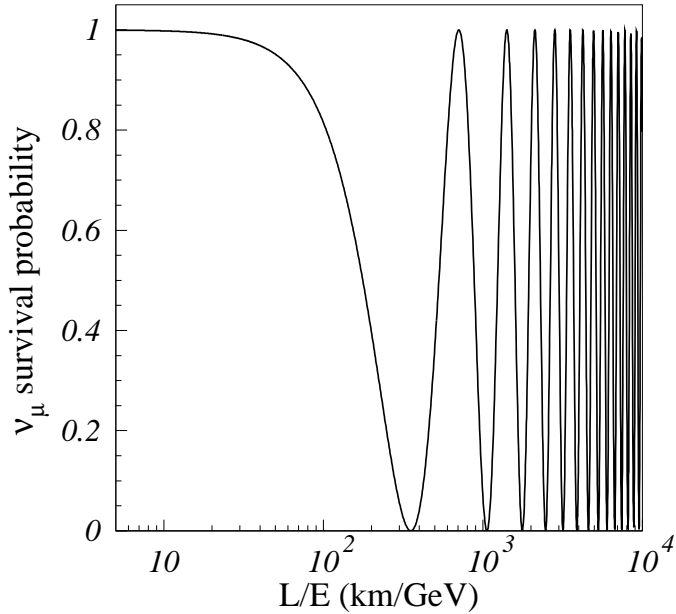,width=10cm}}
\caption{\small Variation of the neutrino survival probability 
as a function of $L/E$
for $\Delta m^2 = 3.5 \times 10^{-3}$eV$^2$ and maximal mixing.}
\label{fg:survival}
\end{center}
\end{figure}

A recent analysis~\cite{superk} reported by the Super-Kamiokande 
collaboration outlines evidence for ${\nu}_{\mu} \leftrightarrow
{\nu}_x$, where ${\nu}_x$ may be  ${\nu}_{\tau}$ or a ``sterile'' neutrino,
with $\sin^2 2 \theta > 0.82$ and 10$^{-3} < \Delta m^2 <
8 \times 10^{-3}$\,eV$^2$ at 90\% confidence level. The most probable
solution is $\Delta m^2 = 3.5 \times 10^{-3}$eV$^2$ and 
$\sin^2 2 \theta = 1.0$ (maximum mixing). Given these values, the 
survival probability $1-P$ vanishes for 
$L/E = (2n+1) \times 353$\,km GeV$^{-1}$, where $n$ is a non-negative integer. 
Figure~\ref{fg:survival} shows the variation of the  
survival probability as a function of $L/E$.


Several points remain to be clarified. First of all, there are three different
regions of $\Delta m^2$ which have been explored with solar
neutrinos, atmospheric neutrinos and short baseline beam neutrinos 
respectively, all of which
indicate evidence of non-zero $\Delta m^2$. This cannot be supported
in a three-flavour neutrino scheme. Moreover, the present result found
in the Super-Kamiokande experiment is only marginally compatible with
the result of Kamiokande~\cite{kamioka} ($\Delta m^2 = 1.8 \times 10^{-2}$)
and is in disagreement with the analysis of 
stopping upward-going muons performed by the IMB collaboration~\cite{imb-stop} 
which
excluded the region presently favoured by Super-Kamiokande on the 
$(\sin^2 2 \theta, \Delta m^2)$ diagram. The results reported by the
Fr\'ejus~\cite{frejus1} experiment exclude the region $\Delta m^2 > 4 \times 
10^{-3}$\,eV$^2$ at 90\% confidence level.

An independent measurement of the region $60<L/E<1250$\,km/GeV 
could resolve the uncertainties in the interpretation of these different
experiments. This region contains the principal oscillation (first dip)
for the $\Delta m^2$ values in question: 
$1~\times~10^{-3}~<~{\Delta}m^2~< 2~\times~10^{-2}$~eV$^2$. 
ANTARES is perfectly suited to this task. The peak sensitivity of the ANTARES
detector is very near the most probable value reported by Super-Kamiokande.
 Accurate measurement
of the position of the principal oscillation would provide a precise measurement
of $\Delta m^2$.

Determination of the neutrino survival probability 
shown in figure~\ref{fg:survival} requires a measurement of the energy of the
incident muon neutrino. For the isotropic $\nu q$ and 
$\overline{\nu}\overline{q}$
charged-current interactions, half of the neutrino energy goes to the hadron
shower. For the $\overline{\nu}q$ and $\nu\overline{q}$ interactions,
a much larger fraction of the energy goes to the muon, but these interactions
are three times less frequent than the isotropic interactions. The energy of the
hadron showers is difficult to estimate accurately, so the measurement of the
oscillation parameters in ANTARES depends principally on the measurement of the
muon momentum. 
The energy of the muon is determined by its range and
the precision depends on the
vertical spacing of the PMTs. The scale is set by the energy loss for vertical
muons between groups of PMTs: 2~GeV for a vertical spacing of 
8~m, and 4~GeV for a spacing of 16~m. Reconstruction inefficiencies
degrade the energy resolution in an energy-dependent way, depending on 
the reconstruction algorithms employed. The current status of the reconstruction
effort is described in chapter~\ref{chap:detector_performance}.

\subsection{Tau neutrinos}

Although the contribution of tau neutrinos to the atmospheric neutrino
flux is negligible, their interactions must be taken into account 
when studying neutrino oscillations. 
If oscillations of the type $\nu_\mu \rightarrow \nu_\tau$ occur, 
the charged current interactions of $\nu_\tau$ will produce charged 
$\tau$ leptons 
that can contribute to the signal observed.
Most of the hadronic and electronic $\tau$ decays
will escape detection, 
but the  muonic 
decays $\tau^- \rightarrow \mu^-\bar\nu_\mu\nu_\tau$ can be seen in the
ANTARES detector, 
and these events might be mis-identified as $\nu_\mu$ interactions.

The muonic branching ratio of the $\tau$ is 17\%. 
Furthermore, the charged-current interactions are 
considerably suppressed by the limited phase space
due to the mass of the $\tau$ (1.78 GeV). 
For very large $\Delta m^2$~(1~eV$^2$), the number of $\nu_\tau$ 
would be equal to the number of $\nu_\mu$ over the entire atmospheric flux 
($\leq$~500 GeV), but the number of $\tau^- (\tau^+)$ 
produced with energies above
10~GeV would be only 53\% (64\%) of the number of $\mu^- (\mu^+)$ because of the
limited phase space, and the contamination of the muon sample due to tau decays
would be about 9\% (11\%). 
For smaller values of $\Delta m^2$, the contamination in the region of the main
oscillation dip could be larger.

In~\cite{Halzentau} an attractive possibility for
detecting very high energy neutrinos is proposed.
The Earth is nearly transparent to low-energy neutrinos, but opaque to
neutrinos above 100 TeV.
Nonetheless, tau neutrinos well above 100 TeV can produce a signal in
ANTARES because, unlike the $e^\pm$ and $\mu^\pm$ produced 
in $\nu_e$ and $\nu_\mu$ interactions, the $\tau^\pm$ produced in 
$\nu_\tau$ interactions decay before they are absorbed, producing 
$\nu_\tau$ of lower energy which continue along the original 
$\nu_\tau$ flight path, but with decreasing 
interaction probability. 
Once the $\nu_\tau$ energy has been degraded to about 100 TeV, 
the $\nu_\tau$ can penetrate the Earth and produce an accumulation of
very-high-energy events in the detector. Such an accumulation would be a signal
for tau neutrinos. Moreover, the flux of $\nu_\tau$ from a given source would 
be constant during the Earth's rotation, whereas the flux of $\nu_\mu$ 
would vary with the sidereal day because of the change in elevation
seen from ANTARES. The variation of the $\nu_\mu$ flux 
could not be observed from AMANDA, because it is located at the South Pole.
A comparison of signals from the same sources observed at the different
elevations corresponding to AMANDA and ANTARES could lead to very exciting
results.

\section{Indirect detection of neutralinos}  

\subsection{Neutralinos as dark matter candidates}

There is extensive astrophysical evidence~\cite{cosmology} that most of the 
matter in the universe is non-luminous.  The matter content of the universe is 
normally described in terms of the density parameter $\Omega=\rho/\rho_c$, where 
$\rho_c=3H_0^2/8\pi G$ is the critical
density and  $H_0$, often 
expressed in the dimensionless form 
$h=H_0/\mbox{100 km s}^{-1}\mbox{ Mpc}^{-1}$, is the
expansion rate.
Estimates of galactic halo masses from rotation curves and
of the galactic number density give a value of $\Omega \ge 0.1$, whereas the luminous matter 
in galaxies corresponds to $\Omega < 0.01$.  Studies of the dynamics of clusters 
and superclusters of galaxies increase the required value of $\Omega$ to around 
0.2--0.3, and the popular inflationary paradigm generally requires $\Omega = 1$ 
(although this may include a contribution from a non-zero cosmological 
constant).

The density of baryonic matter is constrained by the abundances of the light 
elements to a range $0.008 \le \Omega_{\rm baryons}h^2 \le 0.024$.  With current 
measurements of $h$ tending to lie in the range 0.6--0.7, this indicates that 
much of the dark matter required on the galactic scale and beyond must be 
non-baryonic.  A prime candidate for non-baryonic dark matter is the Lightest 
Supersymmetric Particle (LSP).

Supersymmetry is a spontaneously broken symmetry between bosons and fermions, 
postulated as a natural mechanism for avoiding GUT-scale radiative corrections to 
the Higgs mass.  The minimal supersymmetric model (MSSM)~\cite{susy} contains 
boson partners 
for every fermion, fermionic partners for all known bosons, and two Higgs 
doublets.  In order to match the experimental limit on proton decay, 
it is generally assumed that a multiplicative quantum number called 
R-parity is conserved: $R=+1$ for ordinary particles and $-1$ for their 
supersymmetric partners.  The natural consequence of this is that the lightest 
supersymmetric particle is stable, having no R-conserving decay mode.

The MSSM has numerous free parameters (105 from the most general soft-breaking terms 
in addition to the 19 parameters of the Standard Model, reducing to 
seven if we make a number of standard --- but not necessarily 
correct --- assumptions about masses and mixing angles), and the identity of the 
LSP is not unambiguous.  However, experimental limits rule out a charged LSP 
over a broad mass range~\cite{LSP}, 
and it is therefore generally assumed that the LSP is 
the lightest of the four neutralinos, the mass eigenstates corresponding to the 
superpartners of the photon, the Z and the two neutral Higgs bosons.  The mixing which 
generates the mass eigenstates depends on the choice of SUSY parameters: the 
LSP, $\chi$,  can be anything from a nearly pure $B$-ino to a nearly pure 
higgsino~\cite{jungman}.  The mass of the LSP is constrained from below by 
non-detection in 
LEP2, and from above by the requirement that supersymmetry fulfil its role of 
maintaining the mass hierarchy between the GUT scale and the electroweak scale.

Neutralinos produced in the early universe will fall out of equilibrium when the 
annihilation rate, $\langle\sigma_A v\rangle n_\chi$, falls below the expansion 
rate $H$.  This condition leads to the estimate that
$$
\Omega_\chi \simeq 
\frac{3\times 10^{-27}\mbox{ cm}^3\mbox{ s}^{-1}}{\langle\sigma_A v\rangle},
$$
where $\langle\sigma_A v\rangle$ is the thermally averaged annihilation 
cross-section times the relative velocity.  (Note that neutralinos are Majorana 
particles, $\bar\chi~=~\chi$.)  If the masses of SUSY particles are close to the 
electroweak scale, the annihilation cross-section is of order $\alpha^2/(100
\mbox{ GeV})^2 \sim 10^{-25}\mbox{ cm}^3\mbox{ s}^{-1}$, indicating that 
$\Omega_\chi\sim 1$ is a realistic possibility.  This heuristic argument is 
confirmed by detailed calculations, which indicate that a neutralino LSP has a 
cosmologically significant relic abundance over large
regions of the MSSM parameter 
space~\cite{jungman}.

\subsection{Neutralino detection}

If neutralinos make up a significant fraction of the Galactic dark halo, they 
will accumulate in the core of bodies such as the Earth or the 
Sun~\cite{jungman,bottino,edsjo} and in the Centre of our
Galaxy~\cite{gondolo}.  
A neutralino passing through such a body has a 
small but non-zero probability of scattering off a nucleus
therein, so that its 
velocity after scattering is less than the escape velocity.  Once this happens, 
repeated passages through the body will generate additional scatters and the 
neutralino will sink relatively rapidly to the centre.  Equilibrium will be 
reached when the gain of neutralinos from capture is balanced by the loss from 
annihilation, $C = C_AN^2$ where $C$ is the capture rate, $N$ is the number of 
captured neutralinos. The quantity $C_A$ depends on the WIMP annihilation
cross-section and the WIMP distribution.

The capture rate $C$ depends on the local halo mass density $\rho_\chi$, the 
velocity dispersion of neutralinos in the halo $\bar v$, and the elastic 
scattering cross-section, which depends on the mass of the neutralino and on the 
effective volume and chemical composition of the Sun or Earth.  Representative 
values of $\rho_\chi$ and $\bar v$ are 0.3 GeV cm$^{-3}$ and 270 km s$^{-1}$ 
respectively.  Jungman et al.~\cite{jungman} quote capture rates of
\begin{eqnarray*}
C_\odot = 2.4\times10^{37}\mbox{ s}^{-1}\frac{\rho_\chi}{0.3\mbox{ GeV cm}^{-3}} 
f_\odot(m_\chi) f_p(\mbox{GeV}^2)^2,\\
C_\oplus = 2.4\times10^{28}\mbox{ s}^{-1}\frac{\rho_\chi}{0.3\mbox{ GeV 
cm}^{-3}} f_\oplus(m_\chi) f_p(\mbox{GeV}^2)^2,
\end{eqnarray*}
where $f_p$ is the neutralino-nucleon scalar coupling and the dependence on the 
neutralino mass is given by $f_\odot$ and $f_\oplus$ for Sun and Earth 
respectively.  For the Sun, $f_\odot$ is a smoothly varying function of mass, 
decreasing from ${\cal O}(5)$ at $m_\chi=10\mbox{ GeV}/c^2$ to ${\cal O}(0.05)$ 
at 1 TeV$/c^2$; for the Earth, resonances occur whenever the neutralino mass 
equals the mass of the target nucleus, so there is a good deal of structure in 
$f_\oplus$ below $m_\chi = 100\mbox{ GeV}/c^2$, with a main
resonance at 56~GeV ($^{56}Fe$).

The annihilation rate can be calculated~\cite{jungman, edsjo} from the 
thermally-averaged 
cross-section, $\langle\sigma_A v\rangle$, in the limit of zero 
relative velocity; at equilibrium it is simply half the capture rate.  The time 
taken to reach equilibrium, $\tau = (CC_A)^{-1/2}$, is longer for the Earth than 
for the Sun: over a wide range of MSSM parameter space, the Earth's neutralino 
population has not yet reached equilibrium, and the observed annihilation rate 
will be suppressed by some model-dependent amount compared to the equilibrium 
rate.

\subsection{Neutrinos from neutralino annihilation}

Neutralinos annihilate into a fermion-antifermion pair or into various 
two-body combinations of W, Z and Higgs bosons~\cite{jungman, edsjo}.  Direct 
decay into neutrinos is zero in the non-relativistic limit, but decays into c, b 
and t quarks, $\tau$ leptons, Z, W and Higgs can all produce a significant flux 
of high-energy neutrinos (light quarks and muon pairs do not contribute, as they 
are stopped before they decay).  The typical neutrino energy produced is thus 
around one-half to one-third of the neutralino mass, with a broad spectrum whose 
detailed features depend on the branching ratios into the different channels 
(which in turn depend on the neutralino composition---gaugino vs higgsino---as 
well as its mass) and are modified, especially in the Sun, by hadronisation and 
stopping of c and b quarks, and stopping, absorption and possibly oscillation of 
neutrinos~\cite{jungman, edsjo, ellis}. 

Since both the neutrino-nucleon 
cross-section and the range of the produced muon scale with the neutrino energy, 
the resulting muon rate is approximately proportional to $E_\nu^2\mbox{ 
d}N_\nu/\mbox{d}E_\nu$, indicating that this method of neutralino detection is 
most likely to be competitive for higher mass neutralinos. 

\begin{figure}[h]
 \begin{center}
\mbox{
\epsfig{file=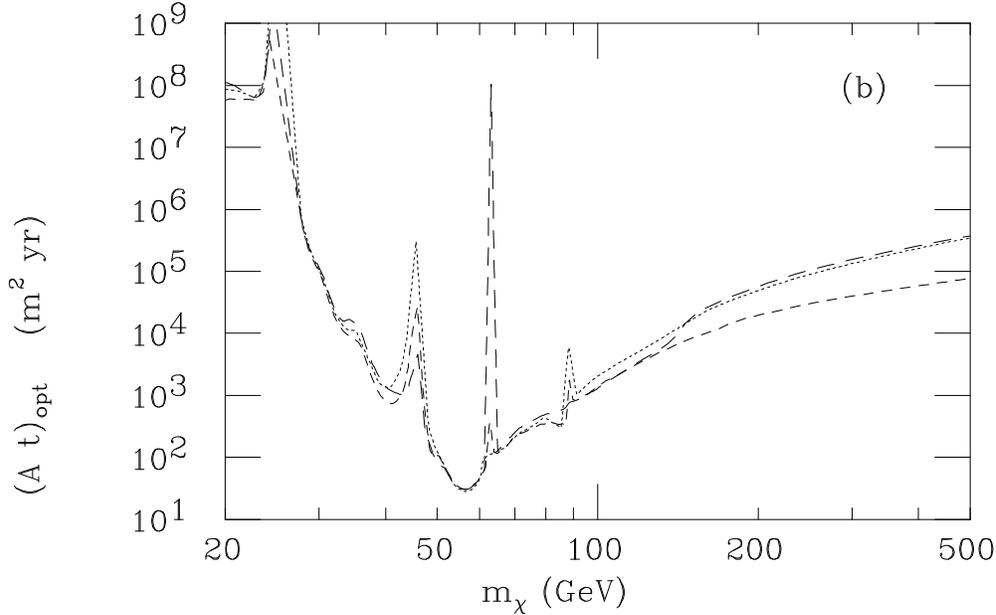,bbllx=50pt,bblly=250pt,bburx=500pt,bbury=500pt,
width=\linewidth}}
\caption{\small Exposure required to detect a neutralino signal. 
  The figure is taken
  from Bottino et al.~\cite{bottino} and the 3 curves
  correspond to different choices of supersymmetry
  parameters.}
\label{fig:bottinofig}
\end{center}
\end{figure}

The expected muon flux depends on the neutralino mass and the assumed MSSM 
parameters.  The results of representative calculations~\cite{bottino} 
are shown in figure~\ref{fig:bottinofig}.  
In these models, the flux from the Sun exceeds that from the 
Earth at high neutralino masses (in others, e.g.~\cite{edsjo}, 
the solar flux is 
larger throughout).  
The exposure required to detect a signal (defined in~\cite{bottino} 
as a 4$\sigma$ effect above background
with at least 4 events) is of the order of 
$10^4$ to $10^5$~m$^2\cdot$yr.

\begin{figure}[h]
 \begin{center}
 \mbox{
\epsfig{file=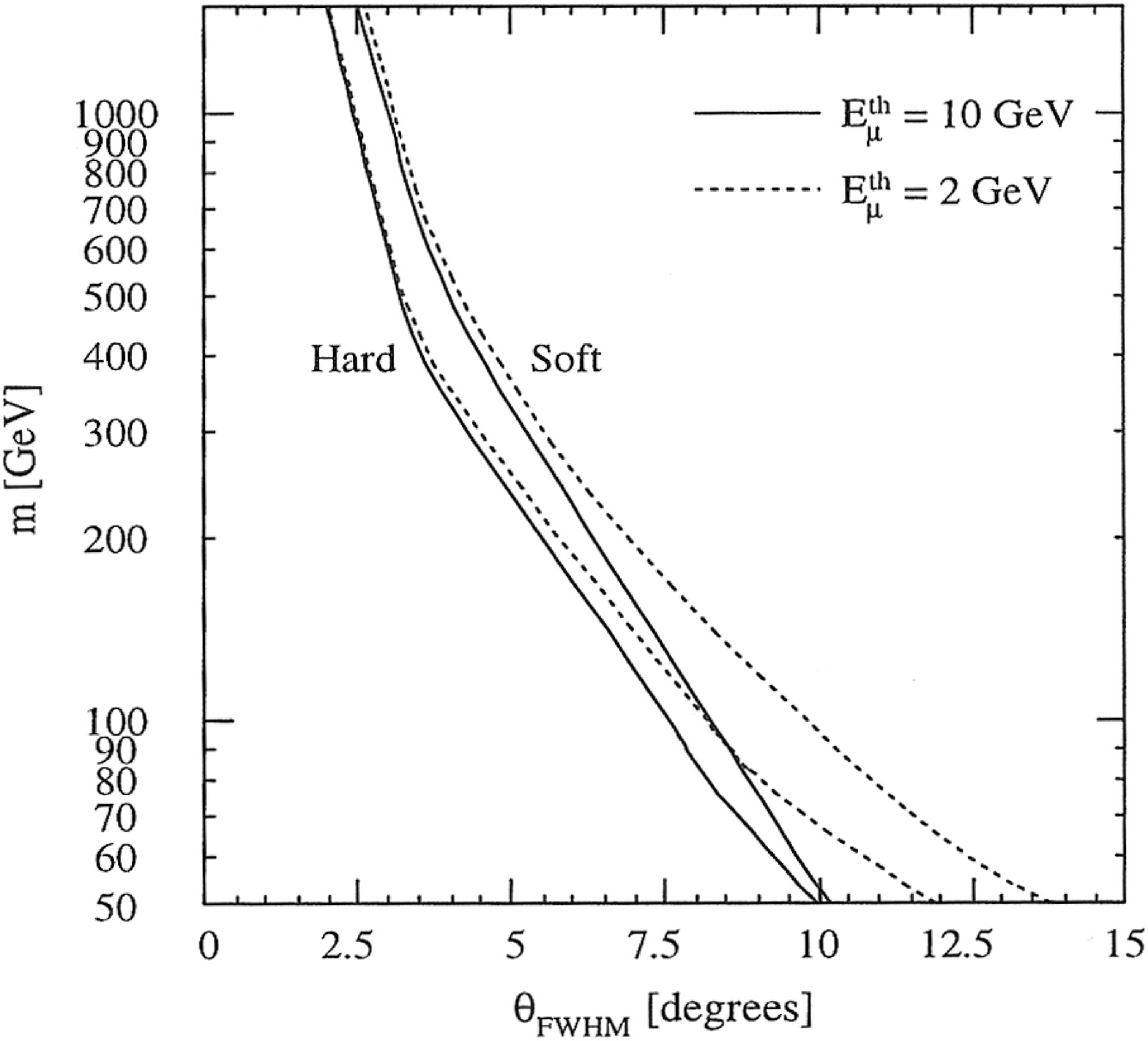, bbllx=0pt,bblly=0pt,bburx=1192pt,bbury=940pt,
width=\linewidth}}
\caption{\small Mass of the neutralino versus expected angle for different
supersymmetry parameters. The figure is taken 
from Edsj$\ddot{o}$~\cite{edsjo}.}
\label{fig:edsjofig}
\end{center}
\end{figure}

The Sun is a point-like source of neutrinos from neutralino annihilation, but the 
Earth is not, especially for lower mass neutralinos.  Despite the smearing 
produced by measuring the muon flux rather than the neutrinos, a detector with 
good angular resolution would be able to use the observed angular distribution 
to constrain the neutralino mass, as shown in figure~\ref{fig:edsjofig} 
from~\cite{edsjo}.  

Indirect detection explores a region of parameter space which is somewhat 
different from that investigated by direct techniques, although there is also 
considerable overlap. Indirect detection is particularly useful when the axial coupling of the 
neutralino dominates~\cite{jungman}, as axially-coupled neutralinos are captured in 
the Sun due to their interaction with hydrogen. Detection of 
neutralinos by both direct and indirect techniques would provide information on 
neutralino couplings and MSSM parameters, as would indirect detection of signals 
from both the Sun and the Earth.

%

\section{Other exotic phenomena}

  \subsection{GUT monopoles}
  
 Grand Unified Theories (GUT) of the electroweak and strong interactions 
 predict the existence of massive magnetic monopoles. It is assumed that they
 were produced shortly after the Big Bang and would have cooled down to very low
 velocities ($\beta \leq 10^{-3}$). They may be at the
 origin of baryon number violating processes~\cite{Rubakov} and have been searched for 
 in proton decay experiments. 
 
 The best limits to date are provided by the MACRO~\cite{MACROMon} and 
 BAKSAN~\cite{BAKSANMon} experiments.
In the  region $10^{-4} \leq \beta \leq 10^{-3}$, they have achieved sensitivity
below the Parker bound~\cite{Parkerbound}. 
Baikal has obtained limits which could be comparable with a limited number of
photomultipliers (36 and 192)~\cite{BaikalMon}. 
A 1000 PMT detector could bring appreciable
improvement on the results. 
ANTARES is studying a special trigger similar to that of Baikal. The slow
moving monopole produces sequential Cherenkov flashes along its track via the
proton decay products. The trigger is based on counting rate excesses 
during short time periods of 100 microseconds to 1~ms. 

\subsection{Relativistic monopoles}

According to~\cite{Weiler}, the Ultra High Energy Cosmic Ray Events 
above 10$^{20}$~eV may be relativistic monopoles. In this case, the basic
mechanism for light generation is Cherenkov radiation. 
The intensity of the light due to relativistic monopoles is giant, similar 
to that of PeV muons, and they can be obtained with the muon trigger. 
Preliminary estimates~\cite{BaikalMon} show that they can be distinguished
from high energy muons, down to monopole velocities $\beta =0.1$.    

\subsection{SUSY Q-balls}

 Supersymmetry predicts the existence of non-topological solitons, often called
 Q-balls. It has been realized that solitons with a large number of baryons
 are stable and can be produced copiously in the early Universe, which means that
  they are an interesting dark matter candidate~\cite{Qballs}.
 The signature and rate of occurrence  of a Q-ball through a detector
 depend on the parameters of the theory. In the case of large baryon numbers of
 the order of $10^{24}$, emissions of the order
 of 10 GeV/mm are possible, a spectacular signature, similar
 to those expected from monopole searches~\cite{BaikalQB}.

\subsection{New phenomena} 

Whenever a new window is opened on the Universe, unexpected
phenomena are observed. This has been verified in
numerous cases throughout history, 
from the observation of the moons of Jupiter by the first Galilean telescopes to 
the discovery of pulsars with radio-based astronomy. Neutrino astronomy
provides an exciting new probe of the Universe to the highest possible red-shifts
in an energy window so far not observable by other
techniques.
There is plenty of scope for surprises.


%% file: chap3.tex

\begin{figure}[h]
 \begin{center}
  \mbox{
     \epsfig{file=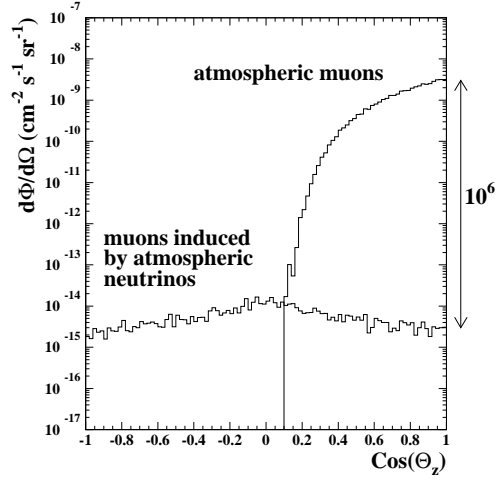,width=0.5\linewidth }}
\caption{\small Zenith angular distribution of the muon flux
above 1~TeV from atmospheric muons and atmospheric neutrino induced 
muons at 2300~m water equivalent depth.}
\label{fig:fluxmu_angle}
\end{center}
\end{figure}

Since the Earth acts as a shield against all particles 
except neutrinos, a neutrino telescope uses the detection of 
upward-going muons as a signature of muon neutrino interactions
in the matter below the detector.
 The muon detection medium may be a natural body of water
or ice through which the muon emits Cherenkov light. Its detection 
allows the determination of the muon trajectory. 
 This detection technique requires discriminating upward going muons against
the much higher flux of downward atmospheric muons 
(figure~\ref{fig:fluxmu_angle}). 
To simplify the discrimination, the detector should be 
installed in a deep site 
where a layer of water or ice would shield it.

In order to correlate the measured muon spectrum with the
original neutrino spectrum, it is necessary to understand the
dynamics of neutrino interactions, the opacity of the
Earth, the energy loss of muons and the resolution of the detector
over a wide range of angles and energies.

\section{Neutrino interactions}

The inclusive deep inelastic charged current cross-section for 
$\nu_l + N \rightarrow l^- + X$ (where the lepton mass is neglected)
is given by \cite{ghandi} 
\begin{displaymath}
\frac{d^2\sigma_{\nu N}}{dxdy} = 
\frac{2G^2_Fm_NE_\nu}{\pi}\frac{M^4_W}{(Q^2+M^2_W)^2}
 \left[ xq(x,Q^2)+x(1-y)^2\bar q(x,Q^2)\right] 
\end{displaymath}

\noindent 
where $G_F$ is the Fermi constant, $m_N$ and 
$M_W$ are the nucleon and $W$-boson masses, and
$Q^2$ is the square of the momentum transfer between the neutrino
and muon.
The Bjorken variables $x$ and $y$ are 
$x = Q^2/2m_N\nu$ and $y = (E_\nu-E_l)/E_\nu$, where
$\nu=E_\nu-E_l$ is the lepton energy loss in the laboratory frame.

\begin{figure}
\begin{center}
\mbox{\epsfig{file=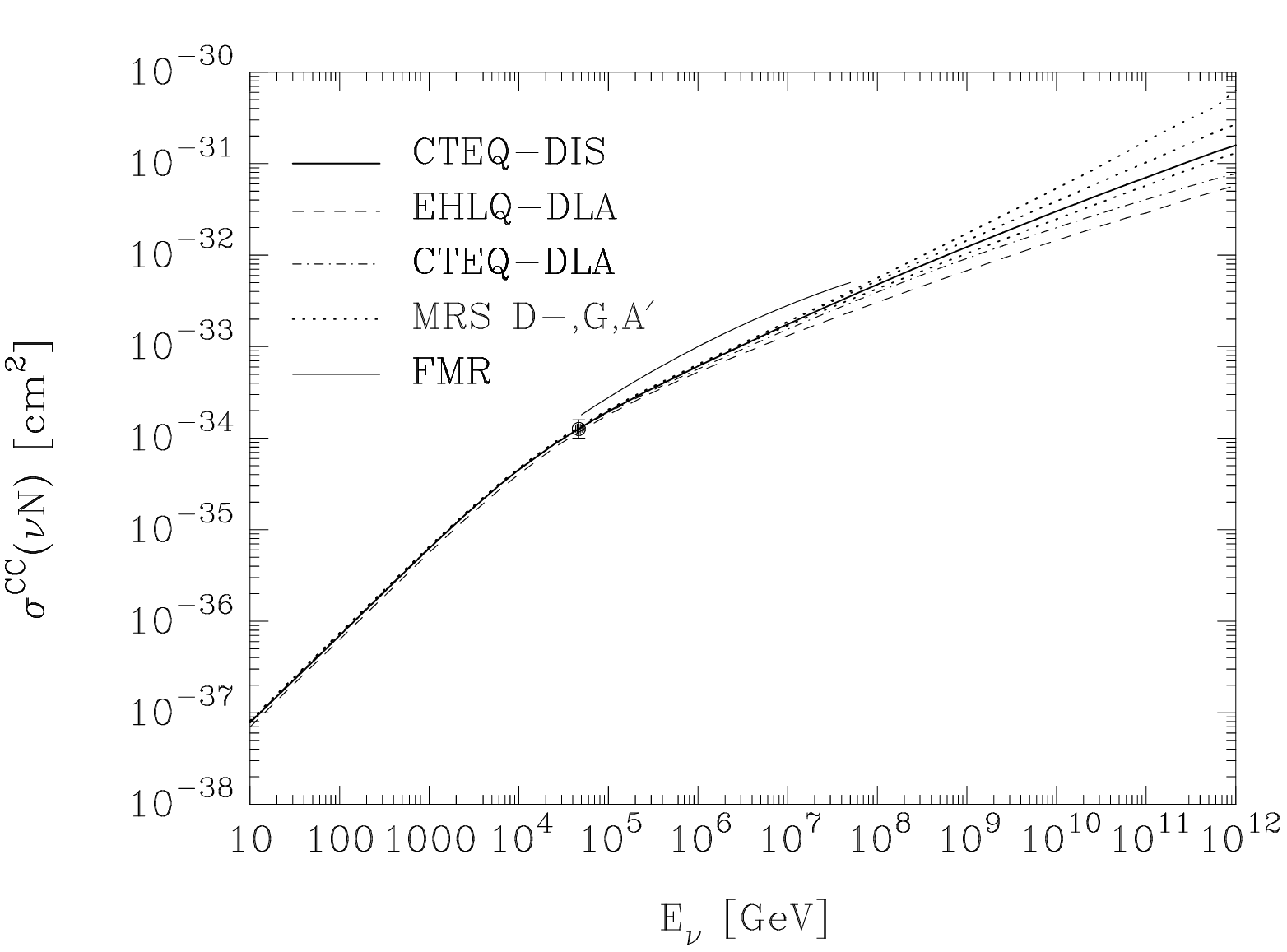,width=9cm}}
\caption{\label{h1zeus}\small Average charged-current 
cross-section for 
$\nu$-N interactions for different sets of parton
distribution functions. The data point corresponds to the average
of the measurements by H1 and ZEUS at HERA 
(taken from \protect \cite{ghandi}).}
\end{center}
\end{figure} 

\subsection{Interactions at low energies}

At energies such that $E_\nu \ll M^2_W/2m_N \approx 5$
TeV, $Q^2$ can be neglected in the $W$ propagator.
In this case the average deep-inelastic $\nu N$ cross-sections grow linearly 
with the neutrino energy
(see figure~\ref{h1zeus}):
\begin{displaymath}
\begin{array}{l}
\sigma_{DIS}(\nu N) \simeq 0.67\times 10^{-38} E_\nu[\mathrm{GeV}]~\mathrm{cm}^2 \ \\
\sigma_{DIS}(\bar\nu N) \simeq 0.34\times 10^{-38} 
E_\nu[\mathrm{GeV}]~\mathrm{cm}^2\ \\
\end{array}
\end{displaymath}

At even lower energies ($E_\nu <$ 100~GeV),
quasi-elastic and resonant contributions 
to the charged-current interactions have to be considered.
In this case $Q^2$ has to be small enough 
to allow a coherent interaction with the
complete target nucleon,
so these cross-sections are
essentially constant with energy. 
A fit to data above 10~GeV gives the following sums 
for the quasi-elastic and resonant
production processes \cite{rn}:

\begin{displaymath}\begin{array}{l} 
\sigma_{QE}(\nu_\mu N) + \sigma_{res}(\nu_\mu N) = 1.50\times 
10^{-38}\,\mathrm{cm}^2 \\
\sigma_{QE}(\bar\nu_\mu N) + \sigma_{res}(\bar\nu_\mu N) = 1.58\times 
10^{-38}\,\mathrm{cm}^2
\end{array}
\end{displaymath}

\noindent
Between 10~GeV and 500~GeV the vertical $\nu_\mu$ flux is approximately
proportional to 1/$E_{\nu}^3$.
%
%
Integrating the cross-sections given above over this flux 
for $E_\nu > 10$~GeV results in a
quasi-elastic and resonant contribution of 11\% for $\nu_\mu$ and
21\% for $\bar\nu_\mu$ interactions.

\subsection{Interactions at high energies}

At energies such that $E_\nu \gg M^2_W/2m_N \approx 5$
TeV, quasi-elastic and resonant contributions are completely negligible.
On the other hand, the $W$ propagator limits the growth of $Q^2$ to 
$\langle Q^2\rangle \sim M^2_W$ and so the cross-section is dominated by 
the behaviour of distribution functions at small 
$x$.

The H1 and ZEUS collaborations at HERA measured the proton structure
function $F_2(x,Q^2)$ via charged current $e-p$ scattering, for $Q^2$
in the range from 1.5 to 5000 GeV$^2$ 
with $x$ down to $3 \times 10^{-5}$ at
$Q^2 = 1.5$ GeV$^2$ and $x$ down to $2 \times 10^{-2}$ 
at $Q^2 = 5000$ GeV$^2$
\cite{H1ZEUS}.

These measurements can be translated into a neutrino-nucleon interaction
cross-section at $E_\nu \simeq 50$ TeV and can also be used as a guide
to extrapolate the parton densities beyond the measured ranges in
$x$ and $Q^2$ to those required for higher neutrino energies. 
Figure \ref{h1zeus}
shows the behaviour of the average $\nu N$ cross-section
for different sets of parton distribution
functions. At very high energy, the cross-section calculated 
with the new parton distribution functions
(CTEQ3-DIS~\cite{cteq})
is more than a factor of 2 larger than previous
estimates.

\subsection{Different types of $\nu$ interactions in ANTARES}

Charged-current $\nu_e$ interactions 
give rise to electromagnetic and hadronic 
showers with longitudinal dimensions of no more than a few metres, 
because the radiation length and the 
nuclear interaction length 
of water are below 1~m. On the scale of ANTARES, these are nearly 
point-like events. 
At energies above 100 GeV, 
the energy resolution of these events is expected to be better than
for muonic events because they leave all of their energy inside the
detector volume. On the other hand, 
their angular resolution will be poor 
compared to muonic events, 
due to the point-like character of the showers.

Charged-current $\nu_e$ interactions will be contaminated by 
neutral-current interactions of both $\nu_e$ and $\nu_\mu$
(and $\nu_\tau$, if present).
The number of neutral-current interactions is about 1/3 of the number of
charged-current interactions. The neutrino type is not identified in the
neutral-current interactions, the energy resolution is poor due to the missing
final-state neutrino, and the angular resolution is poor due to the point-like
character.

Charged-current $\nu_\mu$ interactions produce $\mu^\pm$ 
leptons as well as a point-like hadronic shower. 
The $\nu_\mu$ energy can be estimated from the measured $\mu^\pm$ energy.
In $\nu_\mu d \rightarrow \mu^- u$ interactions, 
the average $\mu^-$ energy is 1/2 of the $\nu_\mu$ energy;
in $\bar\nu_\mu u \rightarrow \mu^+ d$ interactions, 
the average $\mu^+$ energy is 3/4 of the $\bar\nu_\mu$ energy.
The $\mu^\pm$ energy can be determined from the range 
for $E_\mu < 100$~GeV, or from $dE/dx$ for $E_\mu > 1$~TeV
(see below).
For $\nu_\mu$ interactions inside the detector, additional information on the
$\nu_\mu$ energy is available from the hadronic shower.
The ANTARES detector is designed for the detection of these 
charged-current $\nu_\mu$ interactions.

Charged-current $\nu_\tau$ interactions produce $\tau^\pm$ leptons with
electronic, muonic and hadronic decay modes. 
The  $\nu_\tau$ interaction vertex and the $\tau^\pm$ decay vertex
cannot be separated for energies below $\sim$100~TeV. 
The electronic and hadronic modes will look like $\nu_e$ charged-current or
neutral-current interactions. 
The muonic decays $\tau^- \rightarrow \mu^- \bar\nu_\mu \nu_\tau$,
with branching ratio 17\%, will be visible in ANTARES,
but they cannot be distinguished from $\nu_\mu$ interactions.

\section{Cherenkov light emission}

Charged particles emit light under a characteristic angle 
when passing through a medium if their
velocity exceeds the speed of light in the medium. The Cherenkov
angle $\theta$ is related to the particle velocity $\beta$
and the refractive index of the medium $n$:
\begin{equation}
\nonumber
\cos{\theta} = 1/n\beta
\end{equation}
In the energy
range interesting for ANTARES ($E > 10$~GeV), particles will generally be
ultra-relativistic with $\beta=1$. The refractive index of sea water
is $n=1.35$ for a wavelength of 450~nm therefore the Cherenkov light
is emitted under $42^\circ$ for this wavelength.
This easy geometrical pattern of light emission allows a precise 
reconstruction of tracks from the measurement of only few hits
at different space points.

The number of photons produced along a flight path $dx$ in a wave length
bin $d\lambda$ for a particle carrying unit charge is 
\begin{equation}
\nonumber
\frac{d^2N}{d\lambda dx} = 2\pi \alpha \sin^2\theta / \lambda^2
\end{equation}
At wavelengths of 400-500~nm the efficiency of the photomultipliers 
as well as the transparency of the water are maximal.
Within 1~cm flight path 100 photons are emitted
in this wavelength bin. Between 285-400~nm twice as many photons
are emitted, however they contribute less to the detected signal. 
At a perpendicular distance of 40~m from a charged track the 
density of photons between 400-500~nm is still 1 per 340~cm$^2$, 
neglecting absorption and scattering effects. The effective area of the
photomultipliers being considered is in the same 
range (300-500~cm$^2$). This gives an indication of the active
detector volume around each photomultiplier.

For $\beta=1$ the Cherenkov light yield is independent of 
the energy of the charged particle.
This means the
light output of a single particle does not allow its energy to be measured.
However when hadronic or electromagnetic showers are produced
(which might occur at the neutrino vertex as well as for radiative 
processes along a muon track) the total light yield of the shower will
be proportional to the total track length in the shower and therefore
to its initial energy. This allows some calorimetric measurements if
the neutrino vertex is inside the active detector volume or for
muon tracks above 1~TeV where radiative processes dominate its energy 
loss.

\section{Light propagation in sea water}

\begin{figure}[htp]
\begin{center}
\mbox{\epsfig{file=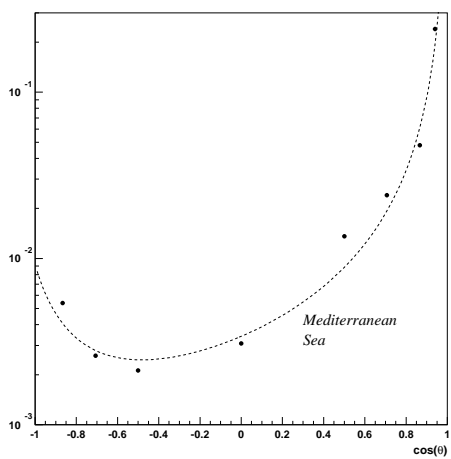,width=0.5\linewidth}}
\end{center}
\caption{\small The angular distribution of scattering in the
deep Mediterranean Sea~\cite{mobley}.}
\label{fig:dis}
\end{figure}

The processes of absorption and 
scattering characterise the transmission of light in water.  They are
parametrised by the absorption length $\lambda_{a}$, the scattering
length $\lambda_{s}$, and the scattering function $\beta(\theta)$
which describes the angular distribution of the scattering~\cite{mobley}. 
The relevant window of
wavelengths for a sea water Cherenkov detector is centred on blue light. Deep
sea water transparency is maximal in the blue, with typical values of 60 m
for $\lambda_{a}$ and $\lambda_{s}$, and a scattering function peaked in the
forward direction with an average value for the cosine of the scattering
angle $\langle \cos(\theta)\rangle  \simeq 0.9$, as shown in figure
\ref{fig:dis} taken from \cite{mobley}. Seasonal variations are expected to affect
these values, especially the scattering parameters which are governed
by the amount of suspended particulate matter. 
In section \ref{sect:trans_data} 
we present results of {\it in situ} measurements of optical properties at the
ANTARES site.

\section{Detector response}

The physical processes involved in neutrino and muon interactions place
limits on the angular and energy resolution possible with a neutrino 
telescope.  These limits must be taken into consideration when 
optimising the detector design.

\subsection{Angular response for $\nu_\mu$ interactions}

The angular response of the detector with respect to the
incoming neutrino direction is crucial for the 
identification of point sources of neutrinos. 
Three factors determine this response:
the angle between the neutrino and the muon in the neutrino
interaction, the deviation of the muon direction
due to multiple scattering 
and the angular resolution of the detector with respect to the muon.

\begin{figure}[h]
 \begin{center}
  \mbox{
     \epsfig{file=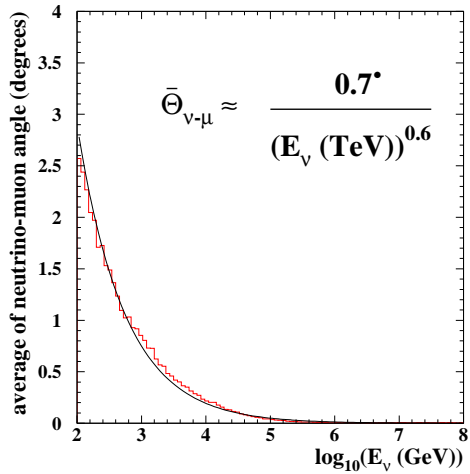,width=0.5\linewidth }}
\caption{\small Angular difference between the 
initial neutrino direction and that
of the muon track at the detector; the functional form shown reproduces 
the observed energy dependence well (solid curve).}
\label{fig:Dangle_Enu}
\end{center}
\end{figure}

The effect of the first two factors is illustrated in 
figure~\ref{fig:Dangle_Enu}.  At 1~TeV the
average difference between the $\nu$ direction and the $\mu$ 
is about $0.7^\circ$.
The difference decreases with increasing $\nu$ energy.

The detector resolution will be determined by the quality of the
alignment of the detector components, the time resolution of the
photomultipliers, the global timing of the readout system 
and the quality of the muon reconstruction.
The reconstruction will be affected 
by light coming from secondary particles
and by scattered light. Monte Carlo studies show that an angular
resolution of $0.1^\circ$ is possible. This means that
above 100~TeV the total angular resolution is dominated by detector
effects whereas below 10~TeV the resolution is dominated by the
angular distribution of the neutrino interactions.

\subsection{Energy response for $\nu_\mu$ interactions}

The energy response is determined by 
the energy fraction transferred to the muon
in the neutrino interaction, the energy lost by the muon outside
the detector and the energy resolution of the detector.
The muon energy determination requires different techniques in
different energy ranges.

Below 100~GeV, the muons are close to minimum-ionizing, 
and the energy of contained events, with start and end points
measured inside the detector,
can be determined accurately from the range.
The threshold for this method is about 5-10 GeV 
for vertical tracks, depending on the vertical distance between groups
of optical modules, and about 15 GeV for more isotropic events, 
depending on the horizontal distance between lines.

Above 100~GeV, the range cannot be measured 
because of the limited size of the detector, but the visible range 
determines a minimum energy that 
can be used for the analysis of partially-contained events:
starting events in which the vertex point is measured inside the detector, 
and stopping events in which the endpoint is measured.

\begin{figure}[p]
 \begin{center}
  \mbox{
     \epsfig{file=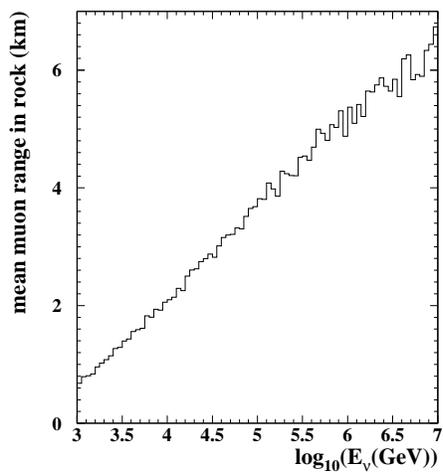,width=0.5\linewidth }}
\caption{\small Average range of the muon in standard rock as a function of the 
initial neutrino energy.}
\label{fig:Rmu_Enu}
\end{center}
\end{figure}

\begin{figure}[p]
 \begin{center}
  \mbox{
     \epsfig{file=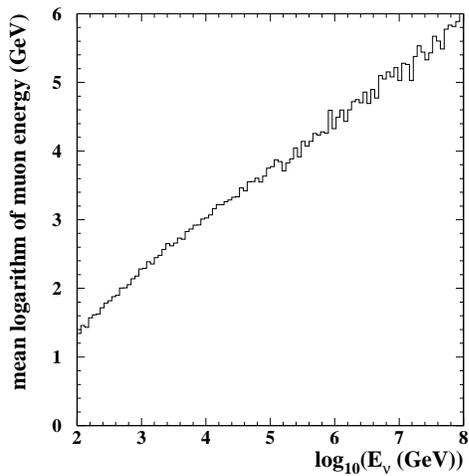,width=0.5\linewidth }}
\caption{\small Muon energy at the detector as a 
function of the parent neutrino
energy.}
\label{fig:fracmu_Enu}
\end{center}
\end{figure}

Above 1~TeV, stochastic processes (bremsstrahlung, pair production,
$\delta$-rays) are dominant, and
the muon energy loss becomes proportional
to the energy. 
The muon range above 1~TeV increases only logarithmically with the muon
energy (figure~\ref{fig:Rmu_Enu}).
On the other hand, the detection efficiency increases with energy
because of the additional energy loss.
The correlation between measured muon energy and
neutrino energy is shown in figure~\ref{fig:fracmu_Enu}.
Monte Carlo studies have shown that the neutrino energy can
be determined within a factor 3 above 1~TeV from the average energy loss.

Above 1 PeV, the Earth becomes opaque to upward-going vertical neutrinos.
Higher energies are accessible closer to the horizon, however.
Very high-energy tau neutrinos can be observed because the 
$\tau^\pm$ produced in $\nu_\tau$ interactions decay 
before they are absorbed, producing $\nu_\tau$ of lower energy 
which continue along the original $\nu_\tau$ flight path, 
but with decreasing interaction probability, 
resulting in an accumulation of events 
at the highest detectable energies.

\section{Observable sky}

\begin{figure}[h]
 \begin{center}
  \mbox{
\epsfig{file=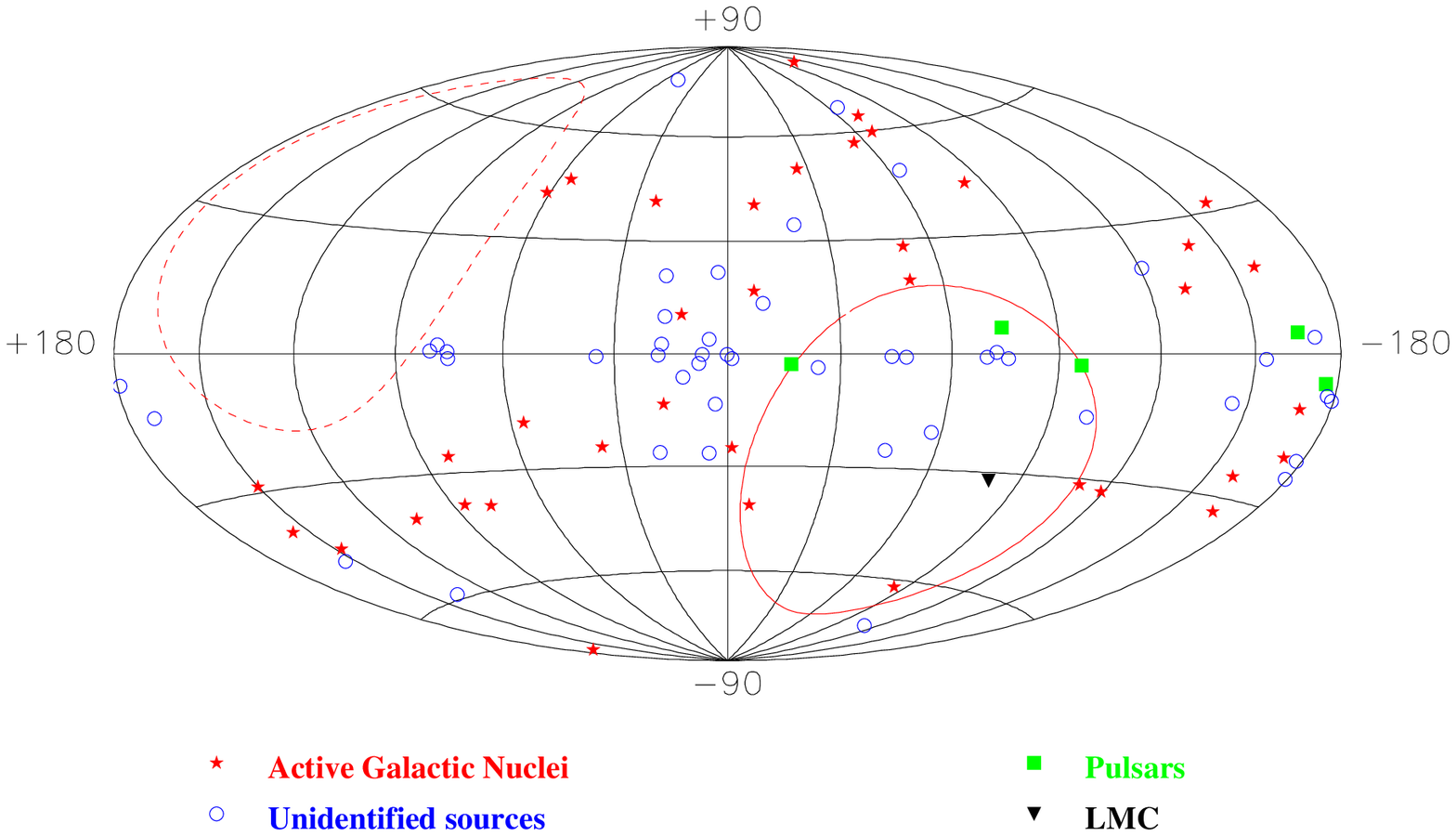,bbllx=30pt,bblly=230pt,bburx=570pt,bbury=580pt,width=\linewidth}}
\caption{\small Visible sky in Galactic coordinates for a 
detector situated at $43^\circ$ North latitude.  The area within the dashed 
line is not observable; the area within the solid line is observable 24 hours
a day.  The sources shown are from the EGRET catalogue.}
\label{fig:sky}
\end{center}
\end{figure}

The ANTARES neutrino telescope, 
situated at a latitude of $43^\circ$ North, can observe 
upward-going neutrinos from
most of the sky (about $3.5\pi$~sr), due to the rotation of the Earth. 
Declinations below $-47^\circ$ are always visible,
while those above $+47^\circ$ are never visible. 
Declinations between
$-47^\circ$ and $+47^\circ$ are visible
for part of the sidereal day (figure~\ref{fig:sky}). 
Most of the Galactic plane is visible, 
and the Galactic centre is
visible most of the  sidereal day. 
Since the AMANDA
telescope at the South  pole
is sensitive to positive declinations, 
the two detectors will have a reasonable area 
in common for cross-checks (about $1.5\pi$~sr). 

\begin{figure}[h]
 \begin{center}
  \mbox{
     \epsfig{file=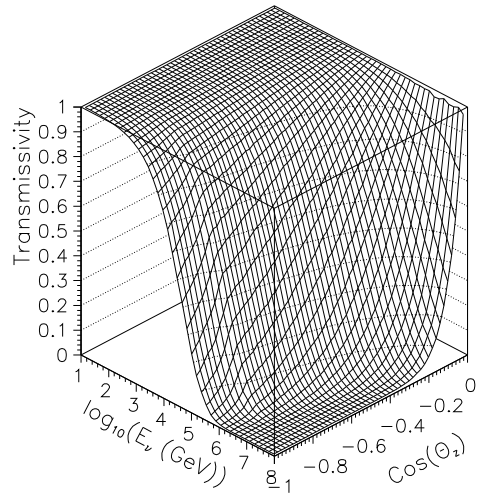,width=0.5\linewidth }}
\caption{\small Transmittivity of the Earth as a function of incoming neutrino
energy and zenith angle.}
\label{fig:transm}
\end{center}
\end{figure}

At energies greater than $\simeq 40$~TeV, the interaction
length becomes smaller than the Earth's diameter for $\nu_\mu$
traversing the dense core of the Earth. Above 10~PeV, only nearly
horizontal $\nu_\mu$ are visible~(see figure~\ref{fig:transm}).
If the field of view can be extended to $10^\circ$ above the
horizon at these energies where
the background is greatly diminished, a non-negligible fraction
of the sky can be kept observable even at these energies.

Like other underground detectors,
neutrino telescopes can observe the sky 
independently of the time of day, the phases of the moon
or the weather. 
Existing underground experiments have usually reached 80\% duty cycle 
after the initial debugging phase. ANTARES aims for
even higher values for the off-shore facilities because
the access is complicated and time-consuming.


%
%

%% file: chap4.tex



In order to ensure the success of the deployment of a large-scale detector
in an uncontrollable environment such as the deep sea, it is necessary
to perform an extensive programme of site evaluation and prototype testing.

The selection of a suitable site for a neutrino telescope
requires consideration of
water transparency, optical background, fouling of optical surfaces,
strength of the deep sea currents, meteorological conditions,
depth, on-shore support, infrastructure and pier availability.
This chapter reports on the results of site evaluation studies for the
proposed site and on initial experience of prototype string construction and 
deployment.

\section{Site evaluation mooring lines}

\begin{figure}[tb]
\begin{center}
\mbox{\epsfig{file=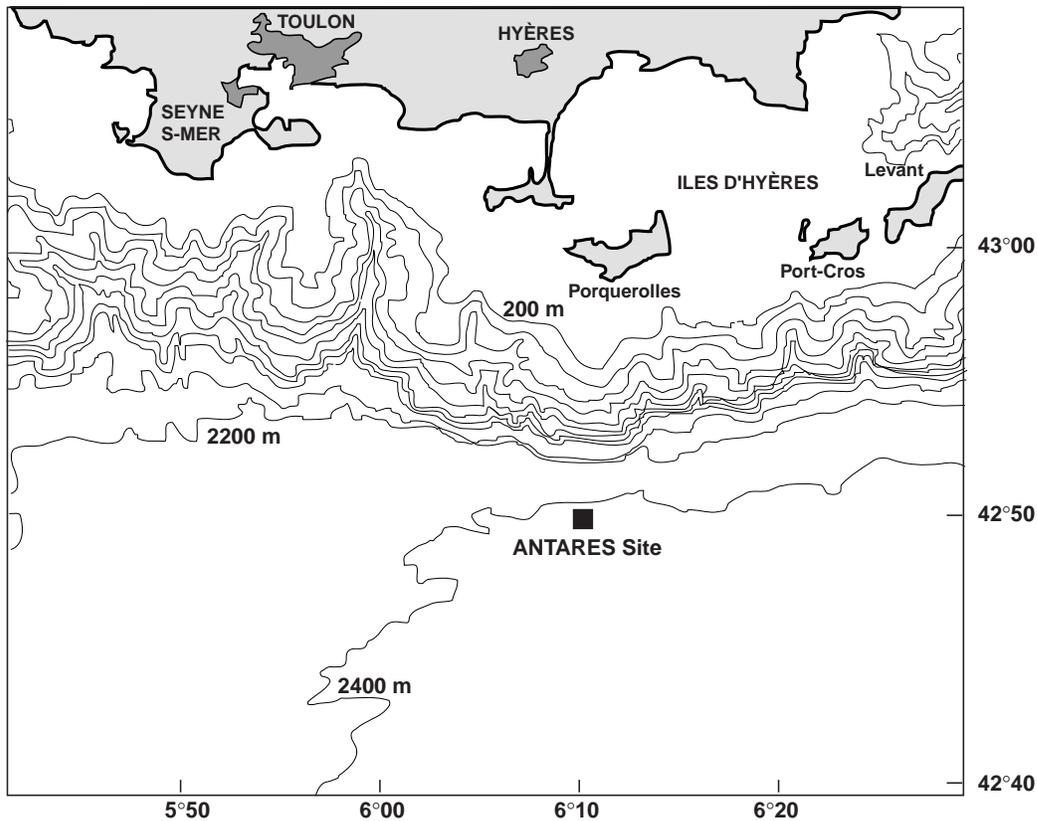,width=\linewidth}}
\end{center}
\caption{\small Map of the ANTARES site near Toulon.}
\label{fig:toulon}
\end{figure}

A detailed programme of {\it in situ} measurements has been undertaken
since October~1996,
most of the data being taken at a site near Toulon 
(42$^{\circ}$50'~N, 6$^{\circ}$10'~E)
at a depth of 2400~m (figure~\ref{fig:toulon}). Optical background data were also taken 
20~nautical miles off Porto, Corsica (42$^{\circ}$22'~N,
8$^{\circ}$15'~E) at a depth of 2700~m.

Autonomous mooring 
lines to measure the site parameters have been developed and
about 20~deployments and recoveries have been successfully performed.
\begin{figure}[h]
\begin{center}
 \begin{tabular}{ccc}
  \subfigure{\epsfig{file=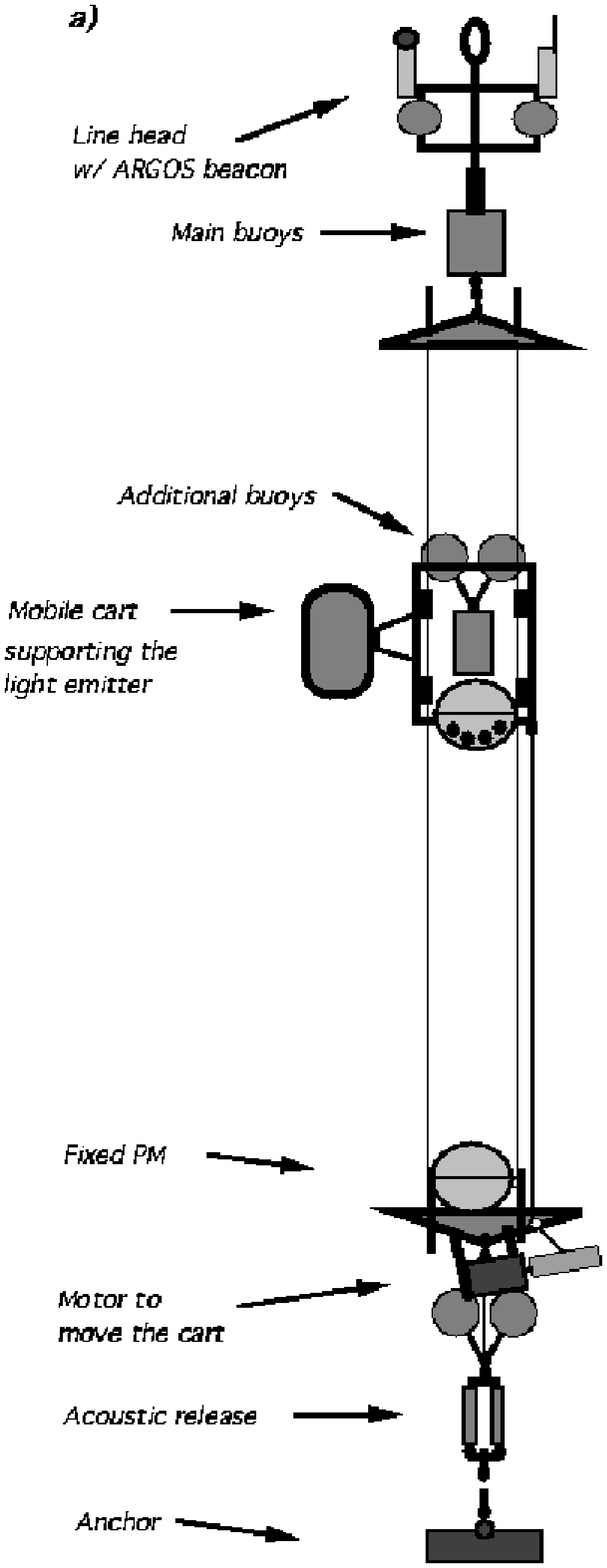,width=0.3\linewidth}}
 &
  \subfigure{\epsfig{file=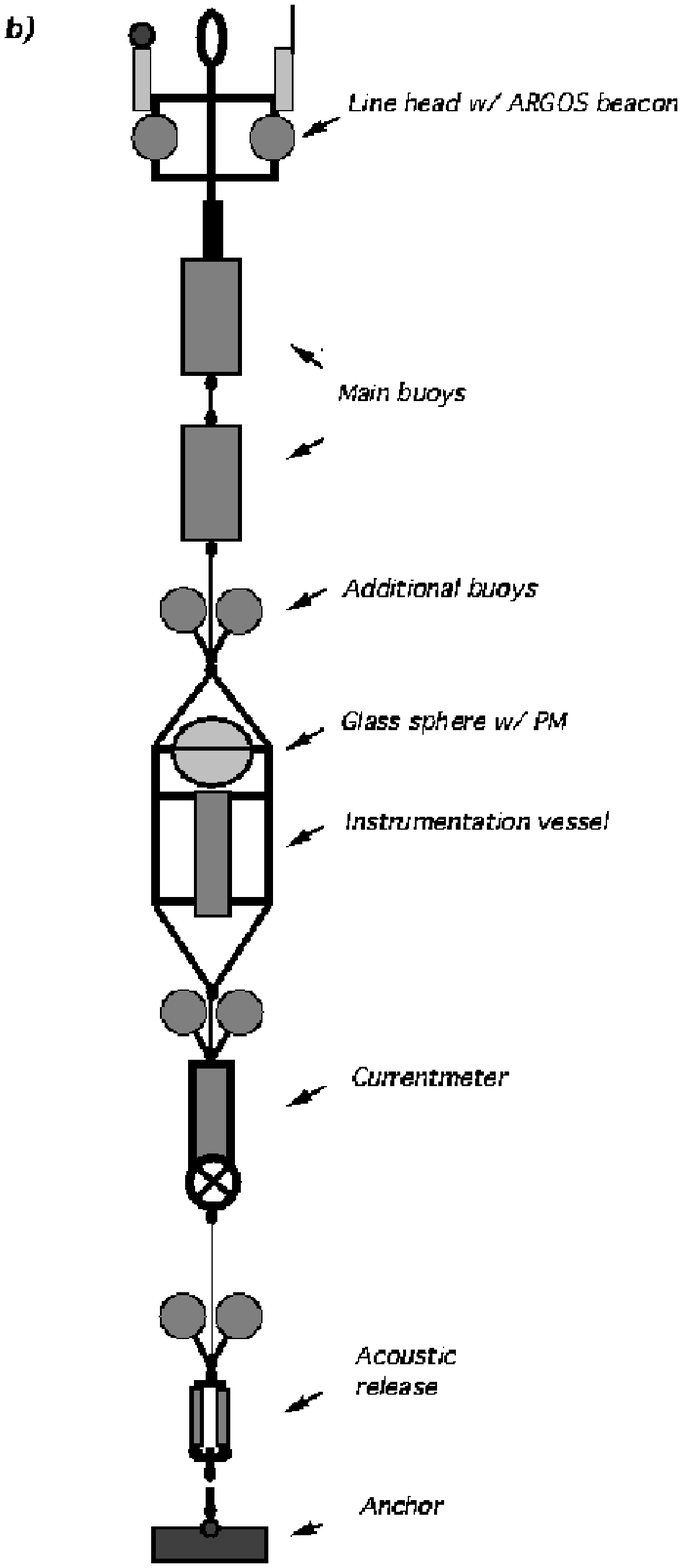,width=0.3\linewidth}}
 &
  \subfigure{\epsfig{file=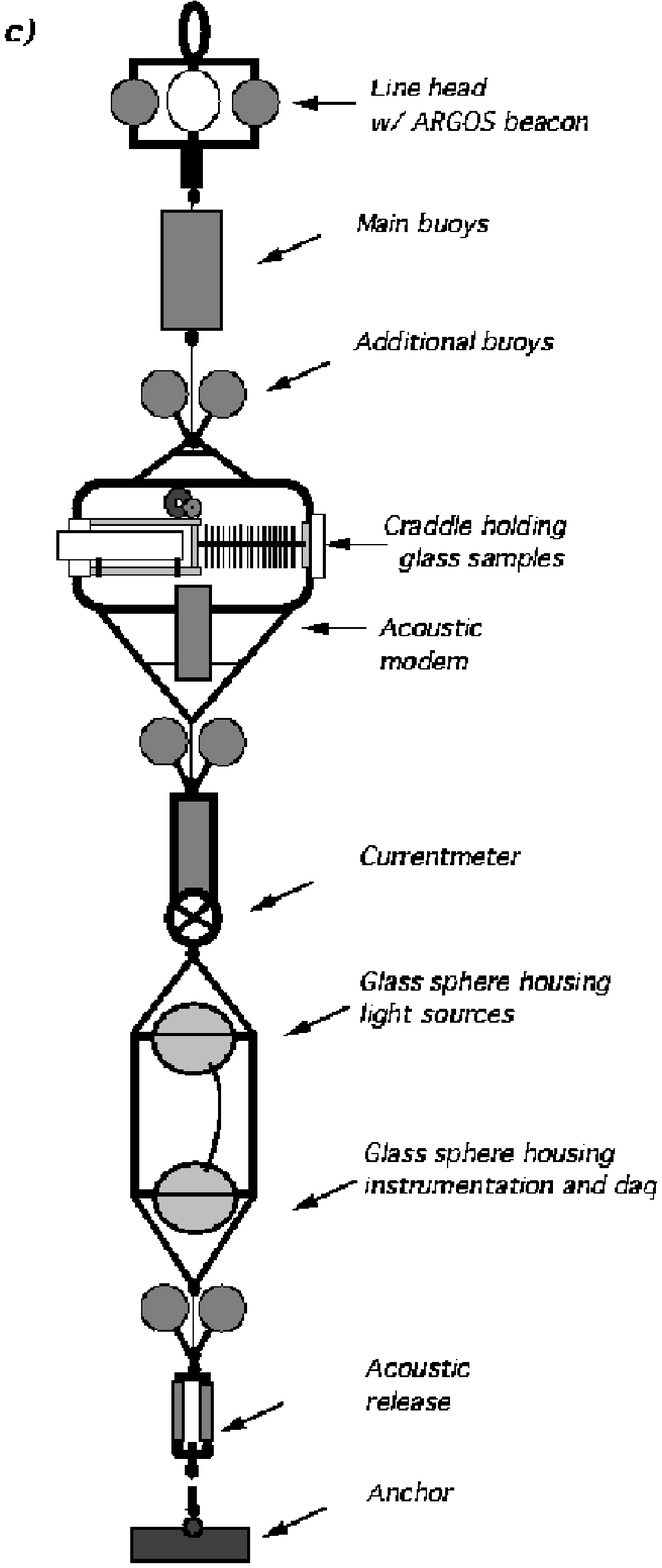,width=0.3\linewidth}}
 \end{tabular}
\end{center}
\caption{\small Test lines to study (a) light attenuation, 
(b) optical background, and (c) optical fouling. }
\label{fig:testlines}
\end{figure}
Three generic test lines exist, as illustrated in figure~\ref{fig:testlines}.  
Each test setup is incorporated in a mooring line anchored at the sea bed
and vertically supported by a buoy.
The measuring system, which can be up to 50~m long, is
placed around 100~m above the sea bed.
A further 30~m of cable and instrumentation sits between the measuring system
and the top of the test line.
The electronics, data acquisition and the detectors are powered by a set 
of lithium battery packs. A 500~cm$^3$ pack  delivers 8~A-h at 26~V.
These compact power sources have a lifetime which permits 
stand-alone tests for periods as long as one year.

The test lines have been deployed using a standard horizontal deployment method.
The buoy at the top of the line is immersed first, so that the line is held
under tension during
the entire operation. The boat continues to move
slowly away from the buoy as the rest of the line is paid out from the boat.
The anchor is the last element to be deployed after which 
the line falls to the sea bed. 

To date, the measurements carried out confirm that the properties of this
site satisfy the constraints of the ANTARES physics programme.

It is planned to explore 
a wider area in the Mediterranean sea, for 
long periods, in order to study 
the variation of parameters as a function of site, depth and 
season. For specific parameters such as water transparency, measurements at 
several wavelengths are also foreseen.
The results obtained so far have been instrumental in steering 
the design of the 
telescope.

\section{Optical properties of the site}
\subsection{Optical background}
\label{subsect:optical_back}

The behaviour of the optical background on site places constraints
on the trigger logic and the electronics 
as well as the mechanical layout of the optical modules. 
A setup has been devised for studying the time dependence of 
the background as well as its spatial extent and its correlation with 
deep sea current. The relevant test line has been immersed ten times in total,
for periods spanning from hours to months. 

The optical modules used for these tests are similar to those discussed in detail
in section~\ref{sect:chap5_om}. In this case 8-inch photomultiplier tubes were used.

Up to three optical modules have been used on the same line, two 
of which, A and B, were 0.5~m to 1.5~m 
apart, while the third, C, was 10~m to 40~m away. A current-meter was installed below 
the optical modules. 
Data consisted of measurements of 
singles rates for all three optical modules and 
coincidence rates for modules A and B within a 
time window of 100~ns.
In order to sample long-term variations,
the system was enabled for a few hours three times a week.

\begin{figure}[h]
\begin{center}
\mbox{\epsfig{file=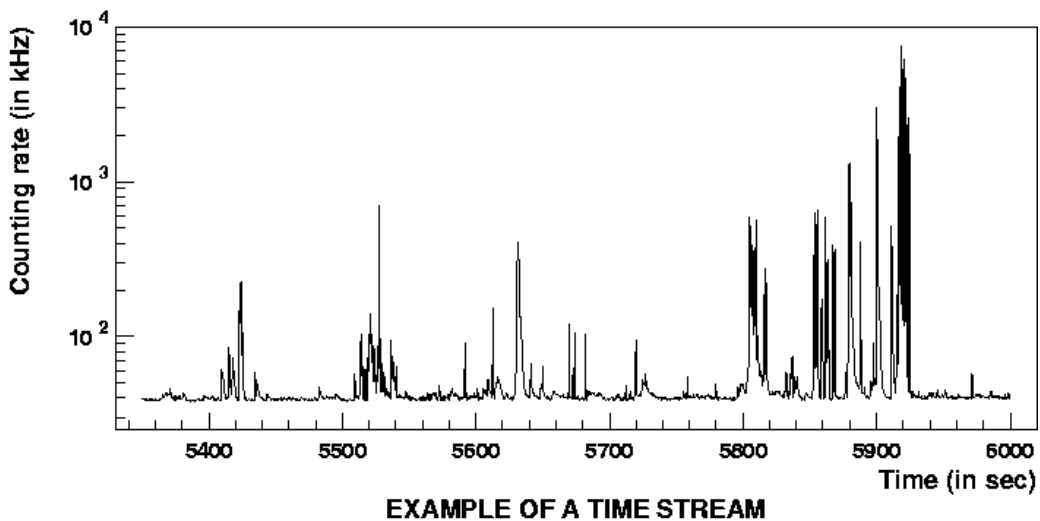,width=\linewidth}}
\end{center}
\caption{\small Time dependence of the counting rate.}
\label{fig:biolum}
\end{figure}

\begin{figure}[h] \begin{center}
\epsfig{file=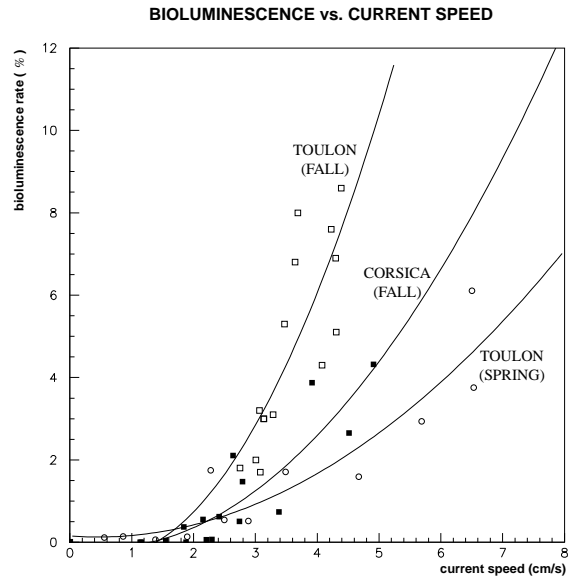,width=.6\textwidth}
\caption{\small Bioluminescence burst activity as a function of current velocity.
A parabolic fit is superimposed on the data.}
\label{fig:currentbiolum}
\end{center} \end{figure}

An example of the observed counting rate is presented in 
figure~\ref{fig:biolum} which exhibits two distinct components: 
a low level background around 40~kHz, and, superimposed on this, 
rapid ($\sim$~1~s) excursions of up to
several~MHz. 
The low level background varies from 17~kHz to 47~kHz over
a time scale of a few hours. This rate changes simultaneously on all optical modules
even when they are 40~m apart. The peak activity is correlated with the
current speed and is limited in spatial extent:
peaks are seen simultaneously by optical modules when they are less than 1.5~m apart,
but not when they are more than 20~m apart. 
The dependence of bioluminescence activity on current velocity is
emphasized in figure~\ref{fig:currentbiolum}
where a correlation is observed between the
two variables. 
This figure also illustrates evidence for site to site and seasonal variations
in this correlation. However, further data are required to understand these effects.

\subsection{Optical fouling}
When exposed to sea water, the surfaces of optical modules are fouled by 
the combination of two processes:
living organisms, mostly bacteria, grow on the outer surface, and sediments fall on the 
upward-looking surfaces. While the bacterial growth is expected to be almost 
transparent,
sediments will adhere to it and make it gradually opaque, 
thus diminishing the sensitivity of the detector.
This phenomenon is expected to be site-dependent as the bacterial growth 
decreases with depth and the sedimentation rate depends on local sources of sediments
such as nearby rivers. A series of measurements has been performed in order to quantify
these phenomena.

Light of wavelength 470 nm from a blue LED source in a glass sphere was 
normally incident on a set of five PIN diodes placed at 50$^\circ$ to
90$^\circ$ from the vertical axis of the sphere
inside a 
second sphere 1~m away. 
Measurements of the light transmission and current velocity were made twice daily.

\begin{figure}[h]
\begin{center}
\mbox{\epsfig{file=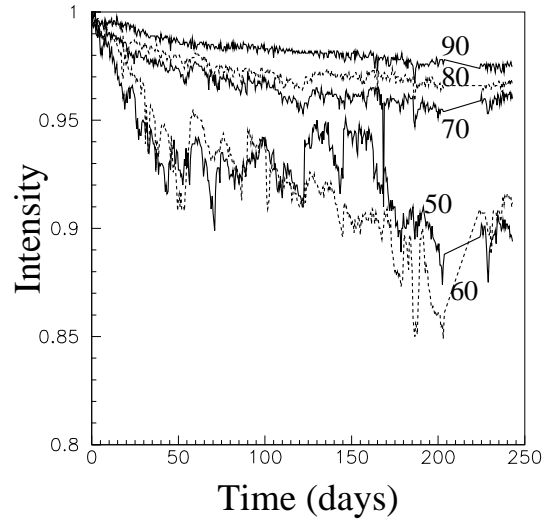,width=0.6\linewidth}}
\end{center}
\caption{\small Light transmission as a function of time and polar angle.}
\label{fig:sedim2}
\end{figure}
Figure~\ref{fig:sedim2} shows the light transmission monitored for 240~days with this
configuration.
For the horizontally-looking 
PIN diode (90$^\circ$), the light source is 
affected by the same fouling as the detector, thus doubling the effect. 
In this case, a transmission loss of 1.2\% per surface is observed after 
8~months of exposure. All optical modules in the ANTARES detector will point 
downward, reducing the loss even further.

\begin{figure}[htb]
\begin{center}
\mbox{\epsfig{file=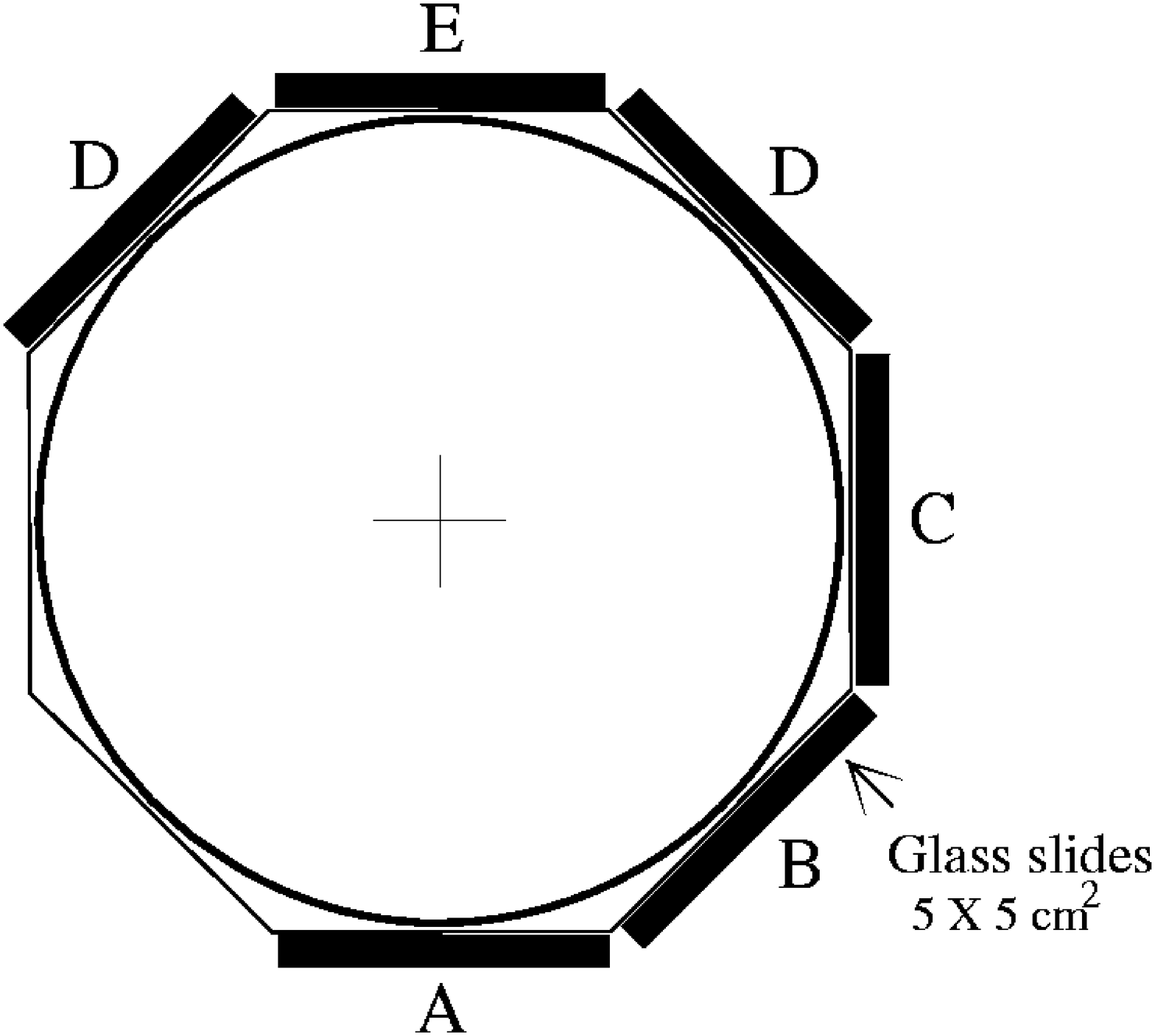,width=0.48\linewidth}
\epsfig{file=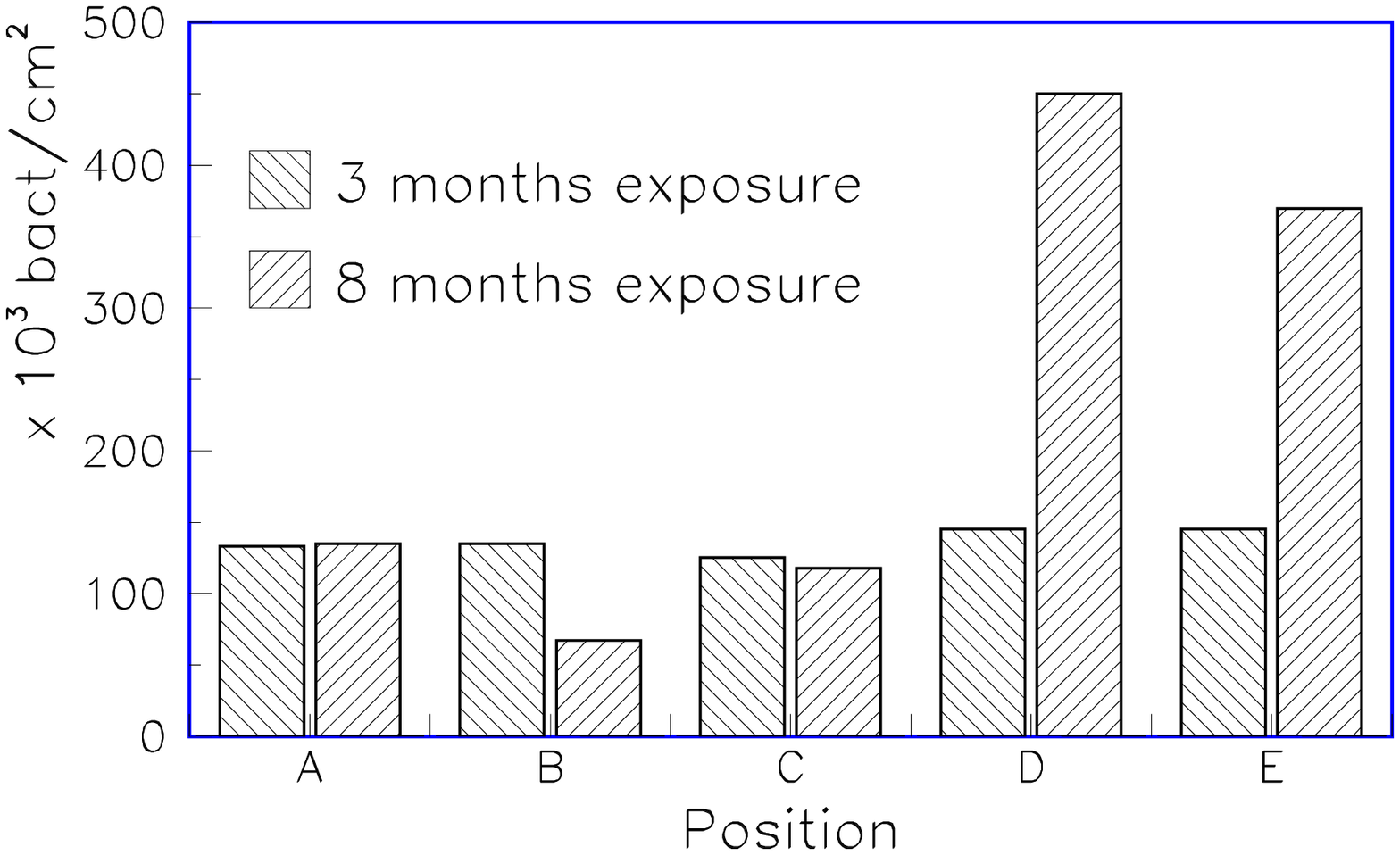,width=0.48\linewidth}}
\end{center}
\caption{\small (left) Position of glass slides on cylinder, (right)
number of deposited bacteria vs. cylinder position}
\label{fig:bacteries}
\end{figure}

In addition, a cradle holding 
glass slides mounted on a horizontal cylinder 
with various orientations around the axis of the cylinder 
(see figure \ref{fig:bacteries}) was incorporated into the mooring line. This 
apparatus was developed by IFREMER \cite{Mazeas}.

Two series of measurements have been  performed
with the apparatus being  
immersed for long periods (3 months and 8 months respectively) 
at 2400~m depth on the same site as the other tests. 
After recovery, a biochemical analysis of the slides 
has been performed by several laboratories 
giving the results shown in figure~\ref{fig:bacteries}.
demonstrating the dependence of the biofouling on the orientation of the slides.
In those cases, where the slides point downward, saturation is observed.
According to marine biologists, these numbers are several orders of magnitude below what 
is observed 
at shallow depths.

\begin{figure}[h]
\begin{center}
\mbox{\epsfig{file=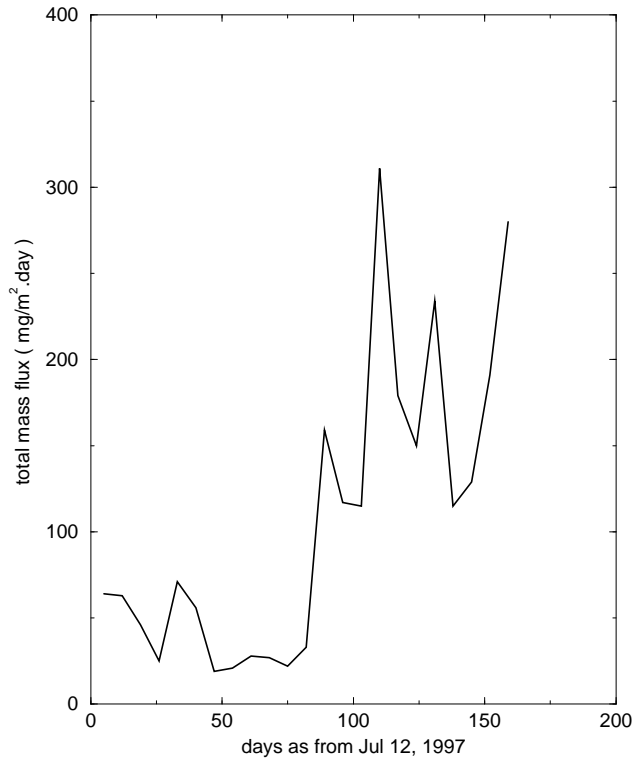,width=0.6\linewidth}}
\end{center}
\caption{\small Total mass flux of sediments at the ANTARES site
as a function of time from July to December 1997.}
\label{fig:massfl}
\end{figure}

The optical modules are housed in pressure-resistant glass
spheres produced by the Benthos company.
The mooring line which measured the fouling of Benthos spheres also included a
sediment trap in order to determine the vertical flux and the composition of 
sediments at the ANTARES site. 
Particulate matter contributes to
 the scattering of light in sea water as well as the fouling of the Benthos spheres. 
 The sediment trap collected samples from July to December 1997 
on a weekly basis. The analysis of the samples
  was performed in the CEFREM (Centre de Formation et de Recherche sur
   l'Environnement Marin) laboratory by the team of
Professor A. Monaco. 
As shown in figure~\ref{fig:massfl}, the total mass flux of sediments 
substantially increases from October onwards, probably as a result of
heavy rainfall draining sediment from the shore.
Indeed, the composition of the sediments shows a large
contribution of material originating from continental river bed. This study will
 be complemented by the analysis of the sediment cores and of the water 
 samples collected during the ANTARES site survey campaign of December 
 1998.
 
\subsection{Transmission properties of the water}
\label{sect:trans_data}

The water transparency affects the 
muon detection efficiency, while the amount of scattered light determines the limit on the 
angular resolution of the detector. These two parameters therefore influence
the detector design and are necessary data for the Monte Carlo simulations
from which the detector response is calculated.  They must be measured {\it in situ},
as water samples may be degraded when brought to the surface. 
Two different experimental setups have been constructed to measure these parameters.

\begin{figure}[htbp]
\begin{center}
\mbox{\epsfig{file=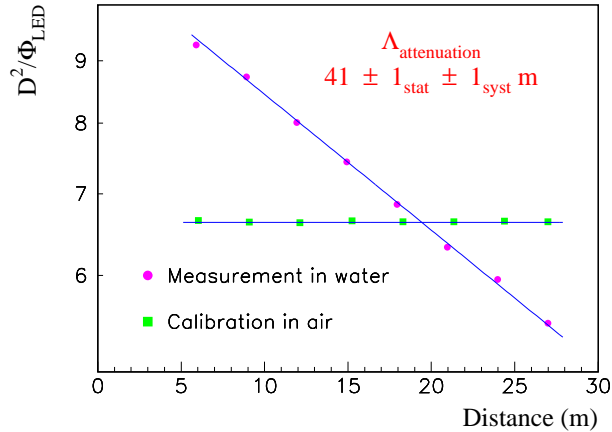,width=0.6\linewidth,bbllx=50pt,bblly=405pt,
bburx=550pt,bbury=725pt}}
\end{center}
\caption{\small Determination of attenuation length. The log of D$^2$/flux is plotted 
against D, the distance between the LED and the optical module.}
\label{fig:test3}
\end{figure}

In December 1997, measurements were performed with a 33~m long rigid
structure holding a collimated and continuous LED source located at a variable
distance from an optical module.
For each selected distance $D$ between the source and the
detector, the LED luminosity $\Phi_{\rm LED}$ was adjusted so as to
yield a constant current $I_{\rm PMT}$ on the photomultiplier
tube. The set-up was calibrated with a similar experiment done in
air.  The emitted and detected intensities in water being related by
\begin{equation}
\nonumber
I_{\rm PMT} \propto \Phi_{\rm LED} /D^2 \times 
\exp(-D/\lambda_{\rm att.\,eff})
\label{eq_lambda_att}
\end{equation}
this test makes it possible to estimate the effective attenuation
length from the dependence of the required LED intensity with the
distance (cf.~figure~\ref{fig:test3}). The agreement of the data with a
decrease following the formula given above
was excellent and yielded an effective attenuation length of
\begin{equation}
\nonumber
\lambda_{\rm att.\,eff} = 41\pm 1\; ({\rm
stat.}) \pm 1\; ({\rm syst.}) {\rm \: m\;\;\;(December\; 1997)}
\end{equation}
This attenuation length results from a combination of absorption and
scattering.  The experimental set-up was unable to separate these, and the
long rail made deployment difficult.  The experiment was therefore redesigned
so that it used a pulsed source, to facilitate scattering measurements, and
a flexible structure to improve ease of deployment.
 
\begin{figure}[htbp]
\begin{center}
\mbox{\epsfig{file=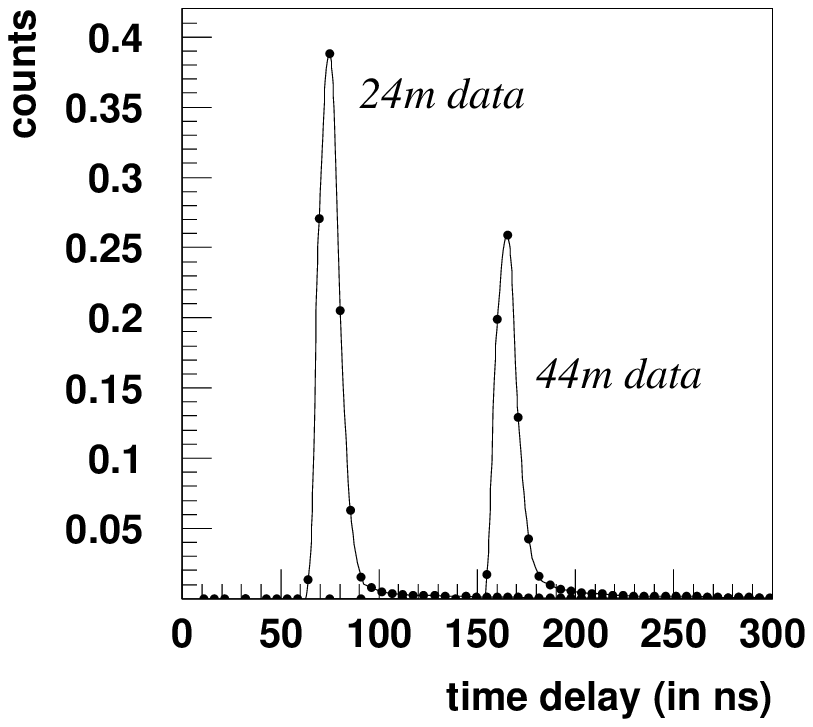,width=0.48\linewidth}
\epsfig{file=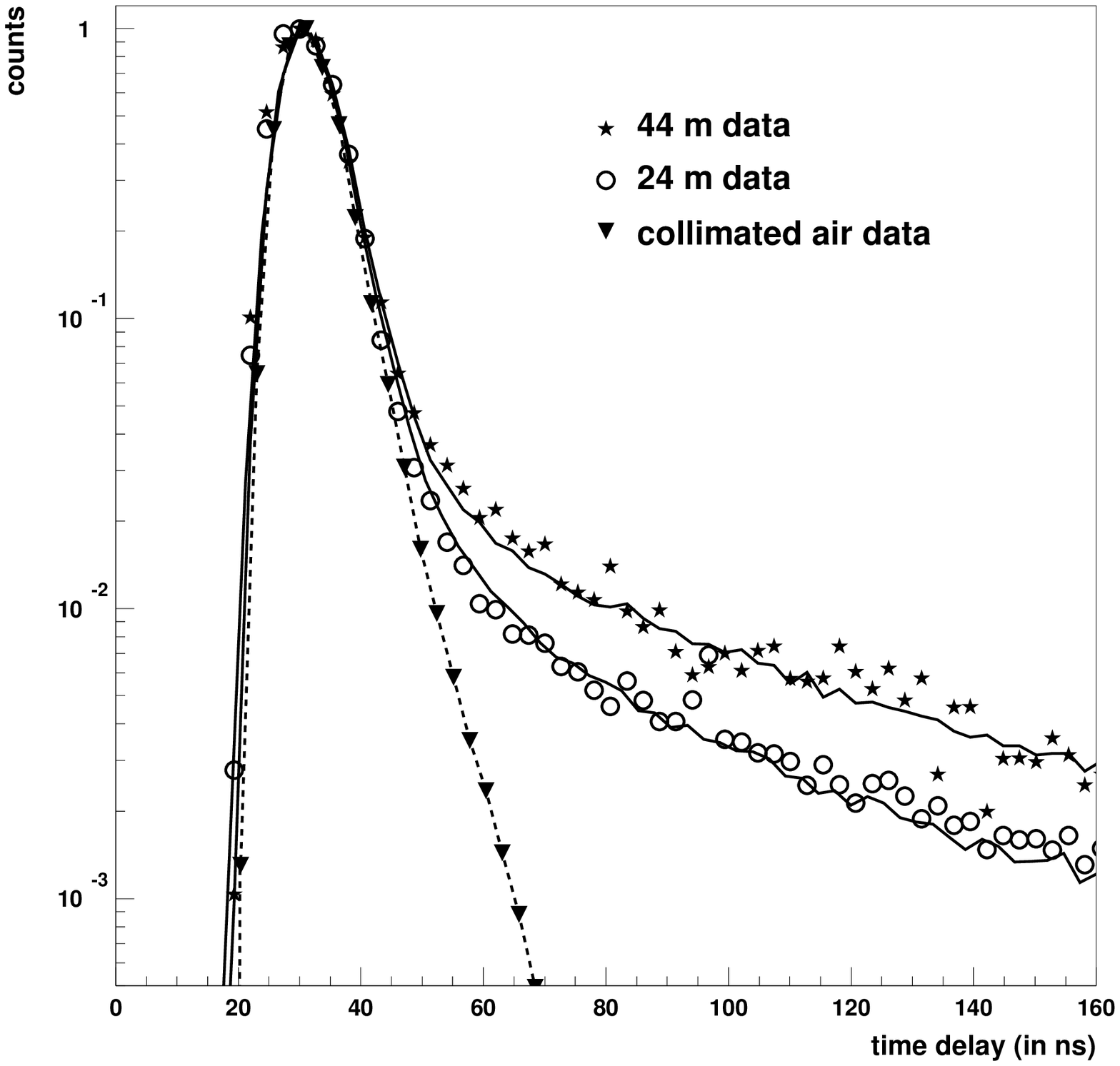,width=0.48\linewidth}}
\end{center}
\caption{\small (left) Distribution of 
arrival times of photons for two distances between
the detector and the source. The 24~m
distribution is normalized to 1, and the 44~m distribution is normalized
relative to the first one. (right) Arrival time
distributions for 24 m, 44 m, and a calibration in air, with
Monte Carlo curves superimposed.}
\label{fig:test3p}
\end{figure}

In July 1998 and March 1999, measurements were performed with a set-up
consisting of a pulsed isotropic LED source located at a distance of
either 24~m or 44~m from a $1''$ fast photomultiplier tube.
An 8-bit TDC measured the distribution of the arrival times
of the photons. The overall time resolution was $\sigma=4.5$~ns.  
Because of this, photons in the tail of the distribution have scattered
with an angle at least $\sim 35^\circ$ for the
24~m spectrum, and $\sim 25^\circ$ for 45 m. Therefore, the
scattering properties of the water are being measured 
for large scattering angles.

The time distributions recorded 
exhibit a peak stemming from direct photons, and a tail extending
to larger delays due to scattered photons. For the 24~m (44~m)
spectrum, 95\% (90\%) of the photons are collected within 10~ns. 
Scattering is
thus a small effect at the ANTARES site, as illustrated in
figure~\ref{fig:test3p}.

An effective attenuation length could be determined from the ratio of
the integrated spectra measured at the two distances, yielding:
\begin{equation}
\nonumber
\lambda_{\rm att,\,eff} = \left|
\begin{array}{rl} 
60.0\pm 0.4\; ({\rm stat.})\:{\rm m} & ({\rm July} \;1998) \\
52.2\pm 0.7\; ({\rm stat.})\:{\rm m} & ({\rm March} \;1999)
\end{array} \right.
\end{equation}
A systematic uncertainty of a few metres might affect these estimates
due to the fact that the LED luminosity is not monitored and yet
assumed to be the same for the time distributions collected at the two
distances. These
measurements indicate more significant attenuation in March than in
July.  The difference between these and the December 1997 measurement is
also significant, although it is partly accounted for by the use of a
collimated source in 1997.  
 
A more detailed analysis can be done by fitting the data with a Monte Carlo
distribution obtained by photon tracking.  The data are well described using
an absorption length in the range 
55--65~m, a scattering length at large angles greater than 200~m 
and a roughly isotropic scattering angle 
distribution.  This is consistent with the effective attenuation length
deduced from the ratio of integrated spectra.

\section{Sea conditions}

Suitable sea  
conditions for periods of up to a few consecutive days are required to perform
deployment and recovery operations.
These conditions depend both on the nature of the operations 
and on the characteristics of the ship. For the single string 
deployment and recovery operations of June-September 1998 with the 
{\it Castor}, a wave height less than 1.5 m 
and wind 
speed less than 25~knots (5 on the Beaufort scale) were specified.

A study has been made incorporating data from a number of
sources, namely:
\begin{itemize}
\item
data on wave height collected by an instrumented buoy moored 
4 nautical miles south of Porquerolles Island from May 1992 to September 1995,
which should be representative of the conditions in the ANTARES site;
\item
data on the wind speed
and direction as recorded by the Porquerolles Island Signal Station 
analysed for the period from January 1992 to March 1996;
\item 
additional information on sea conditions provided by satellite measurements in the area
from 1992 onwards.
\end{itemize}

Preliminary analysis of 
these data has been performed by the Meteomer company.
Periods of three consecutive days with favourable 
sea conditions occur less than five times per month from October  
to April, and more than five times per month from May to September.

\section{Site survey}

\begin{figure}[h]
\begin{center}
\mbox{\epsfig{file=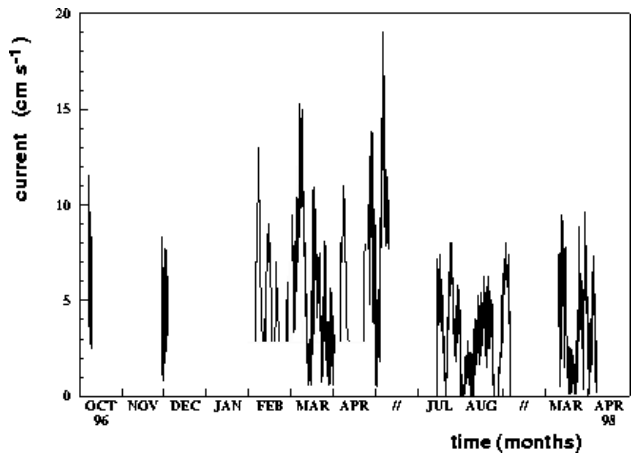,width=0.6\linewidth}}
\end{center}
\caption{\small Summary of the survey of deep sea currents.}
\label{fig:current}
\end{figure}

The strength and direction of
the underwater currents need to be taken into account in the mechanical design of the
detector.
Figure~\ref{fig:current} summarises all the measurements of the deep sea current 
gathered during the test 
immersions. The maximum current observed, namely 18~cm/s,
is accommodated in the 
mechanical design of the strings.

\begin{figure}[h]
\begin{center}
\mbox{\epsfig{file=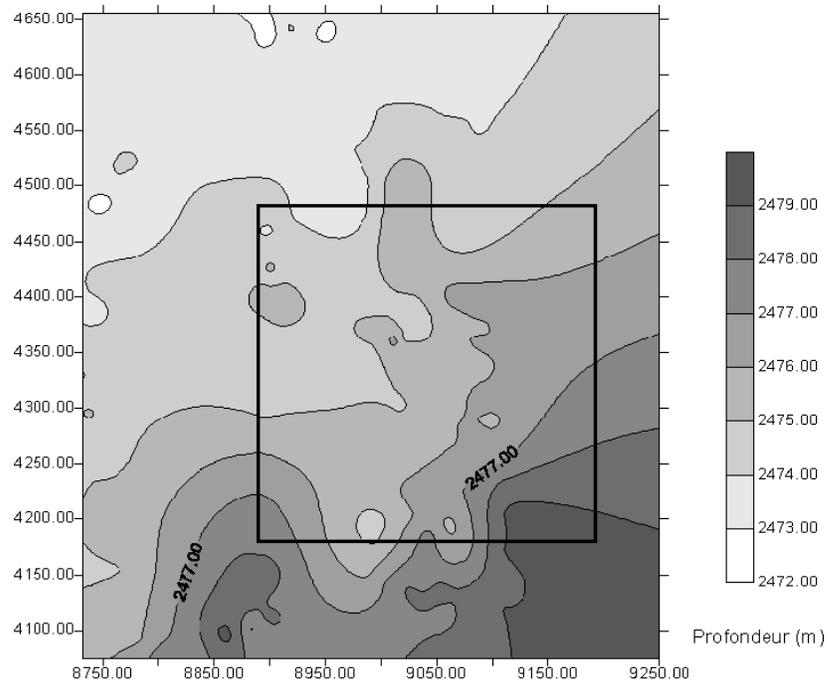,width=0.8\linewidth}}
\end{center}
\caption{\small Site survey map}
\label{fig:site_survey}
\end{figure}

A visual and bathymetric
survey of the sea floor was performed in December 1998, using the
{\it Nautile} submarine.  In the area selected as a potential ANTARES site,
the sea floor is flat and displays no topographical anomalies such as steps or 
rocks, as can be seen in figure~\ref{fig:site_survey}.
During the same series of dives, core samples
from the sea floor were collected.  They consist of solid mud which is
a satisfactory substrate to support the detector. 

\section{Prototype string}
\label{sect:prototype_string}

To gain realistic experience of the issues involved in the
deployment, operation and recovery of a full-scale detector,
a full-size prototype of an ANTARES detector string has been
constructed.
It is equipped with a positioning system, slow control 
network, power distribution, and eight optical modules 
with their associated readout electronics. 
Once deployed, the string is foreseen to
be linked to a shore station by an 
electro-optical cable supplying the power 
to the string as well as enabling the control and readout connection.
The string is adequate for a 
detailed study of the optical background and a 
measurement of the down-going muon flux.

\begin{figure}[htbp]
\begin{center}
\mbox{\epsfig{file=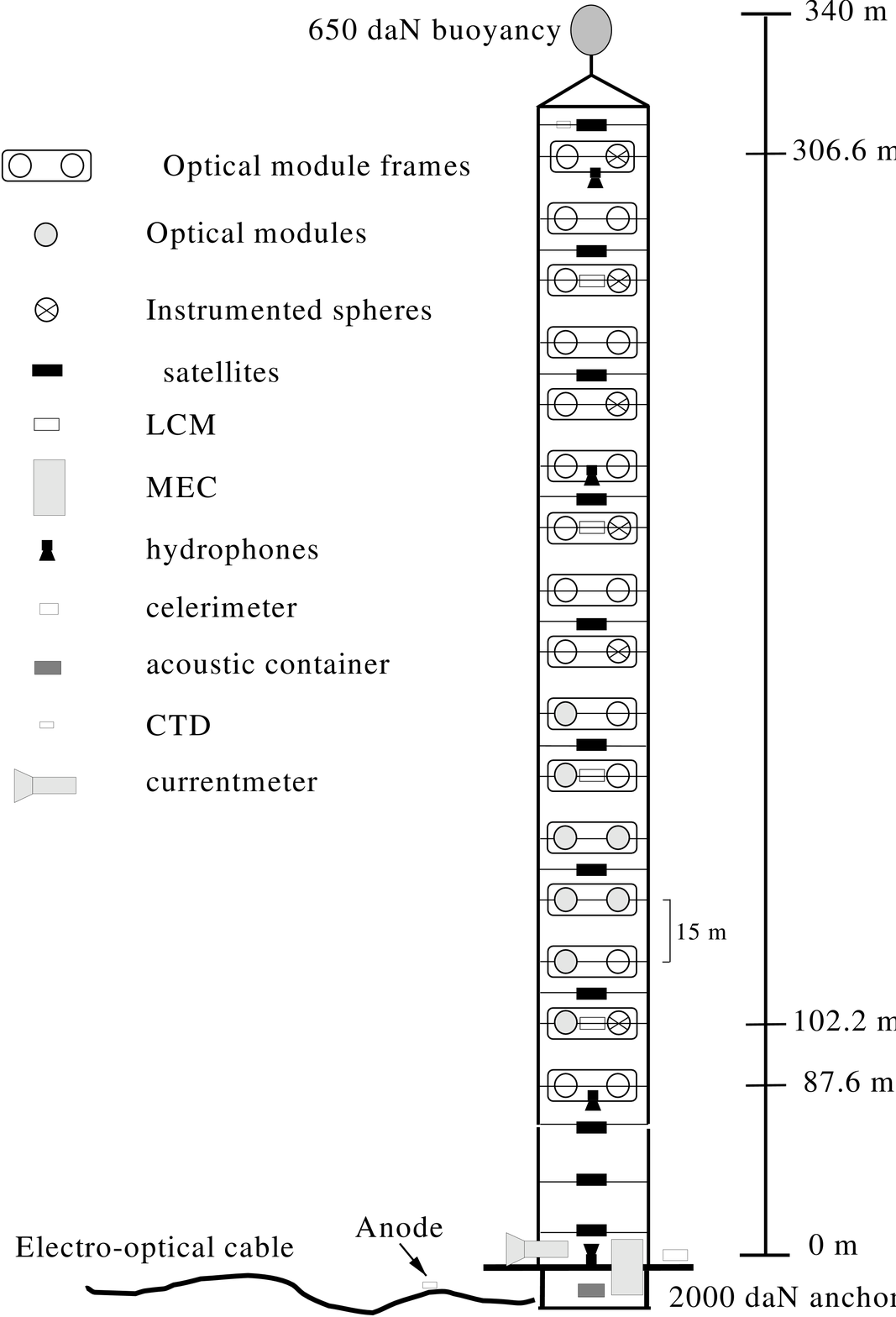,height=17cm}}
\end{center}
\caption{\small Schematic view of the prototype string.}
\label{fig:line5}
\end{figure}

\subsection{Mechanical structure}

The prototype string is 350~m high
(figure~\ref{fig:line5}), anchored 
on the sea floor and vertically supported by a buoy. It is 
composed of two vertical cables 
spaced 2.3~m apart, supporting both ends 
of sixteen Optical Module Frames (OMFs).
The OMFs are placed every 15~m from a height of 95~m up to 
320~m from the sea floor and are constructed from fibre glass
to avoid corrosion. 
Horizontal spacers keep the vertical support cables below
the lowest OMF; a
spacer is also situated at the middle of 
every 15~m segment separating the OMFs. 

Each OMF supports a pair of Optical Modules, 
separated by 1.6~m from axis to axis, while
the central part 
houses a container for electronics made of  corrosion-resistant
 titanium alloy.

\subsection{Slow control system}

A slow control network is 
linked to the shore station through an electro-optical cable
and employs an architecture whereby each electronics
container has a point-to-point connection with the slow control
data acquisition system.
It permits the control of the power distribution electronics, 
optical module motherboard, analogue
readout electronics and the acoustic positioning system. It also handles
the readout and transmission of sensor data, such as the satellites, the
acoustic system and the electronic card temperatures or status. 

This DAQ system is located in the main electronics
container at the bottom of the string.  The main fibre-optic data link
connects this system to the shore station, which receives the
slow-control data and provides a user interface for the slow-control
system. 

\subsection{Positioning}
\label{sect:chap4_pos}

\begin{figure}[h]\centering 
\epsfig{figure=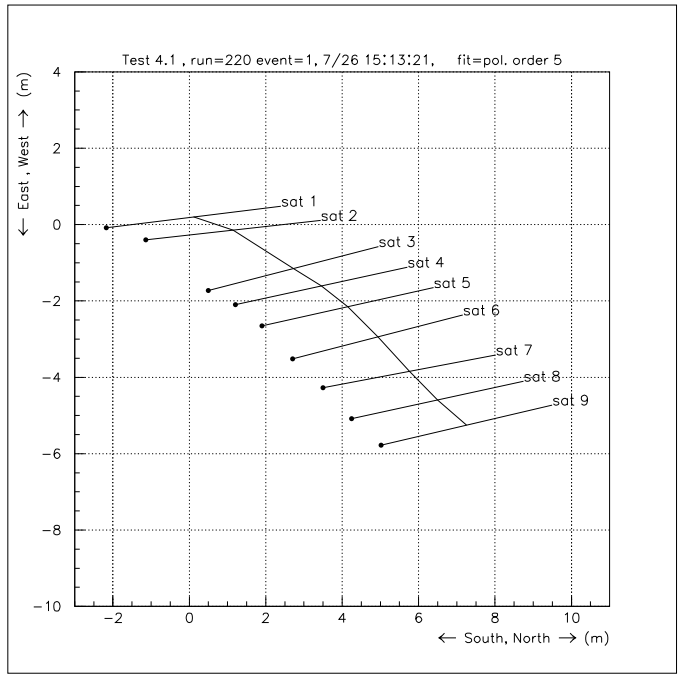,height=3.5in} 
\caption{\small Top view of the reconstructed shape of the prototype string
during deployment, sensors are labelled {\it sat 1} to {\it sat 9}} 
\label{test4}
\end{figure}

A string does not provide rigid support for the optical modules.
Two independent systems have been
incorporated in the prototype string to provide
a precise knowledge of the relative position of each OM at any time.
The first system is based on a set of tiltmeters and compasses which
measure local tilt angles and orientations on 
the string. 
The  reconstruction of the line shape, as distorted by the water current flow, is obtained 
from a fit of measurements taken at different points along the line. 
A successful test of this system was performed
during a deployment of the prototype string which was equipped with several sensors
with a precision of $0.05^{\circ}$ in tilt and of $0.3^{\circ}$ in direction.
Figure \ref{test4} presents a view of the reconstructed line,
the segments indicating the sensor positions and directions
with respect to North. A maximum error of $\sim$~1~m on the reconstructed shape is 
estimated.

\begin{figure}[h]\centering 
\epsfig{figure=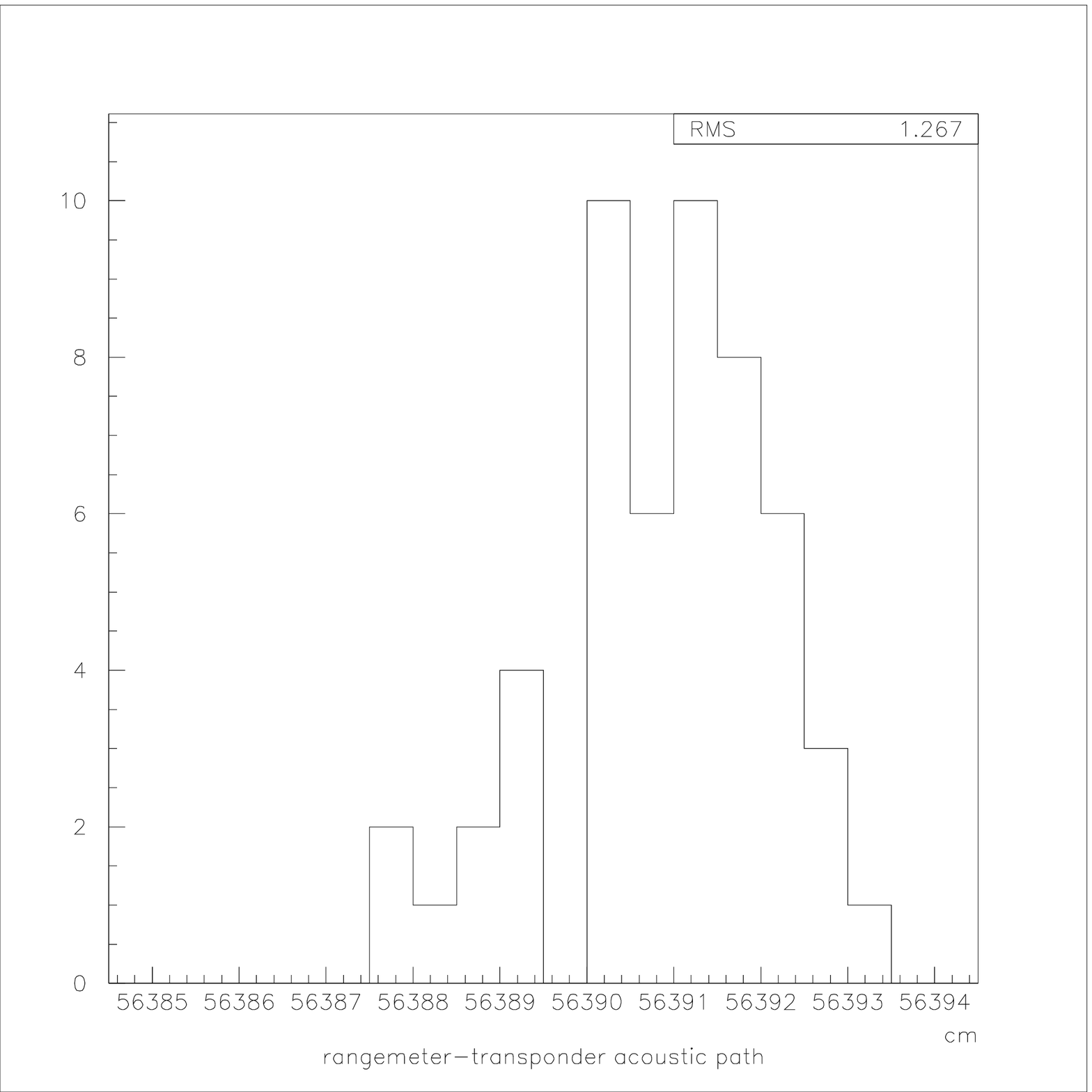,height=3.5in} 
\caption{\small Distribution of the acoustic path measurement between a 
fixed rangemeter and a fixed transponder. } \label{acoustic}
\end{figure}

The second system, based on acoustic triangulation, is more precise but requires more complex and 
expensive electronics. 
In this system, rangemeters placed on the string send an acoustic signal to
a minimum of three transponders fixed to the sea bed. Each transponder replies with its 
characteristic frequency. A global fit of the measured acoustic paths gives the precise
three-dimensional position of the
rangemeters, provided that the positions of the transponders and the sound velocity in water are known.   
The prototype string is equipped with four rangemeters 
(a hydrophone with its electronics container) communicating with four external 
autonomous transponders placed on a fixed structure on the sea floor about 
200~m away from the string. 
The measurement of the communication time 
between one fixed rangemeter and one fixed transponder demonstrates a reproducibility of 
$\sim 1$~cm in the acoustic path length (figure \ref{acoustic}). 

In order to exploit such a system fully, a precise knowledge of the sound velocity in water 
along the acoustic path is required.
This depends strongly on water temperature and also on salinity and depth. The prototype
string is thus equipped with sound velocimeters, which measure the local sound velocity with a 
precision of 
5 cm~s$^{-1}$, and with Conductivity Temperature Depth devices (CTDs) to observe the variations 
of temperature and salinity. 

The systems are complementary: a few points of the line can be measured acoustically
and other points are
obtained by line shape fitting. The tests already performed confirm that 
the desired precision on relative OM positioning can be achieved.

\subsection{String deployment}

The prototype string was deployed, operated and retrieved several times
in Summer 1998, first at 400~m depth and then at 2300~m. 
This string was powered by batteries housed in a 
container at the bottom of the string.
The deployment procedure used was different from
that used for
the mooring lines discussed earlier. In this case, a step-by-step procedure
involving two winches on the boat deck was implemented whereby the
anchor was immersed first, the string being held securely at 2 points, then
each OMF was paid out storey-by-storey. Recovery was performed
in a similar way. Deployment and recovery at 2300~m took 18~hours in total.
This method allowed the equipment to be deployed
in a safe and controlled manner.
The exercise also permitted the handling procedures
for the electro-optical cable 
termination attached to the string anchor to be verified.
The tiltmeter and 
compass data were recorded so that the detector string 
behaviour could be studied during the deployment phase. 
The fully equipped version of the prototype string will be
ready for immersion during Summer 1999.

Successful deployment of the 0.1 km$^2$ array will require strings to be
located in well-defined positions with 80~m spacing or less.
This
was investigated in a multi-string deployment test using the dynamical
positioning ship {\it Provence} and 400~m test strings instrumented with
acoustic beacons.

String positioning precision was investigated by aiming to deploy
a second test string 50~m away from the one which had been deployed earlier.
Relative positions were measured with the aid of three external 
acoustic beacons.

The second string was deployed using a winch, with the ship located at the target
position. Deployment was halted with the string 200~m above the sea floor,
the relative position measured, and the ship moved to adjust as necessary. Using
this technique it proved possible to locate the second string to
within 10~m of the nominal position.

\subsection{Deep sea connection and line recovery}
\label{sect:chap4_connect}

In the full-scale detector, each string will be connected to a common point,
known as the junction box.
A connection scheme from the junction box 
to a prototype string anchor was tested during a ten-day 
sea operation at the ANTARES site, 
using the {\it Nautile} submarine and its support vessel the {\it Nadir} 
from IFREMER.

A reel of cable equipped with deep sea connectors on both ends was 
immersed first. 
Then the {\it Nautile} uncoiled the cable by pulling it 
and plugged the connector of the cable to its counterpart on the anchor. 
The connection procedure was successfully performed twice, using a
free-flight technique which requires only one of the submarine's arms.
This offers the possibility of using IFREMER's second manned 
submarine, the {\it Cyana},
which is equipped with only one arm. 
The advantage of this is that this submarine 
is more freely available than the {\it Nautile}.

\begin{figure}[h]
\begin{center}
\begin{turn}{-90}
\mbox{\epsfig{file=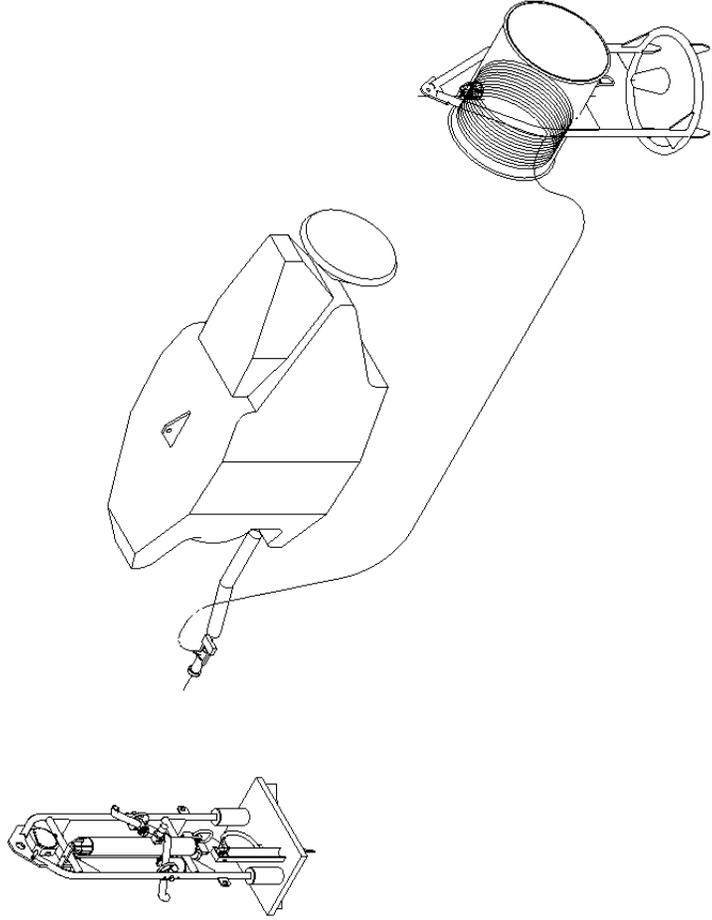,height=10cm,bbllx=0pt,bblly=190pt,bburx=550pt,bbury=700pt}}
\end{turn}
\end{center}
\caption{\small Deep sea connection test using the {\it Nautile} submarine.}
\label{fig:submarine}
\end{figure}

Figure~\ref{fig:submarine} shows a diagram of the {\it Nautile} approaching the anchor,
holding the connector and the cable, before plugging it at the end of the
arm which can be seen in an upright
position at the bottom of the string. Once the connection is made, the {\it Nautile}
pushes the arm down and plugs it into a fork linked to the anchoring weight.
To retrieve the string, acoustic
releases are activated from the surface which disconnect the string from the
anchoring weight. The buoyancy of the string pulls on the connector which is
held back by the fork and is thus unplugged. This system avoids the need for
a submarine for string retrieval. 

The speed of ascent of the string during its trip to the surface is 
around 1~m~s$^{-1}$. 
The distribution of weights and buoyancies 
along the string have to be carefully studied in order to control
the relative speed of each storey so that the string does not become
entangled when it surfaces.

%% file: chap5.tex

The detector design described in this chapter is constrained by the
following factors:

\begin{itemize}
\item the environmental conditions at the site, which influence the 
      spacing of optical modules, the mechanical structure of detector
      strings, and the expected background rates;
\item the practical experience gained from the design and deployment of
      the protoype string;
\item the need to optimise the physics performance of the detector;
\item the requirement for a high level of reliability.
\end{itemize}

The first two issues were discussed in the preceding chapter, and the
physics performance and its optimisation are the subject of
chapter~\ref{chap:detector_performance}.
A key issue throughout the detector design is the question of reliability.
Repair and maintenance of offshore detector elements will involve high
costs and probably long down times, and must be minimised.  For this
reason, the design of each of the detector components discussed in the
subsequent sections will follow a strategy intended to enhance
reliability: reduction of the number of active components, limitation of
power consumption, and avoidance of single-point failure modes.

This chapter outlines the specifications satisfying these constraints.
The design may evolve further as technical details are refined and
additional physics simulations are performed.

\section{Overview}
\label{sect:chap5_gen}

\begin{figure}
\begin{center}
\mbox{\epsfig{file=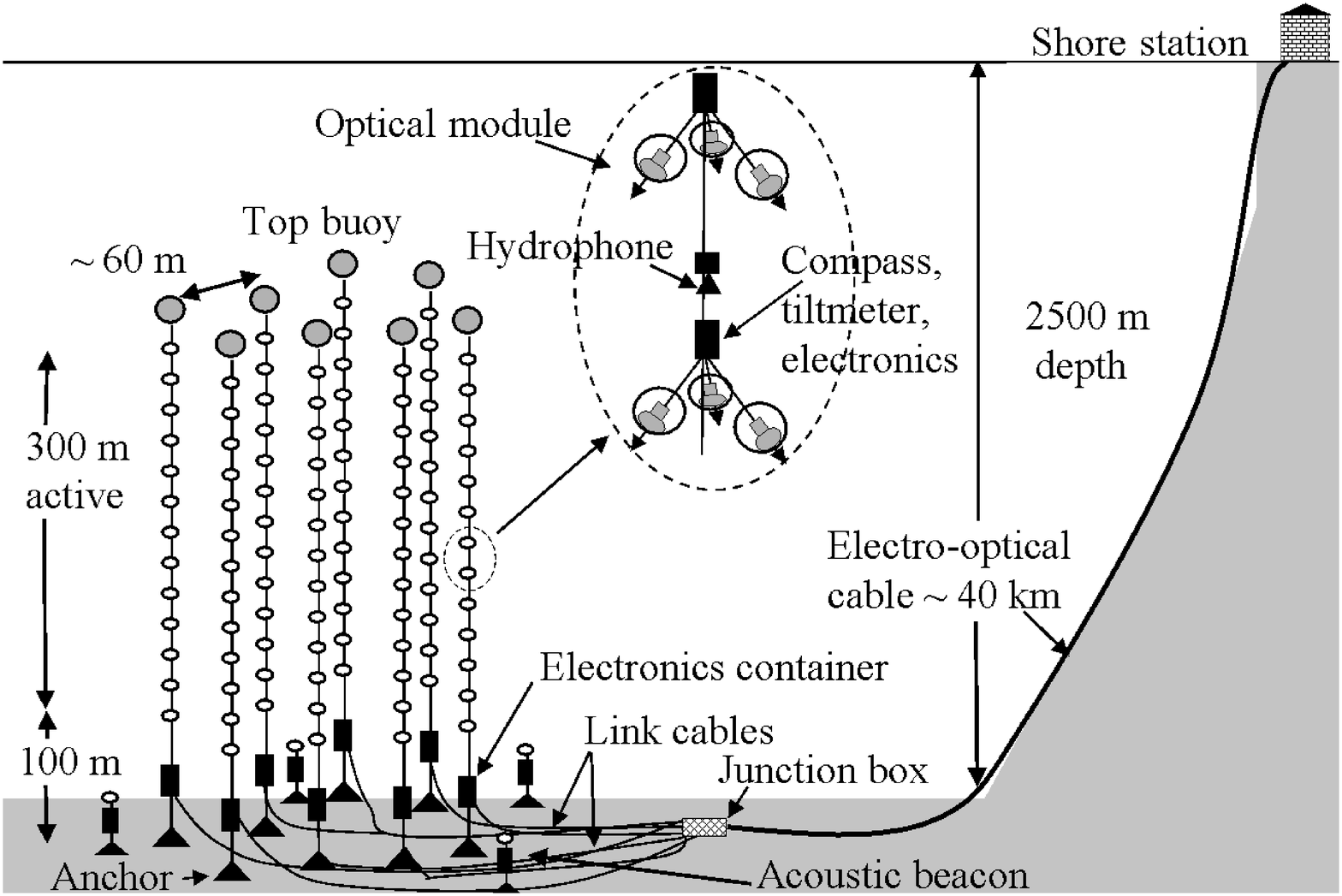,width=\linewidth,height=11.5cm}}
\end{center}
\caption{\small Schematic of part of the detector array;
the magnified view shows two storeys and a hydrophone.}
\label{fg:peb}  
\end{figure}

The detector
consists of an array of approximately 1000 photomultiplier
tubes in 13 vertical strings, spread over an area of
about 0.1~km$^2$ and with an active height of about
0.3~km.  Figure ~\ref{fg:peb} shows a schematic view of part of
the detector array indicating the principal components
of the detector. 

   The basic unit of the detector is the optical module,
consisting of a photomultiplier tube, various
sensors, and the associated electronics, housed in
a pressure-resistant glass sphere.  The electronics
includes a custom-built digital electronic circuit
which captures and stores waveforms,
pulse heights and timing information, as well as the HV
power supply for the photomultiplier tubes and the network
nodes for data transmission and slow control. 

   The optical modules are grouped together in `storeys' of
three modules and interconnected via an electro-mechanical cable.
In the present
design the detector has 13~strings, 
each of which has a total height of about 400~m.  
Of the 13 strings, four consist of 41~storeys spaced vertically by
8~m and nine have 21~storeys spaced
vertically by 16~m. The strings are arranged in a 
randomized spiral (see figure~\ref{fg:fig_4-4-b}) with
a minimum horizontal spacing fixed at some value between
60~m and 80~m.  The smaller spacing gives a higher
detection efficiency for muon energies up to 50~TeV but
could be more difficult to deploy. 

   The optical modules in a storey are arranged
with the axis of the photomultiplier tubes 45$^\circ$
below the horizontal.  The present plan is to use 10-inch
Hamamatsu photomultiplier tubes, but other, possibly
larger, tubes are under development by various
manufacturers.  The angular acceptance of the optical
modules is broad, falling to half
maximum at $\pm
70^\circ$ from the axis (see below).  This means that
the proposed
arrangement of OMs detects light in the lower
hemisphere with high efficiency, and has some
acceptance for muon directions above the horizontal.  
 In the lower hemisphere there is an overlap
in angular acceptance between modules, permitting an 
event trigger based on coincidences from this
overlap.

The relative positions of all optical modules in
the detector are given in real time by a positioning
system identical to that described for the prototype
string in section~\ref{sect:chap4_pos}.

Each string is instrumented with several electronics
containers. At every storey, there is a local control
module (LCM), and at the base of each string there is a
string control module (SCM).  Special containers house
acoustics and calibration equipment.  Each of these
containers constitutes a node of the data transmission
network, receiving and transmitting data and
slow-control commands.  The functions which they support
include reading sensors, adjusting slow-control
parameters, the trigger, and the distribution of power,
master clock and reset signals to the front-end
electronics. 

   The individual SCMs are linked to a common junction
box by electro-optical cables which are connected using
 a manned submarine. A standard deep sea
telecommunication cable links the junction box with a
shore station where the data are filtered and recorded.

   The trigger logic in the sea is planned to be as
simple and flexible as possible.  The first-level
trigger requires a coincidence between any two OMs in a
single storey.  The second-level trigger is based on
combinations of first-level triggers.  Following a
second-level trigger the full detector will be read out. 
A more refined third-level trigger, imposing tighter
time coincidences over larger numbers of optical
modules, will be made in a farm of processors on shore. 
The readout rate is expected to be several~kHz,
and the corresponding data recording rate less than 100
events per second. 

   The following sections of this chapter describe the
various components of the detector in more detail.

\section{Detector string}


 The design chosen for the detector string is similar to that 
used for the prototype string described earlier in 
chapter~\ref{chap:prototypes}, 
a string maintained vertically by its 
own buoyancy and anchored on the bottom of the sea.
Between the buoy and the anchor, the active detector part of the string comprises
a series of elementary detector segments. 
These segments are standardised and so can be mass-produced, and if necessary
interchanged.

\begin{figure}[tbp]
\begin{center}
\mbox{\epsfig{file=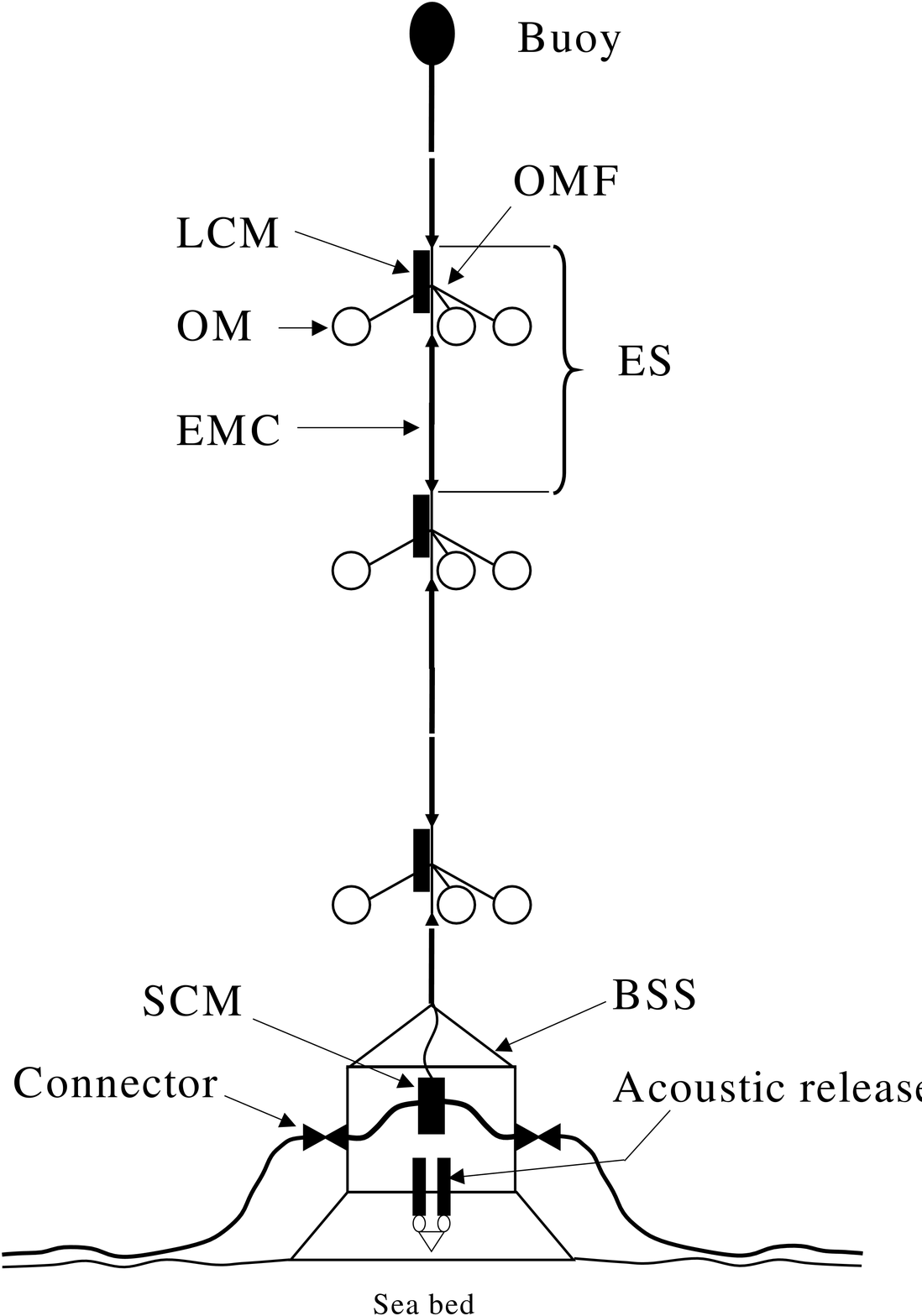,height=12cm}}
\end{center}
\caption{\small Sketch of the detector string, abbreviations are defined
in the text.}
\label{fig:fig_d2}
\end{figure}

In all cases, the design of the string components meets basic specifications 
such as
corrosion resistance, required because of the 
prolonged immersion in salt water,
resistance to high pressure and  water-tightness. Similarly, in certain cases, the
colour and surface properties of detector elements need to be selected so as to 
minimize light reflection. All detector components must remain functional for a 
minimum
lifetime of 10~years.

The principal elements of each string are illustrated in figure~\ref{fig:fig_d2} and
discussed further here.



The bottom string socket (BSS)
anchors the string to the sea bed, facilitates the electrical connection of the string to the
network and permits the release and subsequent retrieval of the string.
To allow precise and 
simple string installation its construction has been optimised for handling on the deployment
ship, resistance to shock, stability during descent.
The BSS is instrumented for acoustic positioning.
Connection of the string to the network is performed by a submarine
and is described in section~\ref{sect:chap4_connect}.

The electro-mechanical cable (EMC) provides mechanical support for the string as well
as enabling the electrical interconnection of the detector string elements.
It must be capable of supporting tensile, torsion and bending stresses 
in order to maintain the string's stability. Its construction must be suitably flexible
to allow the integration of the various configurations, 
as well as supporting the
service, storage, handling and immersion and recuperation 
phases. 
Electrical cables and optical fibres run through the EMC. They 
enable power distribution and the transmission of signals between 
two consecutive electronic containers (LCM or SCM). 

Each elementary segment (ES) consists of  
three optical modules, one local control module and one optical 
module frame. 
The OMF supports the various elements placed within it, namely the optical modules, 
the LCM container and possibly a hydrophone for acoustic positioning.
The OMF design must be sufficiently flexible to 
accept certain changes in the number and positioning 
of the optical modules. 
Major traction forces that work on the substructure during
deployment and recovery must in no way be transmitted 
to those string elements supported by the frame.

The top of the string consists of a buoy. Its 
dimensions and geometry are optimised to minimise hydrodynamic effects such as dragging 
and vibration whilst maintaining a suitable tension in the string.
In addition, the buoyancy should be sufficient to ensure a controlled resurfacing of the string 
during retrieval.

\section{Optical module}
\label{sect:chap5_om}

\begin{figure}[htbp]
\begin{center}
\mbox{\epsfig{file=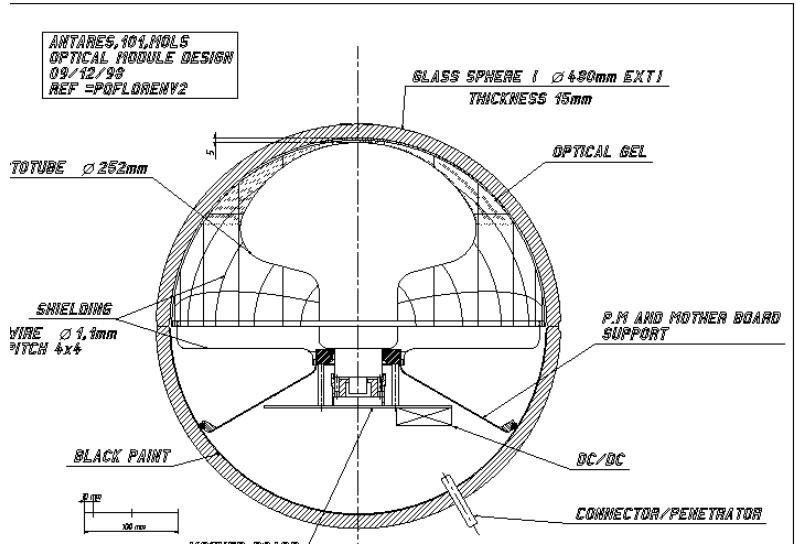,width=0.8\linewidth}}
~\\
\vspace{0.4cm}
\caption{\small Cross section of the optical module.}
\label{fig:chap5_OM}
\end{center}
\end{figure}

   A schematic view of the ANTARES optical module is
shown in figure~\ref{fig:chap5_OM}. Mechanical and
optical aspects of the OMs are reviewed in this section,
as well as the performance of the various
photomultiplier tubes which have been evaluated. 

\subsection{Benthos sphere}

The photomultiplier tube and
its associated electronics are housed in a 
43~cm diameter, 15~mm thick, Benthos sphere, which can withstand
pressures of up to 700~bars. The sphere is made of two halves, one of which is
painted black on its inner surface so as to give the OM
some minimal directionality with respect to Cherenkov
light detection without degrading its acceptance.  
The two halves of the sphere have machined edges which 
form a seal when subjected to an external over-pressure.
Attenuation of light at
$\lambda=450$~nm due to the sphere was measured to be less
than 2\% (see figure~\ref{fig:chap5_optic}). 

\subsection{Optical contact and magnetic shielding}

\begin{figure}[h]
\begin{center}
\mbox{\epsfig{file=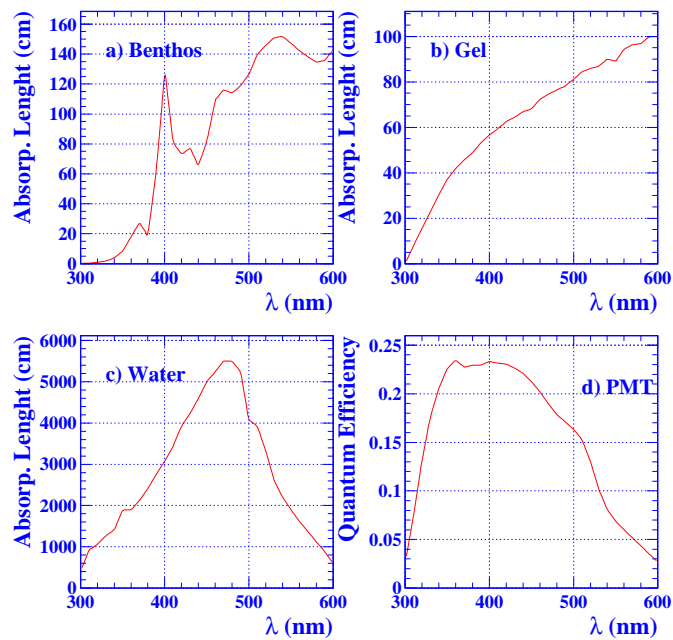,width=0.7\linewidth}}
\caption{\small
Attenuation
length of (a) the Benthos
sphere, (b) silicone gel and (c) sea water; (d)
quantum efficiency for the Hamamatsu photomultiplier tubes.}
\label{fig:chap5_optic}
\end{center}
\end{figure}

Silicone gel ensures both optical coupling and mechanical
support of the PMT. The refractive index of the gel ($n_{gel}=1.40$) does not
exactly match that of the sphere itself ($n_{glass}=1.48$) but is higher
than the refractive index of water ($n_{water}=1.35$) 
and so the amount of light reflected out of the OM is minimized. 
The silicone gel covers the entire photocathode area; its attenuation length 
is given  as a function of wavelength in 
figure~\ref{fig:chap5_optic}~\cite{bib:chap5_transmision_mes}. 


The Earth's magnetic field significantly degrades
the collection efficiency of phototubes by bending
the trajectories of electrons, mainly between the
photocathode and the first dynode.  
A cage made of 1.1~mm thick high-permittivity alloy wire
is used to shield the PMT and to minimise the dependence 
of the OM response with respect to its angle with the
magnetic North. The mesh size of the cage
(6.8~cm) was optimized to reduce the
non-uniformity of the PMT angular response to less than
5\% while minimizing the fraction of light lost due to
the shadow on the photocathode
\cite{bib:chap5_mucage}.  

\subsection{Photomultiplier tube characterisation} \label{sec:chap5_pmt}

During the last three years several test
benches have been set up to measure and compare
the principal features of different PMTs and to provide
detailed characteristics of the OM response. These
include several `dark boxes', in which the PMTs are
exposed to uniform illumination coming from red, green
or blue LEDs or very fast solid-state lasers. This
allows the systematic and precise measurement of a large
number of PMT characteristics~\cite{bib:chap5_pmtstudies}. 
Several PMTs of different diameters
have been studied, and
their performances are summarised here in terms of a 
small number of critical parameters. This discussion only considers a
fraction of the total number of parameters which have actually been measured
(see~\cite{bib:chap5_pmtspecif,bib:chap5_pmtstudies,bib:chap5_gamelle}).

The effective photocathode area ($A_{eff}^{PC}$) is defined as
the detection area of the photocathode weighted by the collection
efficiency.  It
was measured by scanning the entire photocathode
surface with
a collimated blue LED.

Electromagnetic interference in the optical module induces
noise at the PMT anode. It is expected that this will not
exceed 5~mV (rms). This noise governs the gain at which
the PMT is operated. A factor of 10 between the average
pulse height for a single photo electron (SPE)
and the noise is sufficient to ensure
efficient discrimination of the signal. This corresponds to 
an effective working gain of the order of 5$\times$10$^7$.
In view of PMT ageing and possible variations in 
specifications a maximum gain of at least 10$^8$ is required.

The Peak to Valley ratio is computed from the observed
charge spectrum of single photoelectrons with the high voltage
adjusted to give 50~mV amplitude for SPE. The Peak to Valley
ratio is required to be greater than~2. 

Due to imperfections in the electron optics and the finite size of
the photocathode, the SPE transit time
between 
the photocathode and the first dynode has a measurable width,
usually referred to as the 
 transit time spread (TTS).  This
defines the timing resolution of the PMT, which is required
to be comparable to that
from the overall positional accuracy and the timing precision
in the readout electronics, i.e. 1.3~ns rms or 3~ns FWHM.
The measurement of the TTS is performed over the whole
photocathode area with the PMT
operating at a gain of 5$\times$10$^7$.

Four different PMTs are compared in
table~\ref{tab:chap5_PM_comp}. 
The maximum
quantum efficiency, the linearity
and the dark count rate have also been determined.

\begin{table}[htbp]
\begin{center} 
\begin{tabular}{|l||l|l||l|l|}
\hline
Manufacturer 	&\multicolumn{2}{c||}{Hamamatsu}	&\multicolumn{2}{c|}{ETL}  \\
Phototube diameter 		&$8''$ 	&$10''$	
&8$''$ 		&$11''$ \\
\hline
$A_{eff}^{PC}$ (cm$^2$)	        &280  		&440		&240		&620\\ 
Maximum gain 			&$10^9$		&$10^9$		&$10^8$		&5$\times$10$^7$  \\  
Peak to valley ratio		&2-3 		&3-3.5		&$<2$ 		&2\\
TTS (FWHM) (ns)			&2.5		&3.5		&2.5		&3.0\\
\hline
\end{tabular}
\caption{\small
Comparison of relevant properties of
large-photocathode phototubes tested 
to date.}
\label{tab:chap5_PM_comp}
\end{center}
\end{table}

\subsection{Optical module response}

To reproduce the experimental conditions, it is necessary
to investigate the response of an optical module to
the Cherenkov light produced by muons travelling through
water. 
This has been studied by placing an OM
in a 1.5~m high, 1.4~m diameter, 
light-tight cylindrical steel tank 
filled with constantly recycled fresh water~\cite{bib:chap5_gamelle}.
Nearly-vertical atmospheric muons
are tracked in a four plane hodoscope, two planes above the
tank, two below. Between the tank and the two lower planes a lead
shield ensures $E_{\mu} > 0.6$~GeV.  Muons  
traverse the tank and produce Cherenkov light which
illuminates the optical module.  The light yield for muons
of this energy is more than 90\% of the value for a single
relativistic track.
The optical module is able to rotate, so that
its response can be measured at different angles, $\theta_{OM}$, between the 
incident muon and the PMT axis.

\begin{figure}[htbp]
\begin{center}
\mbox{\epsfig{file=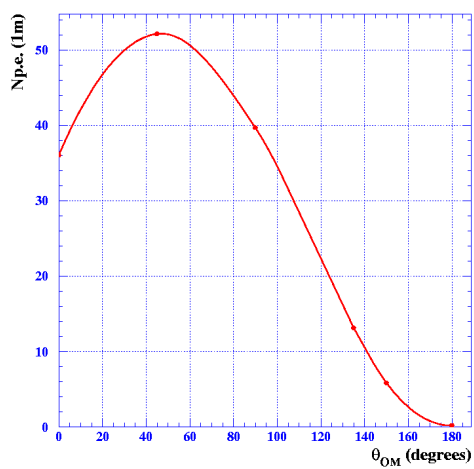,width=0.5\linewidth}}
\caption{\small Number of photo-electrons detected by the
optical module as a function of its zenith angle
$\theta_{OM}$, for 
nearly vertical muons normalised at 1~m distance.}
\label{fig:chap5_gamelle_res}
\end{center}
\end{figure}

Figure~\ref{fig:chap5_gamelle_res} presents
results from this setup using an optical module equipped 
with a 10$''$ tube. 
The angular region where the OM sees more than half the
maximum amplitude 
is of the order of $\pm 70^\circ$. 

\section{Offshore electronics}
\label{off_elec}

The distance between the detector and land precludes a point-to-point 
connection between each optical module and the shore station.  Instead,
an electro-optical cable from the shore station
supplies electrical power to the 
detector array and permits data to flow in both directions.  
The electro-optical cable ends at the junction box to which the strings
are connected.  A star-topology network architecture is used, running from
the string
control module to the optical modules via the local control modules.
A digital scheme has been developed for the necessary data multiplexing.

This network is used to distribute power, collect data, broadcast slow
control commands and master clock signals, and form the trigger.  This
section describes the implementation of the above scheme. 

\subsection{Digital front end}
\label{subsect:ARS_SPE}
\label{sect:chap5_digfe}
The OM electronics must fit in the limited space available
in the OM, consume little power, be reliable and long-lived 
(average lifetime $>$ 10~yr), be inexpensive.

\begin{figure}[h]
\begin{center}
\mbox{\epsfig{file=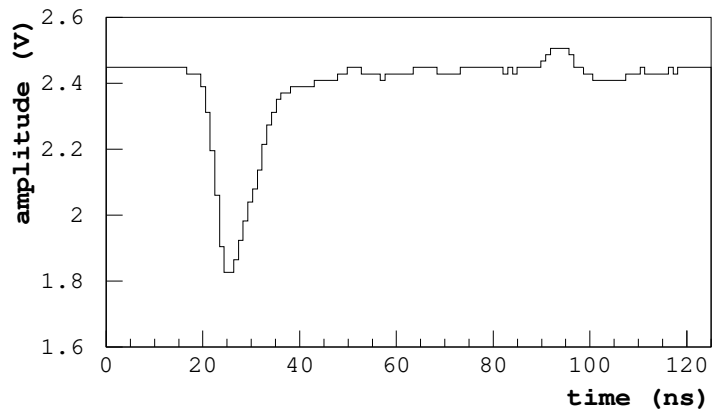,width=0.8\linewidth}}
\end{center}
\caption{\small Charge measured in the 128~switched capacitors of
the ARS memory after the capture of an analogue signal from a photomultiplier
tube.  The reflection of the signal comes from an impedance
mismatch on the test board.}
\label{ars0}
\end{figure}

An Application Specific Integrated Circuit (ASIC) meets these requirements
and can be tailored to our needs.
The ASIC developed for the digital front end is called the 
Analogue Ring Sampler (ARS), similar in design to 
the Analogue Transient Waveform Recorder
\cite{dn}.
The chip samples the photomultiplier tube signal continuously at a tunable
frequency between 300 and 1000~MHz and holds the
analogue information on
128~switched capacitors when a low-level threshold is crossed. 
The information is then
digitized by an external 8-bit ADC. Figure~\ref{ars0} shows the resulting 
histogram for an effective sampling frequency of 1000~MHz.

A 20~MHz reference clock is sampled on one
channel, giving a relative timing of the signals
to better than 1~ns. A time stamp is obtained for each event 
by counting the reference clock cycles. A reset command sent
through the clock stream is used to restart all the
counters of the array synchronously.

Since 99\% of the pulses are single photo-electrons, another
ASIC, dedicated to the treatment of single photo-electron pulses,
has been developed in order to reduce the dead time and the
data flow.
The first part of the ASIC performs pulse shape discrimination (PSD). 
It identifies three types of pulses requiring analysis of the full wave form:
\begin{itemize}
\item large pulses, with pulse heights that cross a threshold 
corresponding to several 
photo-electrons;
\item wide pulses, with time over a low threshold longer than about 15~ns;
\item two pulses separated by less than 50~ns.
\end{itemize}
If none of these conditions is met, only the pulse charge and time of 
arrival are measured, this information, along with the OM address,
is transmitted to shore in 64 data bits. 
The large pulse threshold, the time over threshold, and the time window for 
multiple pulses are adjustable. 
The remaining 1\% of the pulses satisfy one of the three conditions above and
so the pulse shape is transferred for offline analysis, approximately 2000
data bits are required to encode this type of event (waveform event).

\begin{figure}[t]
\vspace{-5cm}
\begin{center}
\mbox{\epsfig{file=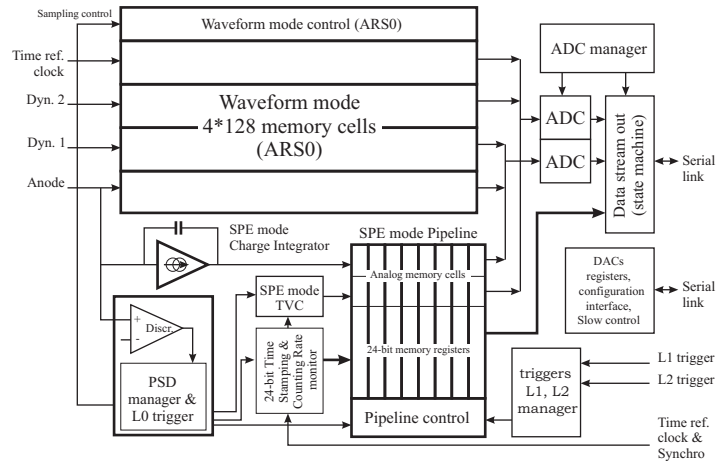, 
width=\linewidth}}
\end{center}
\vspace{-8cm}
\caption{\small ARS1 block diagram.} \label{ars1}
\end{figure}

A new version of the ARS is currently under development. It 
integrates all these functions on the same chip, together with the ADCs,
the DACs and the slow-control interface (figure~\ref{ars1}). 
A pipeline memory is implemented to store
the single photo-electron information long enough
to match the level~2 trigger
propagation and formation time, 
which is around 10~$\mu$s for a 0.1~km$^2$ detector.
It will be possible to use up to four memories per
optical module, in a token ring. This will permit the chip to be used with 
photomultiplier tubes bigger than 10~inches in diameter, 
for which the counting rate may
exceed 60~kHz, and for a km-scale detector, where the trigger formation and
propagation time may reach 30~$\mu$s.

\subsection{Trigger logic and rates}

A level~0 trigger occurs when the output of a
photomultiplier tube crosses a threshold corresponding 
to 30~\% of a single photo-electron amplitude.
In order to deal with the high counting 
rate in the sea (see section~\ref{subsect:optical_back}), 
a level~1 trigger is built out of a tight time coincidence 
between two level~0 triggers from the same storey.
A level~2 trigger can be formed by requiring multiple 
level~1 triggers in a coincidence gate whose width is 
of the order of that needed for 
a track to pass through the entire detector. 
The level~2 trigger condition could be at least two level~1 triggers 
on the same string 
(referred to as the `string trigger') or
at least three level~1 triggers anywhere in the detector
(referred to as the `array trigger'). 
When either the string trigger or the array
trigger conditions is satisfied, a readout request is sent
to the entire array. 
The readout request is received in each OM, 
which starts 
the digitization of all the information within the maximum allowed
time of flight.  Internal delays specific to each OM compensate for the 
trigger formation and readout request propagation times.
When a level~1 trigger occurs in a storey, the two OMs
involved are read out, even if they do not
participate in the subsequent level~2 trigger.
 
The level~1 trigger logic will be installed in each LCM. 
The level~2 trigger logic must be linked to all the LCMs which may participate
in the trigger: therefore the array trigger will be installed in the junction box, 
while the string trigger will be installed at the bottom of each string, and then sent 
to the junction box, from which the readout request will originate.

Trigger rates due to random coincidences from background counting 
rates have been estimated assuming a level~0 trigger rate
of 60~kHz. The level~1 trigger requires a coincidence between
two of the three optical modules on any storey with a pulse width of 20~ns,
yielding a rate about 500~Hz per storey.

Level 2 trigger rates are calculated assuming four dense strings of 41~storeys
with 8~m vertical spacing, 
and nine sparse strings of 21~storeys with 16~m spacing.
The gate width for string triggers is 1.5~$\mu$s, 
the time for a particle to travel the length of the string. For array triggers, which
involve multiple strings, a coincidence width of 2~$\mu$s is used. 
A simple string trigger 
requiring two level~1 triggers on the same string 
leads to a level~2 trigger rate of 
about 3~kHz. A similar array trigger requiring three level~1 triggers within the 353
storeys of the full detector adds 6~kHz, yielding
a total level~2 trigger rate of about 10~kHz.

More restrictive level~2 triggers can be obtained by combining signals from
separate sectors on a single string to yield `proximity' triggers, favouring nearby
hits on the same string. 
The time for light to traverse two sectors is 400~ns. 
Requiring two level~1 triggers within
two adjacent sectors leads to a 
total rate of
300~Hz, down a factor 10 from the simple string
trigger described above.

Even more restrictive level~2 triggers can be defined if necessary. The easiest
way to reduce the trigger rate is to increase the number of level~1 triggers
required within the level~2 coincidence gate. For the simple string triggers, requiring
three level~1 triggers would reduce their contribution to the level~2 trigger to
about 30~Hz, a reduction of a factor 100. For the array triggers, requiring four
level~1 triggers in coincidence would reduce the level~2 rate to about 600~Hz, a
reduction of a factor 10.

During a bioluminescence burst, level 1 triggers from affected storeys provide
no discrimination against noise, because all the OMs in the storey see the burst.
These storeys 
are removed from the trigger logic
in real time, but the OMs are still read out, as the timing resolution of the 
detector is sufficiently precise to allow real hits to be recovered from the noise
when tracks are reconstructed.

\subsection{Data flow rate}

The volume of data transmitted to shore depends on the
trigger rate, the OM background and the
proportion of waveform events. The following
calculation has been performed for a total trigger rate of 10~kHz with a
2~$\mu$s time window, so that 2\% of the overall background activity is read 
out and sent to shore. The noise rate from each OM is assumed to be 60~kHz
for 95\% of the time (quiescent phase) and 300 kHz for the 
remaining 5\% (bioluminescence activity). It is also assumed
that the level~1 trigger is disabled for modules undergoing 
bioluminescence activity.

The three classes of waveform 
events, namely large pulses, wide pulses and events where two pulses occur
within 50~ns are discussed in section~\ref{subsect:ARS_SPE}. Large or wide
pulses originate from decays in or near the OM and amount to a few hundred~Hz,
while closely spaced pulses can result either from
pre- or after-pulsing or from accidental coincidences.
At 60~kHz, the three contributions are of approximately equal importance and
give about 1~kHz of waveform events, while at 300 kHz,
accidental coincidences dominate, yielding about 10~kHz of waveforms.

Given these figures, the data flow rate from 
the level~2 triggers 
is 300~Mb/s.
Reading out 
the OMs involved in level~1 triggers adds a further 
100~Mb/s to give a total data flow rate of about 400~Mb/s.

\subsection{Data handling and transmission to the shore}

A number of electronic functions must be developed in order 
to handle the digital data coming 
from the optical modules and transmit it to shore.  
Functions are required at each storey on the string 
(LCM), at the base of the string (SCM), and in the junction box 
connected to the shore by the electro-optical cable.  
The specific functions are:

\begin{itemize}
\item Organizing data coming from the optical modules and transferring them to the 
      SCM.
\item Processing the level~1 trigger at the LCM and the level~2 trigger at the SCM and at
      the junction box.
\item Slow control and monitoring hardware.
\item Clock distribution from the shore to the LCM containers and on to the optical modules, 
      in order to time stamp all digitized physical events.
\item Handling of the acoustic positioning electronics for each string, distributed over 
      selected storeys throughout the length of the string, performed at the LCM level.
\item Acquisition of mechanical positioning information
      from tilt-meters and compasses
      incorporated in the LCMs.
\item Transmission of data and slow-control commands from the shore to the SCMs
      via the junction box.
\end{itemize}

Experience with the prototype string has led to the adoption of the 
following techniques:

\begin{itemize}
\item Fibre-optics along and between strings, 
      providing reliable, compact and high 
      bandwidth links
      
\item Penetrators as opposed to connectors at the level of each LCM: 
      this permits the use of optical fibres, allows connections to
      be made inside the container, and increases the number 
      of communications channels which can be integrated into a single 
      electro-mechanical cable.  
\end{itemize}

\subsection{Electro-optical cable and junction box}

Existing electro-optical cables designed for submarine telecommunications
applications fulfill the power and data handling requirements summarised
above.  They incorporate a copper coaxial conductor for power transmission
and 16 to 24 optical fibres for digital signals.

The junction box
contains power converters in an oil bath to provide a 
standard 400~V DC voltage, the trigger electronics and the electro-optical
interface for the data, slow control and clock transmission.

The cable and junction box are a potential single point failure of the 
detector.  It is therefore foreseen to have two cables and two junction
boxes, each capable of handling the full requirements of the detector.  

Electro-optical cables link the junction box to each SCM.
since optical fibres are necessary for the data transmission and
the clock distribution. 
In order to use standard undersea electrical connectors, it is
foreseen to convert all optical signals to electrical ones
near the connectors. 

\section{Slow control and commands}
\label{sect:slow_control}

   The system of slow control is intended for the monitoring
of variables which change relatively slowly, and also serves
to control various aspects of detector operation.  PMT
voltage, temperature and power-supply voltages are read from
the optical modules.  Dedicated instruments provide
information on string attitude and orientation, water current
velocity, acoustic positioning information, and other control
data.  Parameters to be adjusted during detector operation
include the PMT voltage, thresholds involved in pulse
detection and triggering, and various calibration systems. 
The slow control system thus gives a user at the shore
station all the information needed to monitor and control the
detector, as well as providing the calibration information
necessary to reconstruct events. 

   Slow-control data acquisition and execution of slow-control commands
are carried out by the processor on the motherboard of the relevant
electronics container (OM, LCM, SCM, or specialized instrumentation
container).

\section{Calibration and positioning}
\label{sect:chap5_calipos}

The pointing accuracy of the detector
is determined largely by the overall timing accuracy of each event. 
This is a quadratic sum of terms due to
\begin{itemize}
\item
the precision with which the spatial positioning and orientation 
of the optical modules is known ($\sigma_{geom}$);
\item
the accuracy with which the arrival time of 
photons at the optical modules is measured ($\sigma_{pmt}$);
\item
the precision with which local timing of individual optical module
signals can be synchronised with respect to each other ($\sigma_{align}$)
\end{itemize}

\begin{displaymath}
\sigma^2_t = \sigma^2_{pmt} + \sigma^2_{geom} + \sigma^2_{align}
\label{sigmat}
\end{displaymath}

Furthermore, {\it in situ}
calibration of the optical module efficiency as a function of time is
necessary in order to measure and correct changes in the response due to
factors such as optical fouling. These issues are now discussed in detail.

\subsection{Positioning}
The reconstruction of the muon trajectory is based on the differences 
of the arrival times of the photons between optical modules. As such, 
it is sensitive to the distances between the optical modules.  In order 
to avoid degrading the reconstruction, it is necessary to monitor the
position of each optical module with a precision of 10 cm (light travels 
22 cm per ns in water). 
The reconstruction of the muon trajectory and the determination 
of its energy also require knowledge of the optical module orientation 
with a precision of a few degrees. The precise absolute positioning of the 
whole detector has to be guaranteed
in order to point to individual sources. 

To attain a suitable precision on the overall positioning accuracy 
constant monitoring of relative positions of the various detector 
elements with respect to absolutely positioned beacons is necessary. 
A full description of the relative positioning equipment on the string 
and its performance during prototype string deployments has already 
been discussed in section~\ref{sect:chap4_pos}. 

The absolute positioning of the detector is performed by acoustic 
triangulation of low frequency acoustic beacons placed on the string 
bottom and a rangemeter on a surface ship, 
equipped with the Dynamic Global Positioning System (DGPS).
A precision of $\sim$~1~m has already been demonstrated for the 
positioning of a prototype string, 
deployed at full immersion depth. 
Precise triangulation necessitates the knowledge of the sound velocity profile
from the sea floor to the surface, which can be strongly distorted
by thermal effects which depend on the season and current.
This has to be measured by using a CTD profiler when the absolute
positioning calibration is performed. 

\subsection{Timing precision and calibration}


A master clock on shore, linked
to Universal Time (UT) through the Global Positioning System network, 
permits to match events to
transient astronomical phenomena such as gamma-ray bursts.
The clock signal is distributed through the array network to
each OM.

A LED is located in each OM, facing upwards towards the PMTs
further up in the string. It can be pulsed synchronously to the
clock signal. This system will enable time synchronization between
adjacent storeys in the same string. 

\begin{figure}[htbp]
\begin{center}
\mbox{\epsfig{file=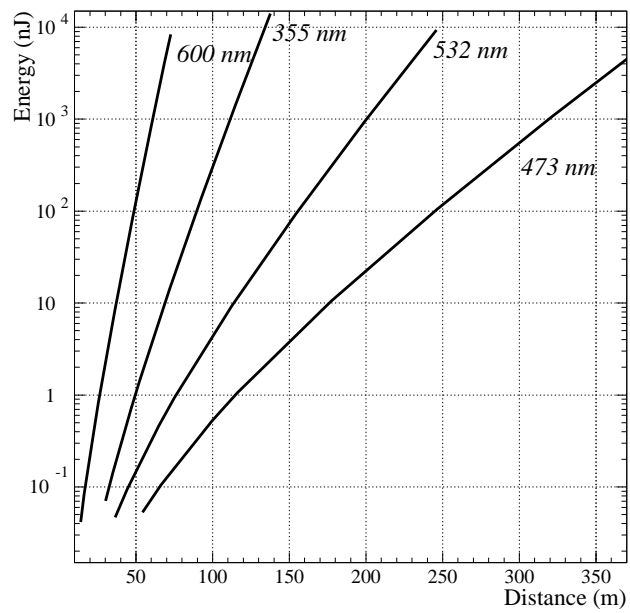,width=0.6\linewidth,bbllx=35pt,bblly=155pt,bburx=535pt,bbury=695pt}}
\end{center}
\caption{\small Distance in water reached by a pulse of light as a function
                of its energy for different wavelengths. At least
		5~photoelectrons are required to be detected in the PMT.}
\label{fig:laser_energy}
\end{figure}

Calibration with external light sources 
will relay the initial calibration
and monitor any possible drifts. 
The proposed system consists of {\em optical beacons} which contain
high intensity pulsed light sources. They
will illuminate several strings simultaneously.
Ideally, the optical beacons should emit light at a wavelength
as close as possible to that which is least attenuated
in water ($\lambda \sim$ 470~nm) and should distribute the light
efficiently throughout the full solid angle with a pulse width of
about 1~ns. 
Pulsed solid-state lasers and blue LEDs are the 
focus of detailed studies.  
Figure
\ref{fig:laser_energy} illustrates the range to be obtained from 
such an optical beacon at various wavelengths, as a function of pulse
energy.

\begin{figure}[h]
\begin{center}
\begin{tabular}{cc}
\mbox{\epsfig{file=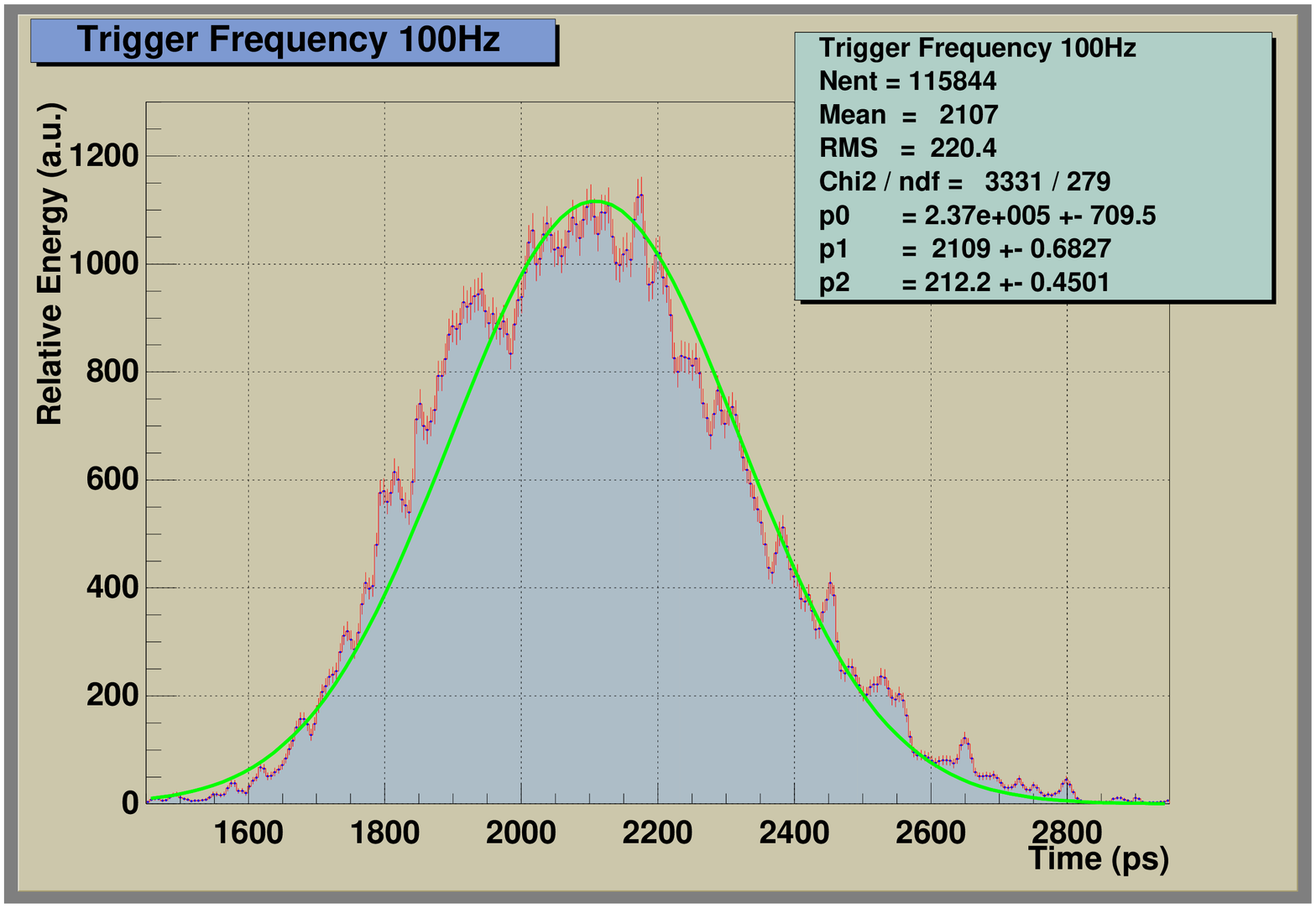,height=0.47\linewidth,
width=0.45\linewidth,angle=-90}} &
\mbox{\epsfig{file=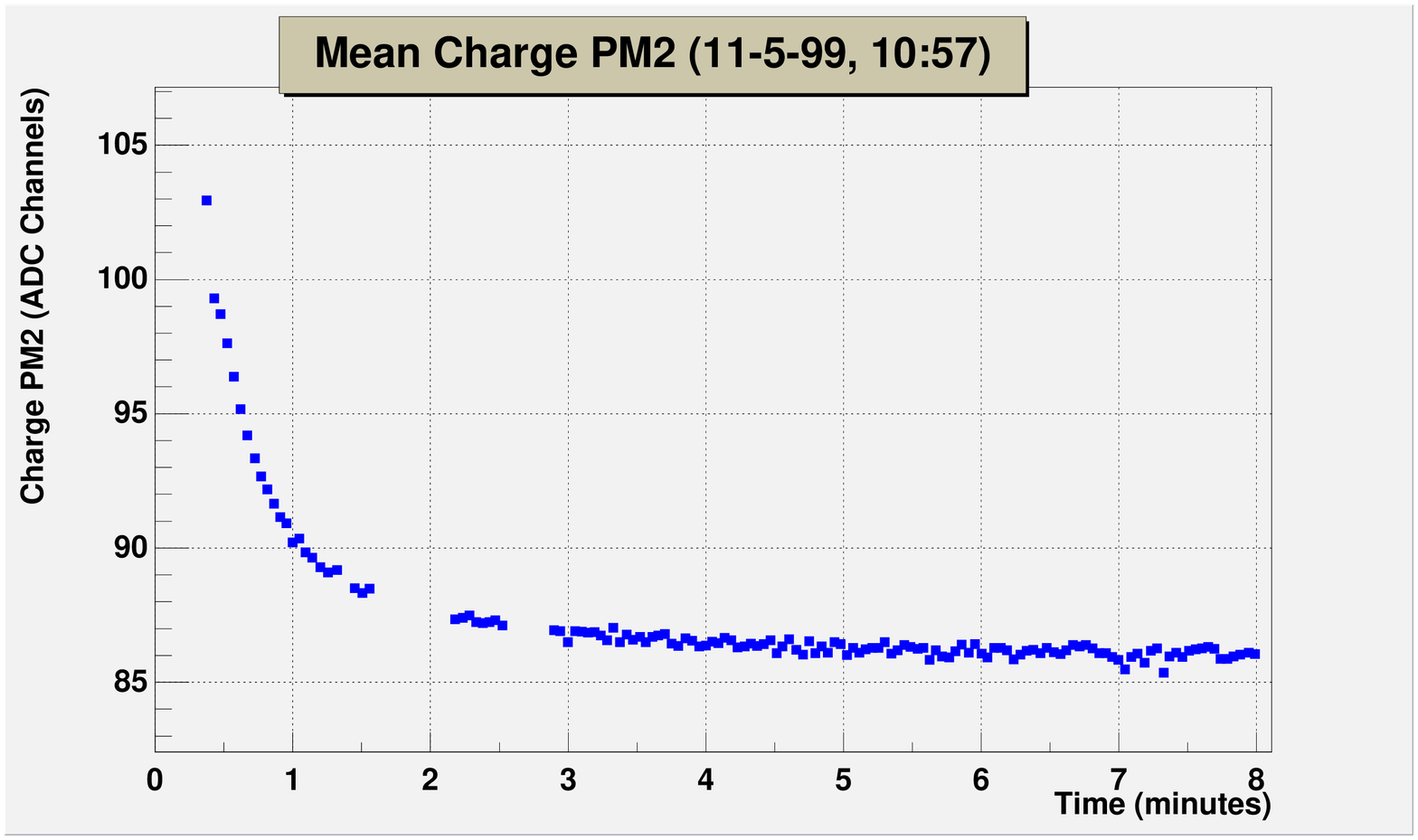,height=0.47\linewidth,
width=0.45\linewidth,angle=-90}}\\
\end{tabular}
\end{center}
\caption{\small Left: Time shape distribution (picoseconds) 
of a Nd-YAG laser pulse as measured by a streak camera. Right:
Warming-up curve (minutes) of a Nd-YAG laser as seen by a photomultiplier.
After a few minutes the output power stability is better than 1\% over
several hours.}
\label{fig:laser_features}
\end{figure}

Doubled Nd-YAG lasers which emit green light ($\lambda$ = 532~nm)
can be passively Q-switched to give sub-nanosecond pulses.
Lasers delivering pulse energies of $\sim$1~$\mu J$
($\sim10^{12}$ photons) and with a time spread (FWHM) of
$\sim$0.5~ns (see figure \ref{fig:laser_features}, left)
have been tested. 
These lasers are small, easy to operate and very robust, complying
with the stringent deployment and installation requirements of
the experiment.
      

After a warming-up period of a few minutes 
Nd-YAG lasers show an output power stability better than 
1\% over several hours (see
figure~\ref{fig:laser_features}, right).

Simulations have been performed to estimate the reach of
optical beacons using lasers together with lambertian diffusers 
($\sim cos\theta$). For 1~$\mu J$ 
Nd-YAG lasers, distances in excess of 200~m can be reached.

\begin{figure}[hbtp]
\begin{center}
\mbox{\epsfig{file=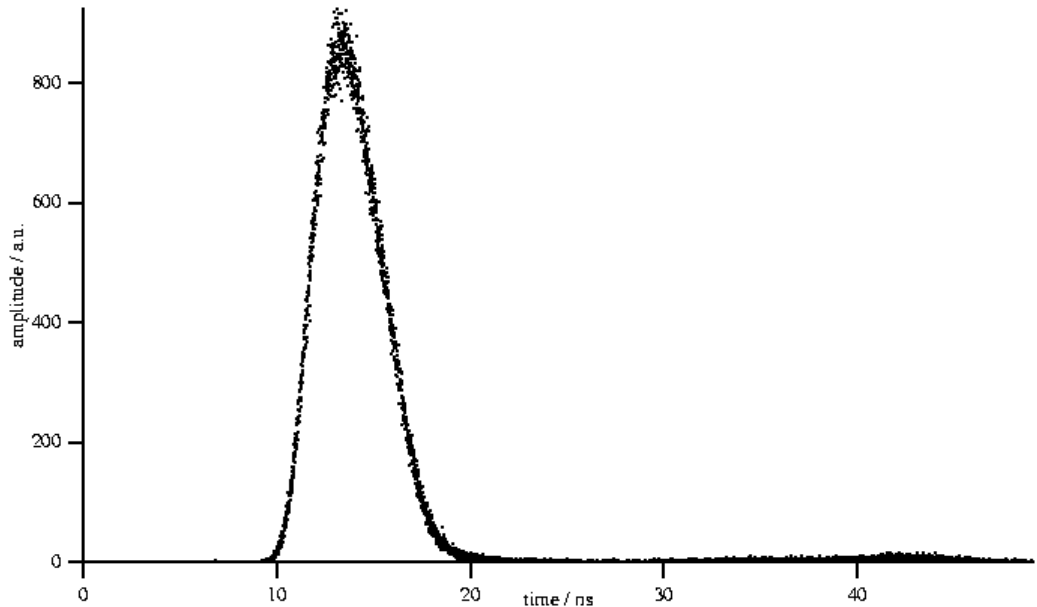,width=\linewidth}}
\end{center}
\caption{\small LED time response}
\label{fig:ledpulse}
\end{figure}

A LED pulser circuit based on that of 
Kapustinsky {\it et al.}~\cite{bib:chap5_led}
has been developed using recent GaAs LEDs from Nichia and Hewlett
Packard which emit in the blue ($\lambda$ =~470~nm).   The light pulse
produced has a risetime of 2.0~ns and a duration of 4~ns FWHM as can be seen
in figure~\ref{fig:ledpulse} which was measured using the single photoelectron
technique.

The intensity of a single flash is 50~pJ ($10^8$ photons).  The timing
jitter between the trigger pulse and the light flash is below 100~ps which
indicates that, it is possible to enslave large numbers of such
flasher
modules together to produce bright flashes.

\subsection{Optical Module efficiency calibration}
\label{sect:chap5_amplical}

The optical module efficiency may evolve with time due to
modifications of the PMT characteristics (gain, quantum efficiency, etc.), 
or to a change in the
transmission of light from the glass sphere surface to the PMT photocathode 
(fouling, ageing of gel, etc.).
As discussed previously, the \potasiu present in sea water, 
is a natural source of single photoelectron signals. 
The distribution of the collected charge from such events
can therefore be used to monitor the gain of the photomultipliers. 
An {\it in situ} measurement of the OM efficiency 
using an external calibrated light source will be a convolution
of three distinct effects: the fouling of the light source itself, 
variations of the light transmission in
sea water, and the OM's intrinsic efficiency, i.e.~the change in the response
of the OM in units of single
photo electrons. 
A relative calibration can, however, be done 
by comparing the response of several OMs to the high power
light sources,
 which will be installed in the detector for the purpose
of timing calibration.
The light from
these sources will arrive simultaneously at many OMs, allowing an accurate
cross-check of the variation in amplitude response of the OMs.
In the case of the Nd-YAG lasers, the accuracy of the measurement of the
amplitude is limited by statistics, not by the source stability.
Since the change in
efficiency is expected to be a slow phenomenon, the statistical
accumulation of calibration events will permit
an accurate determination.  In the same way,
the LEDs on each storey pointing at the OMs further up the string
can be used to measure the change in light transmission due to fouling and
other phenomena.

To avoid problems caused by fouling of the light source, a ship-tethered
calibration system is being considered for absolute efficiency calibration. 
This would consist of
an isotropic calibrated 
light source enclosed in a Benthos sphere and a
rangemeter to permit the accurate positioning of the source using the
acoustic positioning system of the detector.
An accuracy of a few tens of centimetres on distances of a few tens of metres 
between locations of the source in-between strings and the optical modules 
can be achieved and is satisfactory for our purpose. The optical module 
orientation with respect to the  light source is determined from the 
tiltmeter and compass information. Sufficient fluxes of photons can be
obtained by triggering simultaneously a large number of blue LEDs mounted 
inside a Benthos sphere. 
Absolute calibration of the optical modules requires
that the attenuation length of light be known.  This can be deduced directly
from data taken with the light source at a few different positions
with respect to a single optical module.

\section{Onshore data acquisition}


The object of the onshore data acquisition is to apply a level 3 filter to
reduce this rate to a 
reasonable level for archiving on tape, maintain an experiment status database
using the slow control information, and verify the integrity of the data.

The first step of the on-shore processing is event 
building.  Here, time-stamped data from various parts of the detector are 
assembled into events.  Most of the triggers are caused by accidental coincidences, and
a customised version of the off-line reconstruction will be 
used to filter these. 

The event building will associate both digital data from the optical modules 
and slow-control data in the same event.  Fully built events will be used for 
feedback in controlling the detector, as well as for event display and data 
monitoring.  A Unix-based event display has been developed and tested.  The 
programme for monitoring of slow-control parameters and event histogramming
has been written and tested for the prototype string connected to shore 
 (see section~\ref{sect:prototype_string}). It is based on the EPICS
package and will be used in the 0.1~km$^2$ detector.

%% file: chap6.tex

The neutrino sources discussed in the scientific programme fall into two 
distinct categories:
\begin{itemize}
\item high energy neutrinos (above 1~TeV) from astrophysical sources;
\item low energy neutrinos (below a few hundred GeV) for oscillation studies
and from the annihilation of neutralinos. 
\end{itemize}
The two classes behave differently in the detector --- for example, 
muons from high energy neutrinos will pass right through the detector volume, 
whereas the low energy neutrinos
may be partially contained --- so both must be
considered when optimising the detector design.  Extensive simulation
studies have been carried out on the detection of neutrinos from 
AGN, representing the
first class of high energy astrophysical neutrinos, and on the
measurement of neutrino oscillation parameters, which requires a good
understanding of sub-TeV neutrinos.  This chapter describes the
event simulation and reconstruction, the optimisation and performance 
of the detector in both energy regimes, the results of the simulation
studies, and a discussion of relevant systematic effects.

\section{Monte Carlo simulation tools}
\label{sect:MC_sim_tools}

This section
describes the software tools used 
for event generation and detector simulation.
Separate simulation packages are needed to treat signal
events and various sources of background.

\subsection{Event generation}
\begin{figure}
\begin{center}
\mbox{\epsfig{file=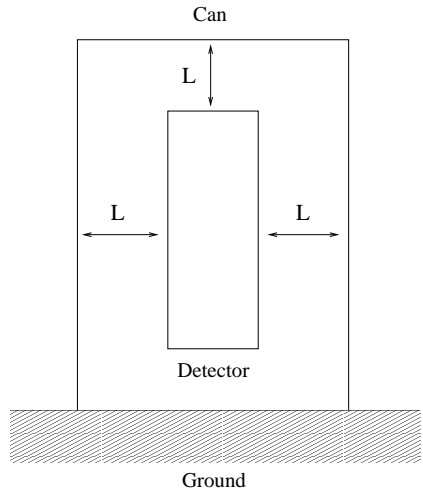,width=0.4\linewidth}}
\end{center}
\caption{\small Scheme of the `can' used for event generation. The distance
  $L$ is taken as 2.5 times the maximum absorption length for light.}
\label{fg:fig_3-2}
\end{figure}

Neutrinos from astronomical sources have been simulated
by generating the kinematic distributions of the neutrino interactions
and
taking into account the absorption and energy loss of the muons in the Earth.
Events are generated with a flat spectrum
in both the $\log$ of the neutrino energy and in the cosine of the
angle of incidence, then weights are applied
according to the flux of the sources.
Events are retained for analysis if there is a muon track
anywhere within a distance $L = 140$~m 
of the instrumented detector, as 
shown in figure~\ref{fg:fig_3-2}.

The main background for upward-going 
astronomical neutrinos 
comes from neutrinos produced by cosmic rays 
interacting in the atmosphere. 
For atmospheric neutrinos interacting
within the volume shown in figure~\ref{fg:fig_3-2},
tracks due to the hadronic showers, 
as well as those due to the muons, are generated
using the LEPTO package~\cite{lepto}.
For neutrinos
interacting outside this volume, 
only the tracks due to the muons are simulated.

Muons originating directly from the cosmic ray showers are simulated
with a single muon component and a multi-muon component.
High statistics samples of single muons have been 
generated with a parametrization of the flux at 
the detector depth~\cite{okada}
and also by propagating the sea level flux to the detector using
the package PROPMU~\cite{propmu}. 
The multi-muon component has been generated using the HEMAS
package~\cite{hemas} with the muons propagated to the detector using
PROPMU. At the present time,  only relatively small samples of multi-muons have
been produced.  

\subsection{Detector simulation}

The response of the detector to the various types of physics events
has been studied mainly using a version of the  DADA program~\cite{dada}
which originated in the BAIKAL collaboration. 
This program has been modified to improve the performance at high energies 
and to adapt it for the ANTARES detector.
In DADA, the muon is tracked using GEANT, and light coming from the secondary
particles generated
is parameterized for energies above 0.5 GeV for electrons,
above 1 GeV for photons, and above 10 GeV for hadrons.
Less energetic secondary particles are taken into account by increasing the
amount of Cherenkov light emitted by the muon 
(+~18\%~+~3.2\%~$\log~E_\mu~({\rm GeV})$).
A parametrization is used for the angular and longitudinal distribution of the
Cherenkov light initiated by electromagnetic and hadronic
cascades~\cite{dada}. 

Checks on some of the simulation results have been made with a new
program, GEASIM, developed for ANTARES. Some of the simulations of multi-muons
have also been  performed with the KM3 program~\cite{km3} which, unlike
the other programs, simulates the scattering of light in the water. 

If not specified otherwise, the simulation was performed 
with 10-inch photomultiplier tubes. All parameters used in
the simulation are based on the measurements described in 
section~\ref{sec:chap5_pmt}. 
The angular response simulated is that 
shown in figure~\ref{fig:chap5_gamelle_res},  
with a timing accuracy of 1.5 ns. 
For each optical module the time and amplitude of the pulse
corresponding to an event is simulated taking into account
the noise coming from the optical background, 
the gain of the photomultiplier
and the effect of the electronics.  All simulated events include 
random hits from \potasiu in the sea water.

Photons are scattered and absorbed by sea water. 
DADA includes the effect of absorption, 
with an absorption length of 55~m unless otherwise stated, 
but does not include scattering.
For a distance of 45 m between source and optical modules,
less than 10\% of the photons are scattered, 
so the approximation of no scattering 
should have little effect on the signal. 
Nonetheless, the scattering must be included for 
the background studies. 
No additional smearing has been incorporated into the simulations 
for the relative positioning of the optical modules.

\section{Track finding and reconstruction}

Before track reconstruction, a pattern recognition step is necessary to
remove hits from the  \potasiu background.
The hardware trigger will make requirements on the number of coincidences
between two of the three optical modules in each storey of the detector.
For the simulations of astrophysical performance a total of
at least four coincidences on at least two different strings were required;
the requirements for the oscillation study are discussed in section
\ref{subsect:tracks_for_oscillations} below.

\subsection{Pattern recognition}\label{pattern}

\begin{figure}
\begin{center}
\mbox{\epsfig{file=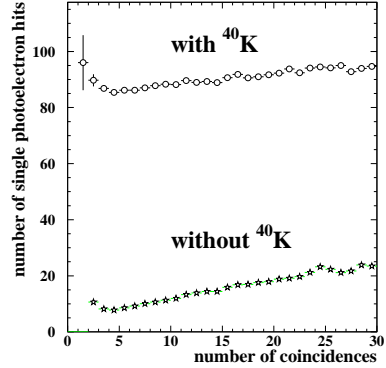,width=0.4\linewidth}}
\end{center}
\caption{\small Average number of single photoelectron hits 
  versus the number of
  coincidences for high energy muons, 
  with and without optical background. The simulated background rate
  is 20 kHz.}
\label{fg:fig_4-5}
\end{figure}

Figure~\ref{fg:fig_4-5} shows the number of
hits with amplitude corresponding to one photoelectron,
for a simulation of high energy muons with and without \potasiu background.
About 90$\%$ of single photoelectron
hits come from \potasiu. They cannot be removed by an amplitude cut without
a large loss in signal hits. 

\begin{figure}
\begin{center}
\mbox{\epsfig{file=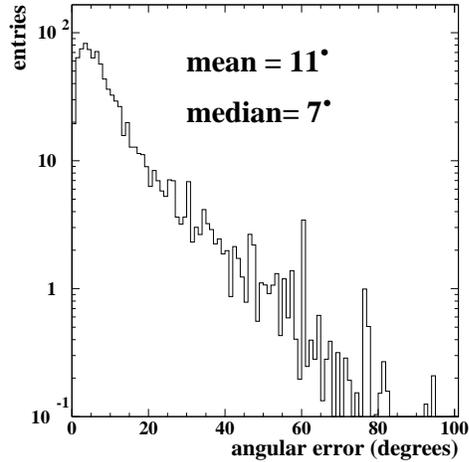,width=0.5\linewidth}}
\end{center}
\caption{\small Angular error for the prefit solution for 
  high-energy events with at
  least 4 coincidences on at least 2 strings, for an initial $1/E^2$ neutrino
  spectrum.} 
\label{fg:fig_4-6}
\end{figure}

The method adopted to remove the \potasiu
hits is to start with a track fit 
which uses only coincidence hits. 
Figure~\ref{fg:fig_4-6} shows the angular error of this prefit 
when four coincidences are required.
The prefit is used to define a road in time 
for the selection of
single hits to be used in a further fit.
The hits are kept
if the time difference between
measurement and expectation is between $-150$ and $+50$~ns
and if the  distance between the track and the optical module is smaller
than 100~m. This selection reduces the number of hits
due to the optical background by a factor of 100, 
while keeping 46\% of the 
single photoelectron hits associated with the muon tracks.

\subsection{Track reconstruction}\label{track_rec}
\begin{figure}
\begin{center}
\mbox{\epsfig{file=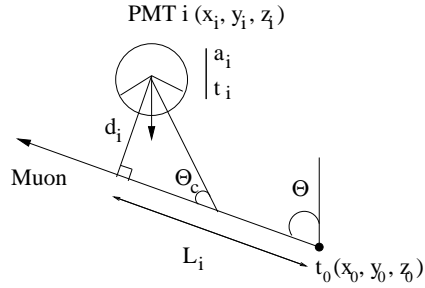,width=0.4\linewidth}}
\end{center}
\caption{\small Propagation of a muon and its associated Cherenkov light.}
\label{fg:fig_3-14}
\end{figure}

For each optical module, the arrival time of the
Cherenkov light is
\begin{displaymath}
(t_i)_0 = t_0 + (L_i + d_i \tan \theta_c)/c
\end{displaymath}
where $d_i$, $L_i$ are defined in figure~\ref{fg:fig_3-14}, $\theta_c$ is 
the Cherenkov angle, and $t_0$ is the time
at a reference point. Then $(t_i)_0$ is
smeared out by the time resolution (PMT transit time
spread, positioning error, scattering) to give the pulse time $t_i$.
Photons from secondary particles will also be detected, 
with a distribution of arrival times that decreases
 exponentially with respect to
$(t_i)_0$. Photons from the optical background 
arrive at random times.

\begin{figure}
\begin{center}
\mbox{\epsfig{file=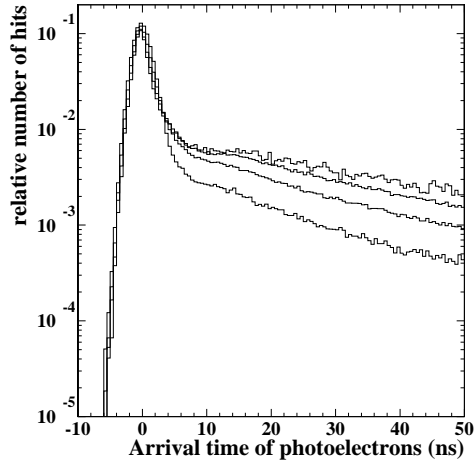,width=0.5\linewidth}}
\end{center}
\caption{\small Unnormalised probability density function
$P(\Delta t)$ for the 
arrival time of hits originating from muon tracks with four different 
energies: 2~TeV, 10~TeV, 50~TeV and 250~TeV.}
\label{fg:fig_3-15}
\end{figure}

Figure~\ref{fg:fig_3-15} shows the expected 
time distribution of the hits for different muon energies. 
This is the probability distribution function $P(\Delta t)$ 
which
depends on the muon energy. In order to simplify the fitting
procedure, the energy dependence is neglected. 
The probability distribution is evaluated 
either at the most-likely energy value or as an average
over the energy range under study.

First a preliminary estimation (prefit) of the track parameters is
made, based on a linear fitting procedure~\cite{DumandPrefit}.  
Then the full fit is made, using a likelihood function (${\cal L}$) 
which is the product of the
probability for each optical module $i$ registered at time $t_i$~: 

\begin{displaymath}
{\cal L} = \prod_i P(t_i - (t_i)_0)
\end{displaymath}

\noindent
The maximisation of  ${\cal L}$ by numerical methods yields the five track
parameters $x$,~$y$,~$z$~(or~$t$),~$\theta$, and~$\phi$, 
as well as a covariance matrix. 
This matrix provides the errors which will be used 
when applying selection criteria.  


\section{Astrophysical neutrinos}

The principal interest of a deep-sea neutrino detector is the study of
astronomical sources such as active galactic nuclei and gamma-ray bursters.
These are expected to produce very high energy neutrinos yielding TeV--PeV
muons which will not stop within the detector fiducial volume (uncontained
events).  This section presents the results of a simulation study for 
different models of neutrino production in AGN.

\subsection{Event selection}\label{cuts}
After the track reconstruction, a selection is required in order to keep 
well-reconstructed events. The main selection criteria are:
\begin{itemize}
\item the errors on parameters of the reconstructed tracks have to be small:\\
      $\delta_x, \delta_y<$ 5~m \hspace{1cm} $\delta_{\theta} < 1^{\circ}$ 
      \hspace{1cm}
      $\delta_{\Phi} < 2^{\circ}$ \hspace{1cm} $\delta_t <$ 6~ns;
\item the distance between the initial muon track given by the prefit and the
      final track is required to be
      less than 150~m and the angle between them less than 45$^{\circ}$.
\end{itemize}

\begin{figure}
\begin{center}
 \begin{tabular}{cc}
  \subfigure[]{\epsfig{file=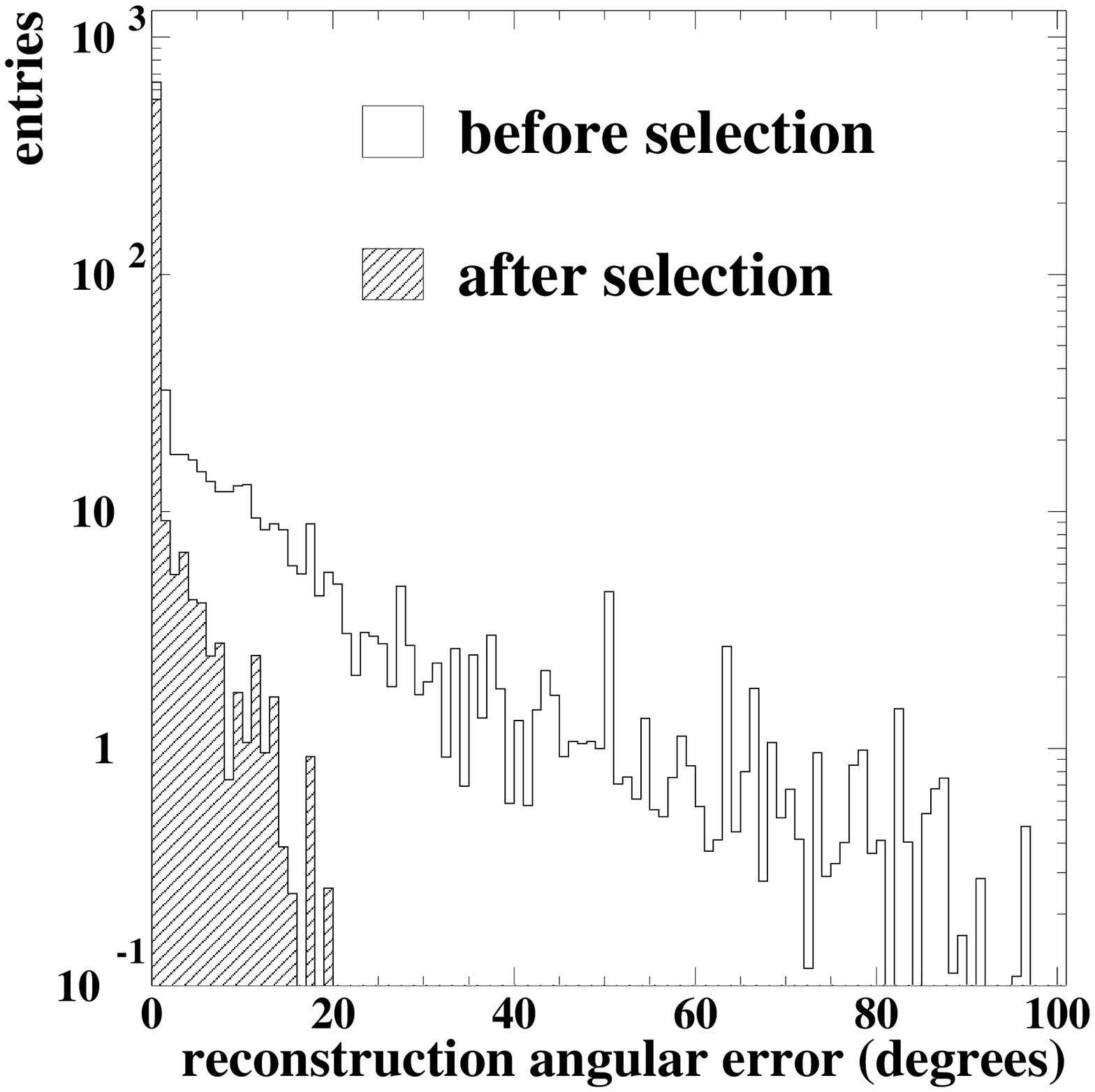,width=0.4\linewidth}}
 &
  \subfigure[]{\epsfig{file=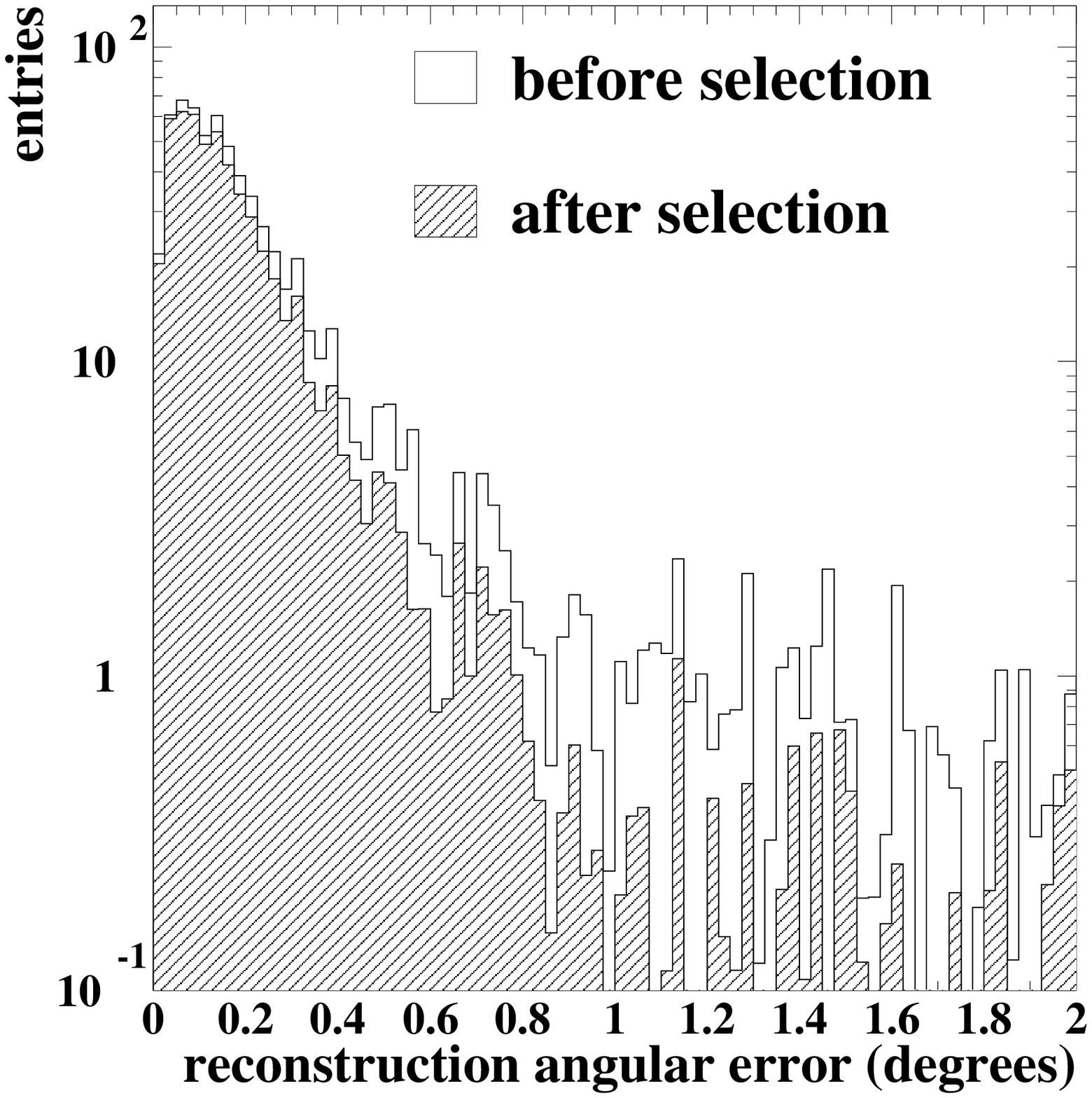,width=0.4\linewidth}}\\
\end{tabular}
\end{center}
\caption{\small Angular error for the reconstruction with and without
  selection (a)~for the full range of errors, 
  and (b)~for the low error range only.} 
\label{fg:fig_4-15}
\end{figure}

These criteria have to be optimised for each physics channel. 
The effects of one set of criteria dedicated to neutrino astronomy are shown in
figure~\ref{fg:fig_4-15}. A dramatic reduction is observed 
in the tail of the distribution.  
These selection criteria 
will be used in the following to obtain the expected detector
performances for diffuse sources of neutrinos.

\subsection{Optimisation for high energy events}

Given an overall conceptual design of a 0.1 km$^2$ detector 
with about 1000 optical modules, a number of detailed design parameters
can be varied to optimise the performance of the detector.
This process is not completely intuitive since
the optimization procedure is dependent on the physical processes under
consideration. The objective has been to define a  
detector with a good angular resolution, a large effective area and a good
rejection against cosmic ray muons. It will be shown that the energy
resolution is not heavily dependent on the geometry.

\begin{figure}
\begin{center}
\mbox{\epsfig{file=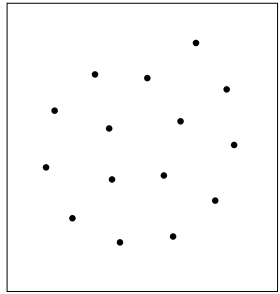,width=0.35\linewidth}}
\end{center}
\caption{\small Strings arrangement proposed to avoid symmetries.}
\label{fg:fig_4-4-b}
\end{figure}

The strings are arranged as shown in figure~\ref{fg:fig_4-4-b} in order to 
avoid potential symmetries which could contribute to ambiguities in the
reconstruction. 

The detector parameters that have been considered are:
\begin{itemize}
\item the number of photomultipliers $n$ in a local cluster ($n$ = 2, 3, 4)
 and their orientation.\\
      $n$ = 2 \hspace{1cm} the PMTs look downwards (2)\\
      $n$ = 3 \hspace{1cm} the PMTs are at 45$^{\circ}$ below the horizon (3)\\
      $n$ = 4 \hspace{1cm} the PMTs are horizontal (4H)\\
      \hglue 1.8cm {\em or} the PMTs are at 45$^{\circ}$ below the horizon (4S) ;
\item the distance between two local clusters~: $d_z$ = 8, 12, 16, 20~m ;
\item the distance between two strings~: $d_H$ = 60, 80, 100, 120~m.
\end{itemize}

Other parameters such as PMT size and the optical
properties of water affect the optimisation. 
It has been assumed that they are decoupled. The
optimisation is made by changing them one at a time.
Checks are made to ensure 
that the angular and the energy resolutions are not affected
by the geometry, to first order. 
The uniformity of the detector response as a 
function of the zenith angle has also been considered.
The following figures show the sensitivity of the results to the various 
detector geometry parameters.  Optimisation for specific physics analyses
will be considered later.

\begin{figure}
\begin{center}
\mbox{\epsfig{file=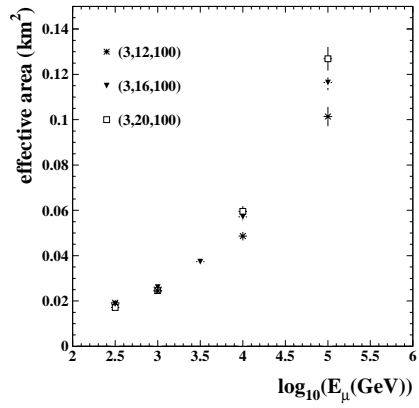,width=0.4\linewidth}}
\end{center}
\caption{\small Effective detector area as a function of the muon energy 
  when varying the vertical distance between two local clusters $d_z$. 
  The area is averaged over all muons coming from the lower hemisphere.}
\label{fg:fig_6-1}
\end{figure}

The effect of varying the parameter $d_z$ is illustrated in
figure~\ref{fg:fig_6-1}. The value of $d_z$ = 16~m has been chosen; the 
corresponding effective area is close to the value found for 
$d_z$ = 8~m, and the zenith-angle response is more uniform.
           
\begin{figure}
\begin{center}
\mbox{\epsfig{file=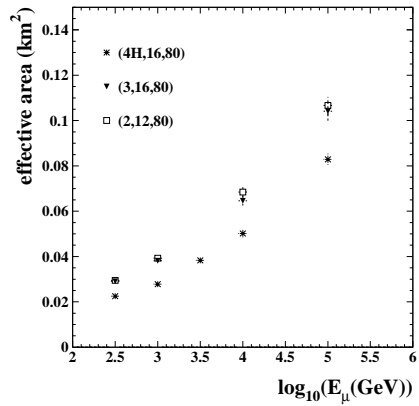,width=0.4\linewidth}}
\end{center}
\caption{\small Effective detector area as a function of the muon energy 
 when varying the number of photomultipliers ($n$) in a local cluster. 
 The area is averaged over all muons coming from the lower hemisphere.}
\label{fg:fig_6-2}
\end{figure}

Figure~\ref{fg:fig_6-2} shows the effect of the number of photomultipliers
in a storey (the total number of optical modules is kept constant). 
The geometry with three photomultipliers oriented at 45$^\circ$ below the
horizontal is preferred to that with four horizontal photomultipliers
on the same argument as above, namely uniformity in zenith angle.

\begin{figure}
\begin{center}
\mbox{\epsfig{file=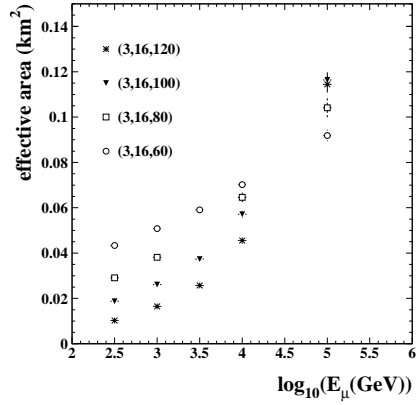,width=0.4\linewidth}}
\end{center}
\caption{\small Effective area as function of the muon energy, varying the
  horizontal distance between two strings $d_H$. The area is
  averaged over all muons coming from the lower hemisphere.}
\label{fg:fig_6-4}
\end{figure}

Figure \ref{fg:fig_6-4} shows
the variation in effective area when modifying the parameter $d_H$; 
the choice of $d_H$ depends on the energy range to be optimised.
A 60~m spacing ensures the best effective area for energies up to
some tens of TeV and becomes less efficient for higher energies.
It should be noted that the effect of scattering in water has not been
considered in this study. By taking this into account
and thus having a smaller effective attenuation length
it is expected that smaller string spacings will be favoured 
even at the higher energies.

The configuration which was chosen for the following studies has 15
strings with an inter-string distance of 80 m.  Each string is equipped
with 21 storeys spaced vertically by 16 m,  each storey consisting of 3 
photomultipliers oriented at 45$^\circ$ below the horizontal. 

\begin{figure}
\begin{center}
\mbox{\epsfig{file=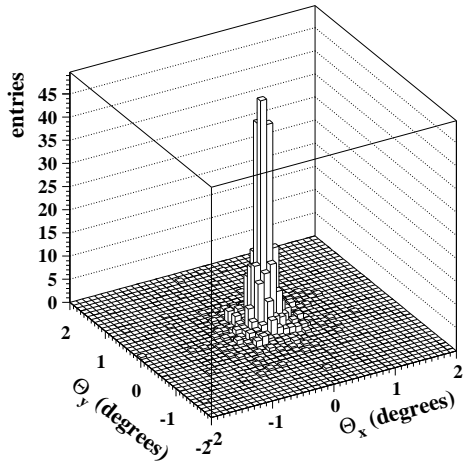,width=0.5\linewidth}}
\end{center}
\caption{\small Angular resolution for a $1/E^2$ neutrino spectrum,
for $\sigma_t = 1.3$~ns.}
\label{fg:fig_4-19}
\end{figure}   
\begin{figure}
\begin{center}
\mbox{\epsfig{file=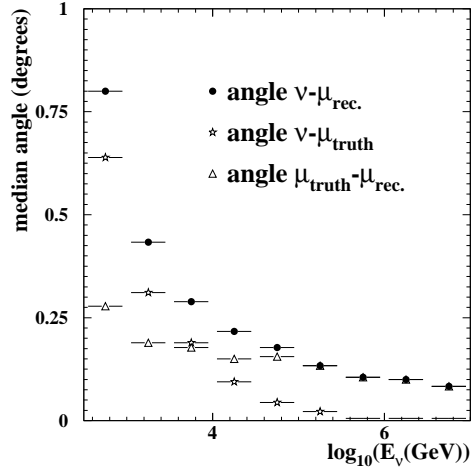,width=0.5\linewidth}}
\end{center}
\caption{\small  Median angles as a function of $\log(E_{\nu})$. The total
  angle, the physical angle and the reconstruction error are indicated.}
\label{fg:fig_4-18}
\end{figure}

\subsection{Angular resolution}
 
The angular resolution depends on the algorithms used, on the
selection applied to the reconstructed events and on the timing accuracy
which was discussed in section~\ref{sect:chap5_calipos}.

The angular resolution is shown in figure~\ref{fg:fig_4-19} for 
$\sigma_t = 1.3$~ns. Half of the events from a simulated point source 
with a $1/E_\nu^2$ spectrum fall inside a circular region of 
radius $0.2^\circ$. 
One hemisphere can be effectively divided into 200,000
pixels.

It is worth noting that above 10~TeV the angular resolution is not dominated
by the physical angle between the neutrino and the real muon, but by the
reconstruction errors (see figure~\ref{fg:fig_4-18}).

\subsection{Energy resolution}
\label{subsubsect:energy_estimation}

For energies above 1~TeV, the muon energy loss is
dominated by catastrophic energy loss 
from bremsstrahlung and pair production,
both of which increase with energy. 
The amount of light produced by the track is proportional to
the energy loss
and this allows an estimation of the muon energy.
An estimator $x$ which is roughly proportional to the energy 
is formed using the number of hits, their 
observed pulse-height amplitudes $a_i$, 
and the amplitudes $a_i(0)$ expected for a minimum-ionizing 
particle~: 

\begin{displaymath}
 x = N_{hit} \left( \frac{\sum_{i} a_i}{\sum_{i} a_i (0)} - 1\right)
\end{displaymath}

\noindent
Since the energy information comes from stochastic processes,
a truncated mean is used to decrease the fluctuations.
Hits are selected for the energy estimation depending on
whether they are lower-energy hits (smaller pulse heights)
or higher-energy hits (larger pulse heights).
Only hits which are within 
4~ns of the arrival time expected from the reconstructed track 
($|t_i-(t_i)_0|~<$~4~ns) are candidates for use in the estimators.
Hits are included in the calculation of 
the low- and high-energy estimators if they satisfy 
one of the following criteria~:

\begin{itemize}
\item $0.1 < \frac{a_i}{a_i (0)} < 100$ for the low-energy estimator,
\item $10 < \frac{a_i}{a_i (0)} < 1000$ for the high-energy estimator.
\end{itemize}

\begin{figure}
\begin{center}
 \begin{tabular}{cc}
  \subfigure[]{\epsfig{file=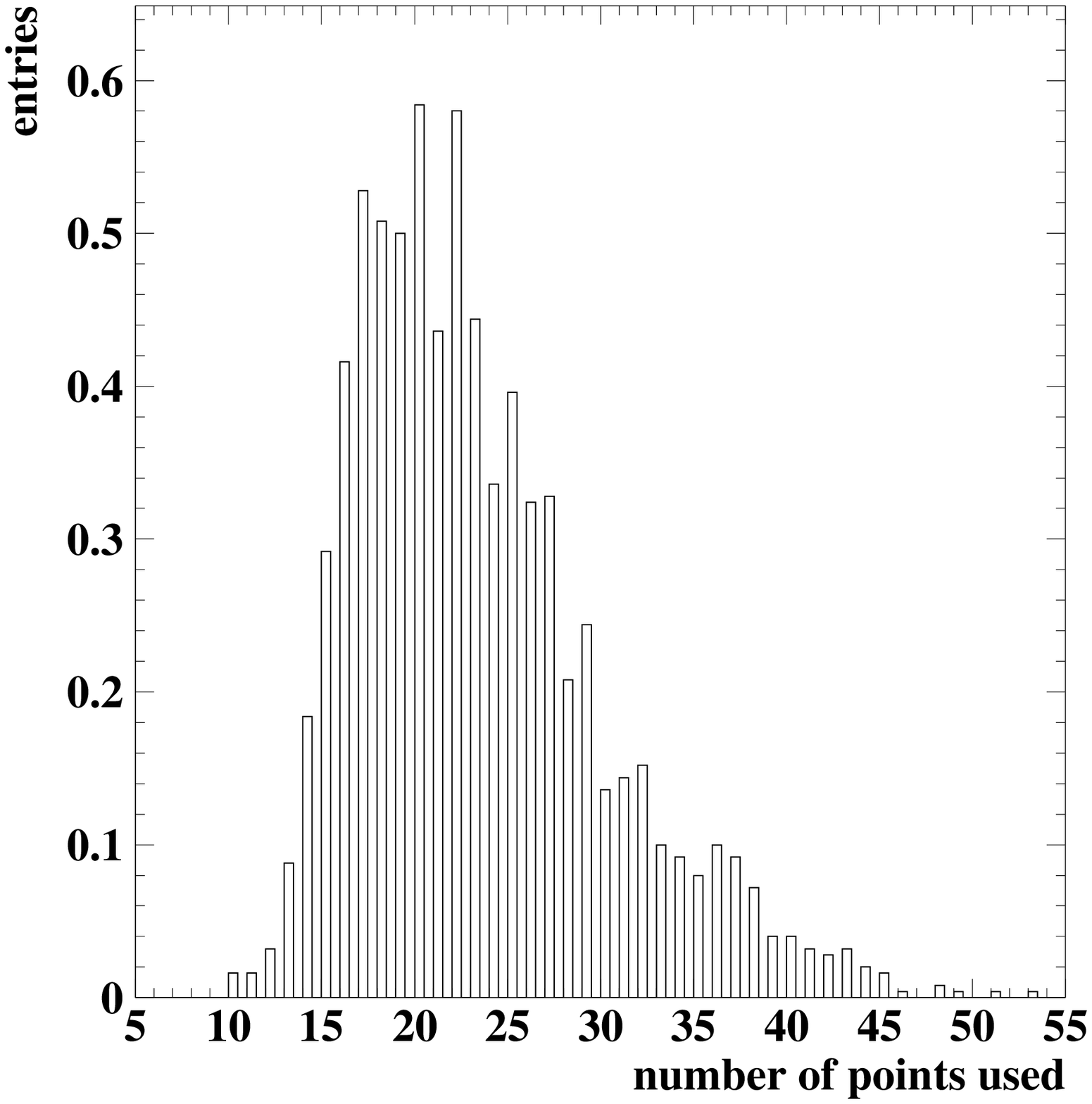,width=0.4\linewidth}}
 &
  \subfigure[]{\epsfig{file=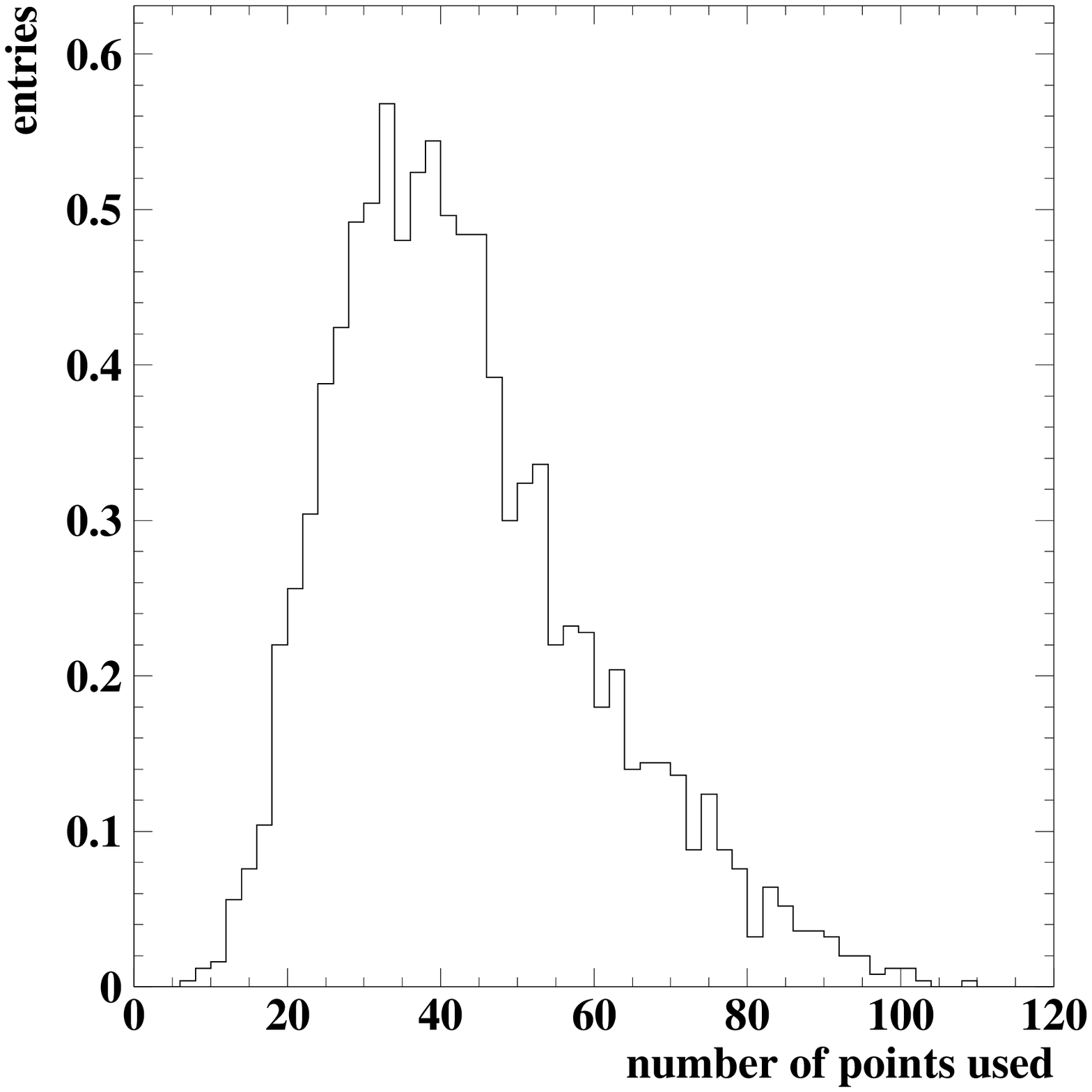,width=0.4\linewidth}}\\
\end{tabular}
\end{center}
\caption{\small Number of hits used for the energy evaluation (a) at 2~TeV,
  using the low energy muon estimator,  
  and (b) at 50~TeV, using the high energy estimator.} 
\label{epoints}
\end{figure}

\noindent
The number of hits satisfying the selection criteria
is shown in figure~\ref{epoints}(a) for a lower-energy muon (2~TeV) 
and in figure~\ref{epoints}(b) for a higher-energy muon (50~TeV).
The low-energy estimator is used if more than 10\% 
of the hits are accepted for that estimation;
otherwise the high-energy estimator is used.

\begin{figure}
\begin{center}
 \begin{tabular}{cc}
  \subfigure[]{\epsfig{file=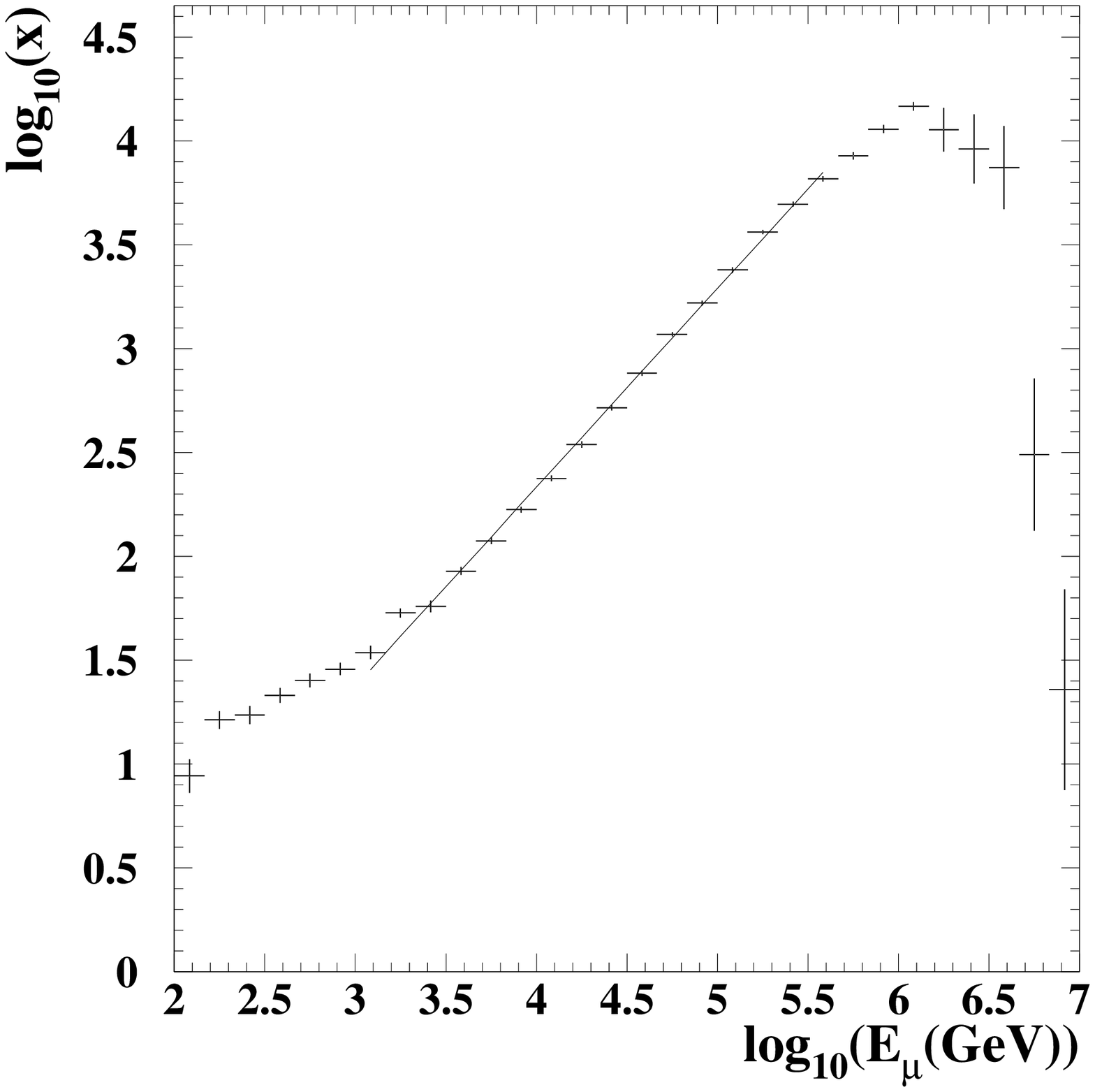,width=0.4\linewidth}}
 &
  \subfigure[]{\epsfig{file=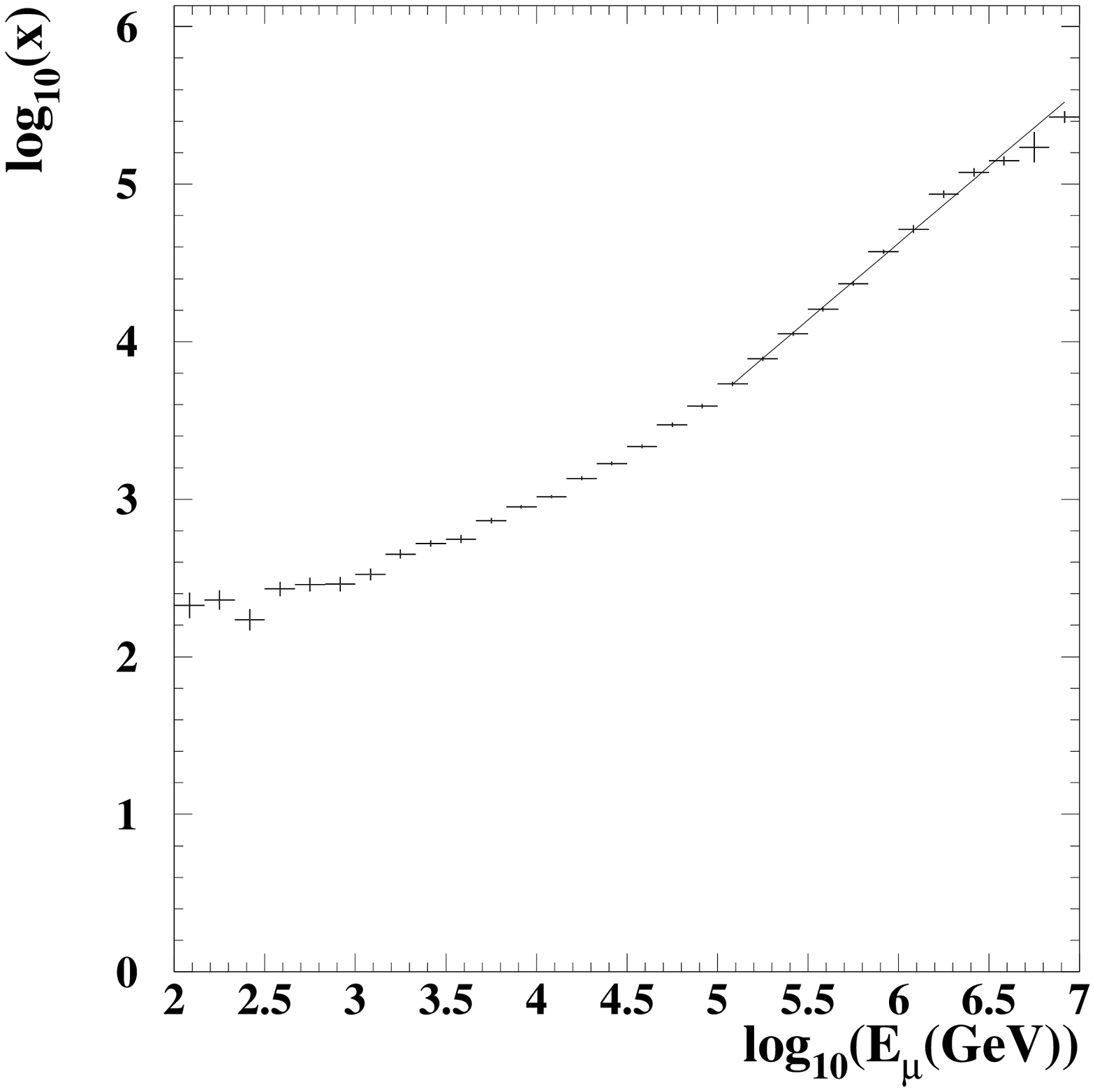,width=0.4\linewidth}}\\
\end{tabular}
\end{center}
\caption{\small  Parametrization of the low energy (a) and high energy (b)
  estimators. }
\label{fg:fig_3-21}
\end{figure}

The average value of log($x$) as a function of 
log($E_\mu$) is shown in figure~\ref{fg:fig_3-21} 
for events selected for the low- and high-energy
estimations.
The linear portions of these curves are used to estimate the
muon energy between 1~TeV and 10~PeV. 
The following parametrizations are obtained~: 

\begin{displaymath}
 \log(E) = 
 \left\{ \begin{array}{ll}
               1.56 +  1.04 \times \log(x) 
               & \mbox{for the low-energy estimator}  \\
               1.26 +  1.02 \times \log(x) 
               & \mbox{for the high-energy estimator}                 
         \end{array}
 \right.
\end{displaymath}

\begin{figure}
\begin{center}
\mbox{\epsfig{file=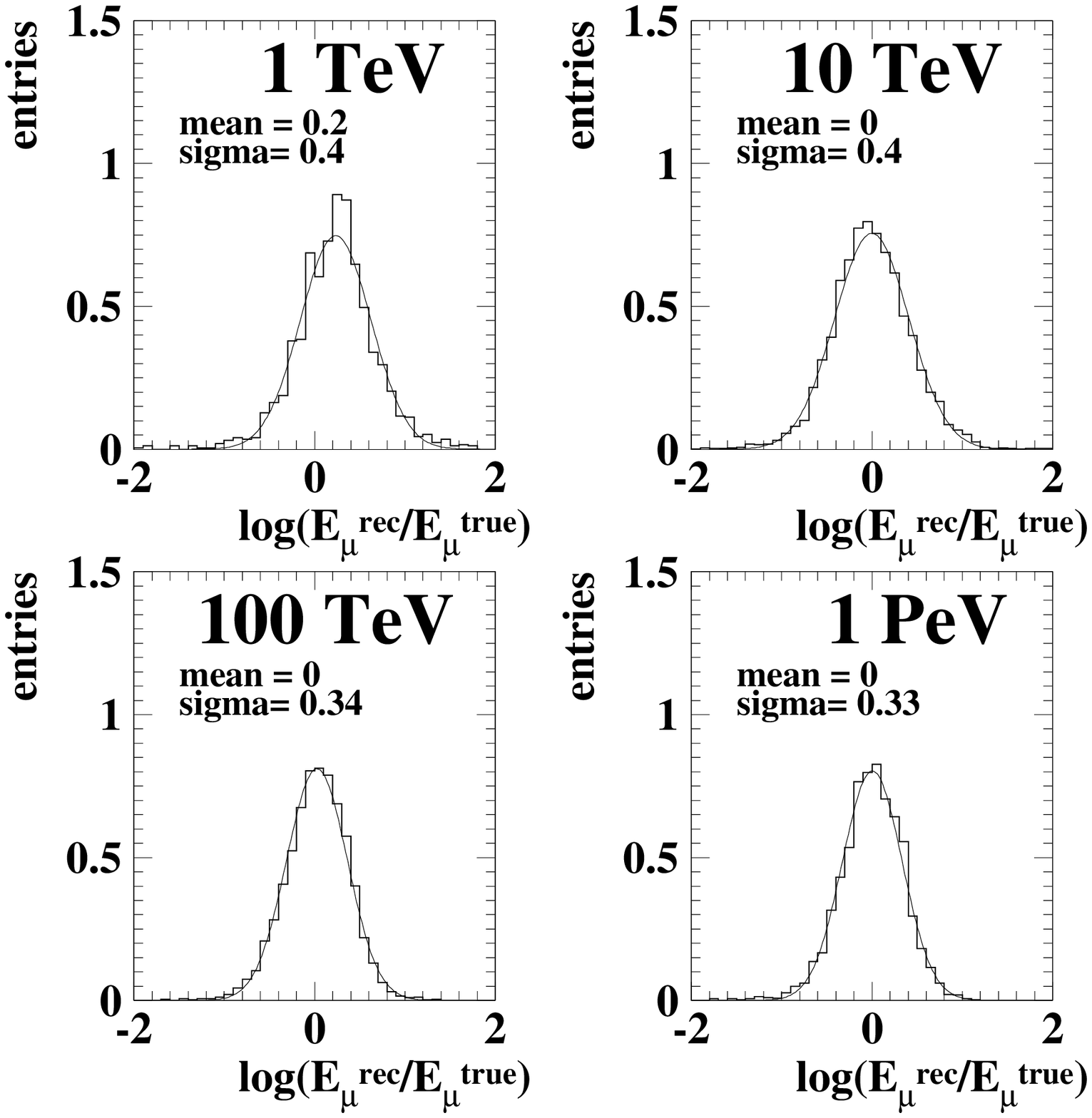,width=0.6\linewidth}}
\end{center}
\caption{\small The reconstructed muon energy for different true energies.}
\label{fg:fig_4-22}
\end{figure}

Figure~\ref{fg:fig_4-22} shows the reconstructed energy spectrum 
for simulated muons with energies from 1~TeV to 1~PeV.
The resolution on $\log E$ is 0.40 at 1~TeV, decreasing to 0.33 at 100~TeV; 
this implies a muon energy resolution $\Delta E / E$ of about a factor 3
below 10~TeV, improving to a factor 2 above 10~TeV.

\begin{figure}
\begin{center}
 \begin{tabular}{cc}
  \subfigure[]{\epsfig{file=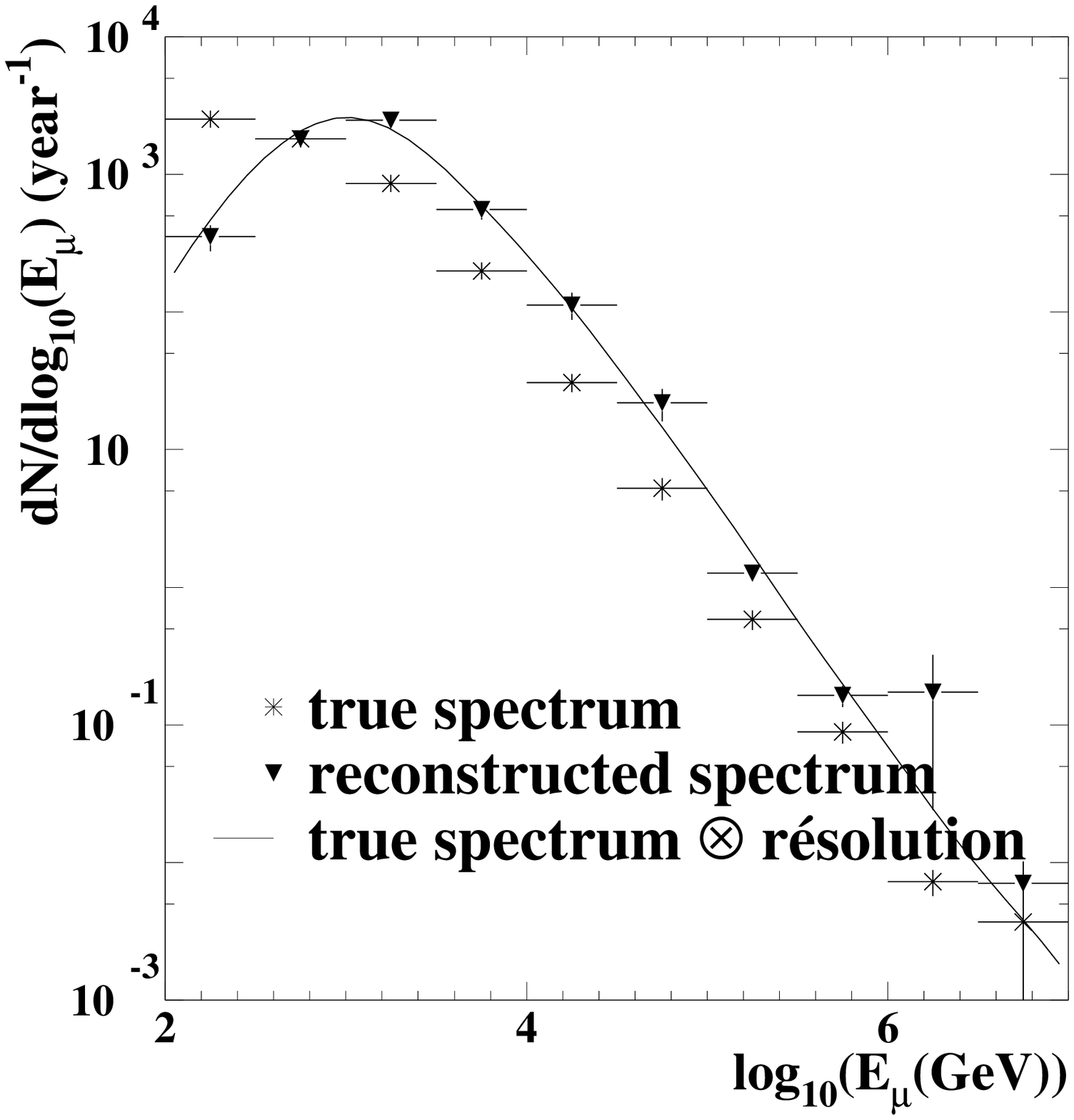,width=0.4\linewidth}}
 &
  \subfigure[]{\epsfig{file=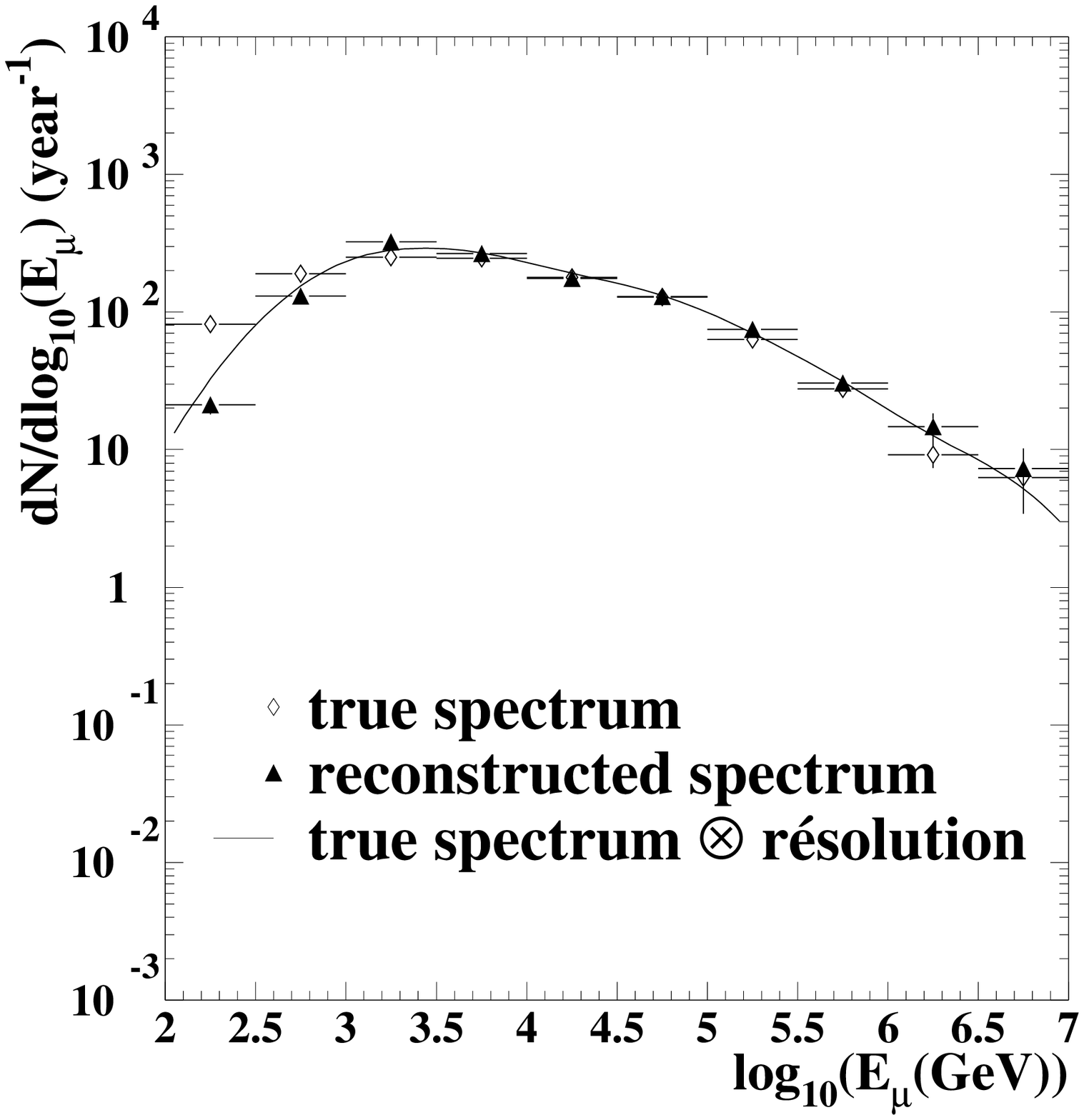,width=0.4\linewidth}}\\
\end{tabular}
\end{center}
\caption{\small True and reconstructed spectra for (a) muons coming from
  atmospheric  neutrinos and (b) muons coming from AGN (NMB model).}
\label{fg:fig_4-25}
\end{figure}

\begin{figure}
\begin{center}
 \begin{tabular}{cc}
  \subfigure[]{\epsfig{file=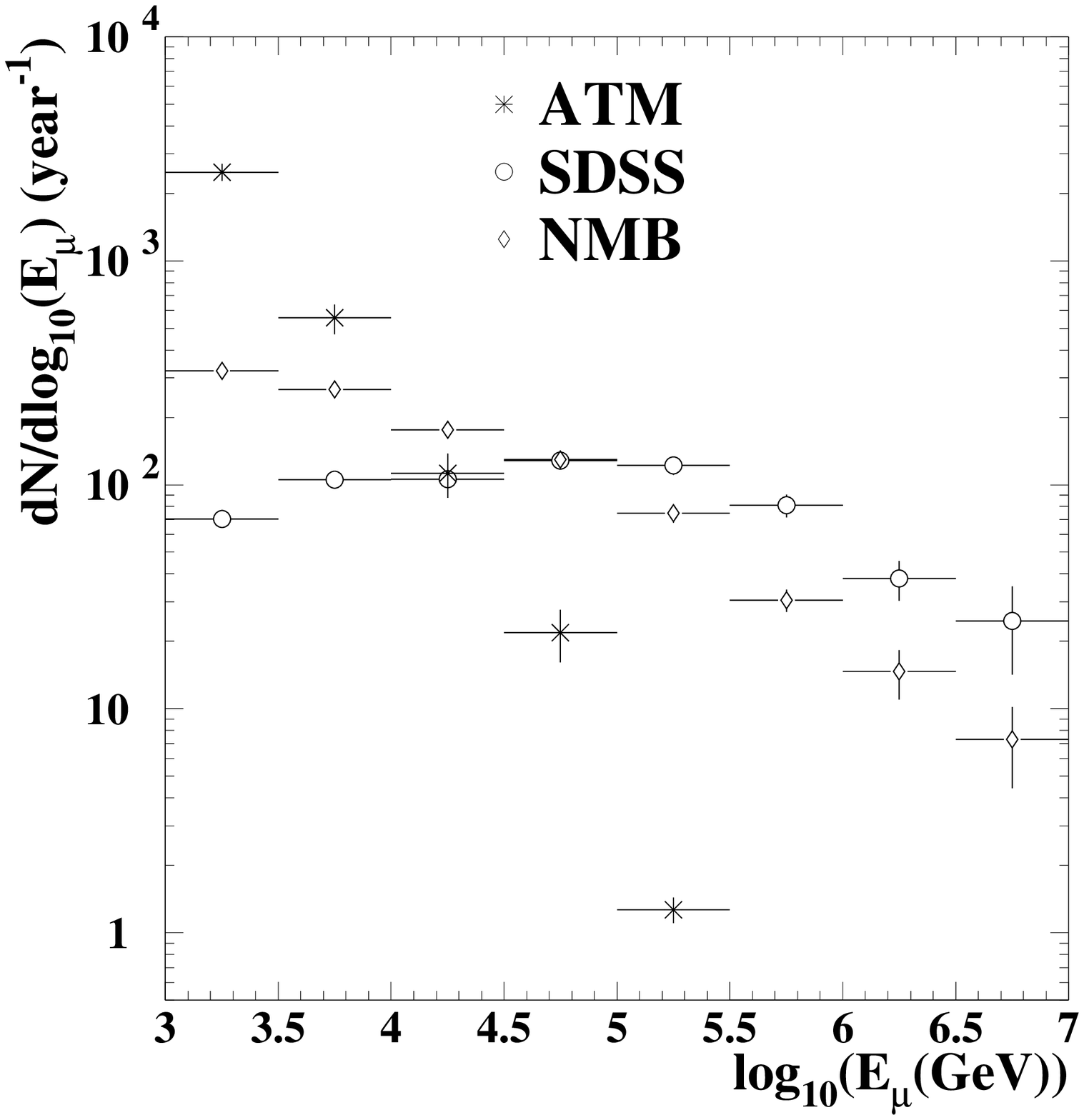,width=0.4\linewidth}}
 &
  \subfigure[]{\epsfig{file=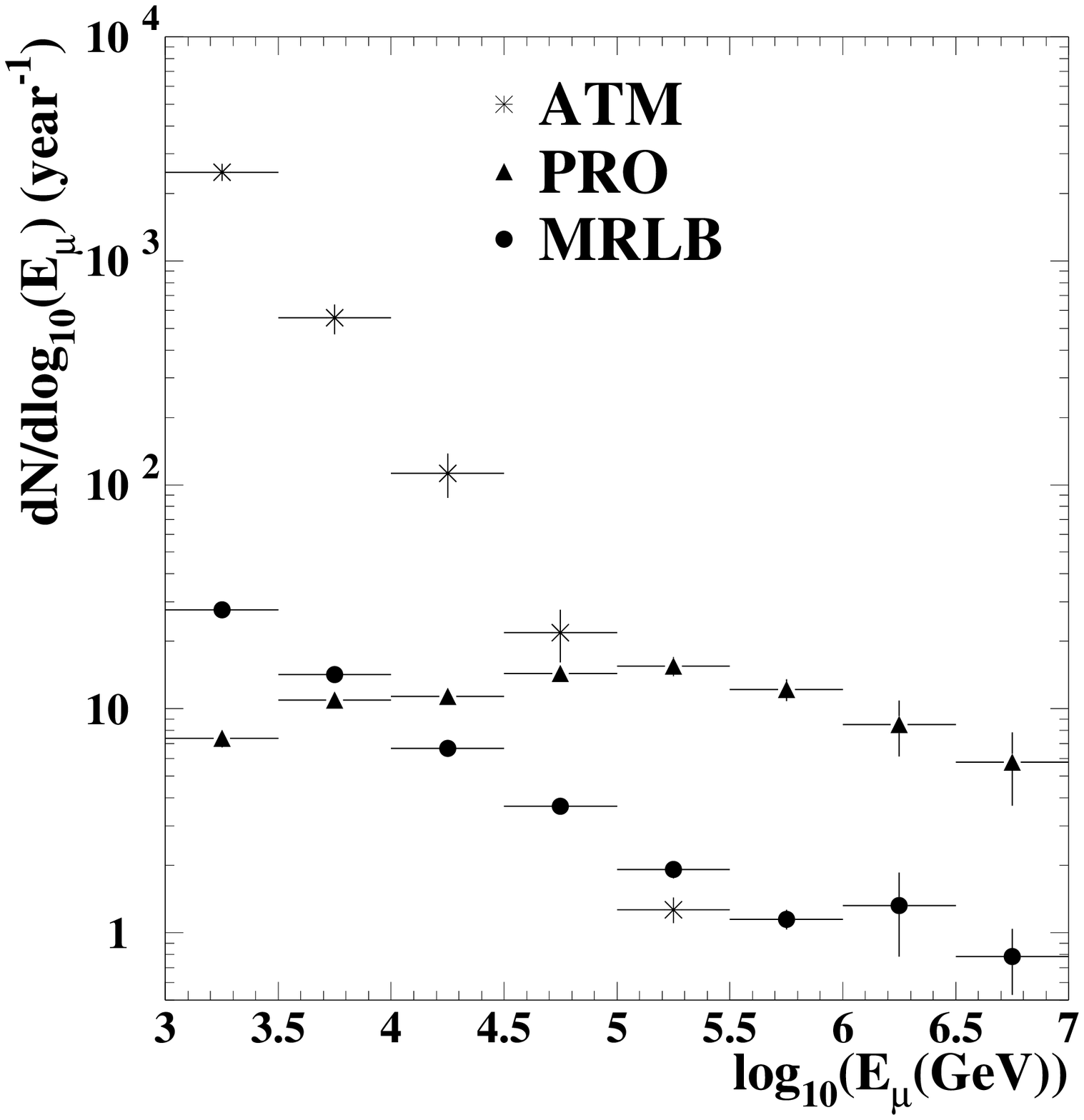,width=0.4\linewidth}}\\
\end{tabular}
\end{center}
\caption{\small Reconstructed spectra of muons coming from 
(a)~generic AGN and (b)~blazars, compared to 
muons coming from atmospheric
  neutrinos~(ATM).}
\label{fg:fig_4-26}
\end{figure}

In figure~\ref{fg:fig_4-25}, it can be seen that taking into account the
energy resolution enables a correct energy spectrum to be unfolded.
By placing a cut on the reconstructed energy (figure~\ref{fg:fig_4-26})  
it is expected that the contamination of
atmospheric neutrinos can be controlled.


\subsection{The diffuse neutrino flux from active galactic nuclei}

The integrated neutrino flux from a large population of distant AGN generates
a diffuse neutrino flux (similar to the diffuse X-ray background)
even if individual sources are not detectable. Calculations of this diffuse
flux have been done for four models of neutrino production by AGN:
\begin{itemize}
\item SDSS~\cite{sdss}: generic AGN model in which the interactions of the
protons with the material of the accretion disk dominate;
\item NMB~\cite{nmb}: generic AGN model in which the dominant interaction is 
with the ambient radiation field;
\item PRO~\cite{pro}: blazar model where the protons are accelerated in the 
innermost region of the jets and interact principally with accretion disk
radiation;
\item{MRLA--B}~\cite{mannheim}: lower and upper limits, respectively, of a
blazar model in which the protons are accelerated in the jets and interact
primarily with synchrotron radiation produced by relativistic electrons
in the jets.
\end{itemize}
\noindent  
Other calculations of diffuse neutrino sources are described 
in reference~\cite{Gaisser95}. 

\begin{figure}[h]
\begin{center}
\mbox{\epsfig{file=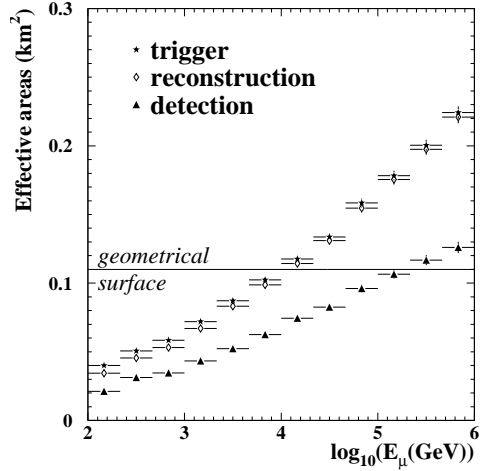,width=0.5\linewidth}}
\end{center}
\caption{\small Effective area for triggering, reconstructed and selected
  events. The areas are averaged for muons coming from the lower hemisphere.}
\label{fg:fig_4-9}
\end{figure}

Figure~\ref{fg:fig_4-9} shows the effective area for the 
events which trigger the detector, for those which are reconstructed and 
those which satisfy the selection criteria (see section~\ref{cuts}). 

The corresponding counting rates of accepted events are summarised in 
table~\ref{tab:4-3} for different diffuse sources (atmospheric
neutrinos and AGNs) in the optimised detector  with different (true) muon 
energy thresholds. For the same sources, the expected number of events above
the  reconstructed muon energy threshold is given in table~\ref{tab:4-4}.

The sensitivity is model dependent, and is affected by the rejection
of cosmic ray muons and by the measurement of the muon energy.
Beyond 10~TeV, if no event rate above that for atmospheric neutrinos is
observed (this corresponds to $\sim$ 40~events per year), 
then a minimum detectable $1/E^2_\nu$ flux at the 5 $\sigma$ level of~ 
\bec
\frac{d\Phi}{dE_\nu} > 6\cdot 10^{4} \left(
\frac{E_\nu}{1 \mathrm{TeV}}
\right)^{-2} \;\; \mathrm{km}^{-2}\;\; 
\mathrm{yr}^{-1}\;\;\mathrm{sr}^{-1}\;\;\mathrm{GeV}^{-1}
\eec
is obtained (assuming that the down-going cosmic
muons can be successfully rejected). 

\begin{table}[p]
\begin{center}
\begin{tabular}{|l||c|c|c|c|}
\hline
 & \multicolumn{4}{|c|}{{\bf Minimum E$^{true}_{\mu}$}}\\ 
\cline{2-5}
{\bf Model} & 10~GeV& 1~TeV & 10~TeV & 100~TeV\\
\hline\hline
\underline{Atmospheric} &   &       &     &       \\
ATM~\cite{volkova}&    2725$\pm$146       & 544$\pm$44  & 18$\pm$2 &
0.34$\pm$0.05 \\
\hline
\underline{Generic AGN model} &          &         & & \\
SDSS~\cite{sdss} &  354$\pm$13      & 334$\pm$12   & 260$\pm$12 &
132$\pm$10 \\
NMB~\cite{nmb}  &   591$\pm$15      & 454$\pm$13   & 206$\pm$9  & 
53$\pm$4  \\
\hline
\underline{Blazars} &          &         &  & \\
PRO~\cite{pro}       &  45$\pm$2    & 43$\pm$2    &   35$\pm$2   &  
21$\pm$2    \\ 
MRLA~\cite{mannheim} &  34$\pm$1  & 18.1$\pm$0.7  &  4.0$\pm$0.2 &
0.52$\pm$0.04 \\
MRLB~\cite{mannheim} &  37$\pm$1  & 21.1$\pm$0.8  &  6.6$\pm$0.4 & 
2.4$\pm$0.3  \\
\hline
\end{tabular}
\caption{\small Diffuse neutrino flux from AGN.  The table shows the
number of events per year 
  with true muon energy greater than E$_\mu^{\mathrm{true}}$
which are reconstructed as 
  upward-going in the optimised detector. The errors are statistical.} 
\label{tab:4-3}
\end{center}
\end{table}

\begin{table}[p]
\begin{center}
\begin{tabular}{|l||c|c|}
\hline
 & \multicolumn{2}{|c|}{{\bf Minimum E$_\mu^{\mathrm{rec}}$ }}\\ 
\cline{2-3}
{\bf Model} & 10~TeV &  100~TeV\\
\hline\hline
\underline{Atmospheric} &   &                  \\
ATM~\cite{volkova}&     68$\pm$13   & 0.8$\pm$0.1  \\
\hline
\underline{Generic AGN models} &          &         \\
SDSS~\cite{sdss}    &  251$\pm$12       &   134$\pm$10      \\
NMB~\cite{nmb}&  217$\pm$9       &   64$\pm$4     \\
\hline
\underline{Blazars} &          &        \\
PRO~\cite{pro}&   34$\pm$2       &   21$\pm$2      \\
MRLB~\cite{mannheim}  & 7.8$\pm$0.4       &   2.6$\pm$0.3      \\
\hline
\end{tabular}
\caption{\small Diffuse neutrino flux from AGN: the number of 
events per year 
  with reconstructed muon energy greater than E$_\mu^{\mathrm{rec}}$ 
which are accepted as 
  upward-going in the optimised detector. The errors are statistical.} 
\label{tab:4-4}
\end{center}
\end{table}

\subsection{Detecting astrophysical point sources}
Due to the extremely low angle between the muon and the parent 
neutrino and to the good quality of the muon direction measurement, the 
atmospheric neutrino background contaminating each individual source can be 
reduced to a very low level by selecting very small angular regions of the 
sky. 
For a point-like source, visible 50\% of the time, evidence for observation
will be possible with 6 events above 1~TeV (3 in the central pixel). This
corresponds to a fake probability smaller than 7\%.

The sensitivity for an individual source would be:

\begin{displaymath}
\frac{d\Phi}{dE_\nu} > 5\cdot10^{4} \left( \frac{E_\nu}
    {1~\mathrm{TeV}}\right)^{-2} \;\;\mathrm{km}^{-2}\;\; 
    \mathrm{yr}^{-1}\;\;\mathrm{GeV}^{-1}
\end{displaymath}

\noindent
The corresponding luminosity in protons emitted at any energy can be derived
with the usual assumption that 5\% of the proton energy is converted
into neutrino energy. This gives a sensitivity of about $10^{30}$\,W
(about 1/10 the Eddington limit for a solar mass) for an object located
at a distance of 10\,kpc.

Another way to estimate the ANTARES potential for the detection of 
point-like sources 
is to calculate the number of events expected for a given flux. 
A flux estimate can be obtained 
by using the measured low-energy gamma-ray fluxes, assuming  
{\it i)} that the low energy gamma-rays are of hadronic origin and 
{\it ii)} that the emitted 
gamma-rays have a differential energy spectrum $E^{-2}$.
With these assumptions the muon neutrino flux is about 40\% of the flux of 
gammas at the production source. Using the second EGRET 
catalog~\cite{egret:cat} 
for sources measured during the P12 period, the derived neutrino flux has been 
extrapolated to the energies where the neutrino detector is sensitive. 

The detection of the most luminous individual sources could require more than
one year of data acquisition. Nevertheless, a statistically
significant effect could be detected in one year
by adding the contributions of all the
extra-galactic sources. The expected number of events is between ten 
and one hundred per year (depending on the value of the differential
spectral index used: 2 or 2.2) to be compared to a total
background of about three events per year.

The sensitivity to gamma ray bursts is improved by the detection of events
correlated in space and in time with optical information. In this case,
the background is negligible.
With the flux given in reference~\cite{WaxBah99} 
and the variations of the effective detector area 
shown in figure~\ref{fg:fig_4-9}, the rate 
expected is $\simeq 10$~signal events per year with 
a background rate of less than 0.001~events per year.

\subsection{Background rejection}

One critical item is the rejection of down-going muons from the
normal cosmic ray flux. At a depth of 2300~m in water, 
down-going muons correspond to a
flux 1.5$\times$10$^5$ times higher than that for the atmospheric neutrinos.
Therefore, a rejection factor of more than 1.5$\times$10$^5$ is required to
have a signal-to-background ratio exceeding 1. 
This is less critical for
point-like sources where the rejection required is decreased 
by the number of pixels defined in the sky (about 200,000).

The rejection is, in principle, easy to study but in practice 
the results are limited by
the number of  triggering cosmic ray events that can be generated.
Simulated single muons and multi-muons reaching the detector
have been studied, 
as discussed in section~\ref{sect:MC_sim_tools}. 
The simulation statistics are summarised in
table~\ref{tab:4-5}. 

\begin{table}
\begin{center}
\begin{tabular}{|l||c||c|c|c|}
\hline
 & {\bf mono-muons} & \multicolumn{3}{|c|}{{\bf multi-muons~: E$_p$}}\\ 
\cline{3-5}
 & & 2-20~TeV & 20-200~TeV & $>200$~TeV\\ 
\hline\hline
 & & & & \\
Flux at 2300~m & $10^{-7}$ & $0.4\cdot10^{-8}$ & $1\cdot10^{-8}$ & $0.4\cdot10^{-8}$ \\
depth (cm$^{-2}$s$^{-1}$)  & & & & \\
\cline{3-5}
 & 85$\%$  & \multicolumn{3}{|c|}{15$\%$}\\
\cline{3-5}
 & & 22$\%$ &56$\%$  & 22$\%$ \\
\hline
\hline
 & & & & \\
No. of simulated  & & $141\cdot10^6$& $28\cdot10^6$& $12\cdot10^6$ \\
primary  & & & & \\
interactions  & & & & \\
\hline
No. of simulated  & & & & \\
events reaching   & $0.3\cdot10^6$& $22\,000$& $0.7\cdot10^6$& $6\cdot10^6$ \\
the detector  & & & & \\
level  & & & & \\
\hline
No. of & & & & \\
triggering & $27\,800\pm167$& 230$\pm70$ & $15\,093\pm183$ & $384\,746\pm620$\\
events& & & & \\
 & & & & \\
\hline
\hline
       \begin{tabular}{l}
        corresponding \\
        real time\\
       \end{tabular}
& 
       \begin{tabular}{cc}
       $<1$~TeV & 1.5 days\\
       1-10~TeV & 0.9 days\\
       $>10$~TeV  & 20 days\\              
       \end{tabular}
      &  30 minutes& 4.5 hours & 3.9 days \\
\hline
{\it no of triggering} & & & & \\
{\it events per hour} & {\it 4698$\pm234$}& {\it 460$\pm140$} & {\it 3290$\pm41$} & {\it 4074$\pm7$}\\
\mbox{}  & & & & \\
\cline{3-5}
 & 38$\%$  & \multicolumn{3}{|c|}{62$\%$}\\
\cline{3-5}
 & & 6$\%$ &42$\%$  & 52$\%$ \\
\cline{3-5}
 & \multicolumn{4}{|c|}{$\underbrace{\makebox[10cm]{ }}$}   \\
 & \multicolumn{4}{|c|}{{\it $12\,522\pm262$}}    \\
 & \multicolumn{4}{|c|}{100$\%$}    \\
\hline
\end{tabular}
\caption{\small Summary of the simulation statistics.
For the multi-muons, E$_p$ is the primary energy.}
\label{tab:4-5}
\end{center}
\end{table}

Approximately three events per second will satisfy the combined trigger and filter
conditions, of which 38\% are single muons and 
the remaining 62\% are multi-muons (half of them coming from primaries of
more than 200~TeV, very few from initial energies smaller than 2~TeV).

For the potentially most dangerous events 
(multi-muons coming from a very high energy primary cosmic),
the equivalent of 4~days of data has been generated. Clearly more statistics
are required for an effective background estimate; however, the existing
sample took
80 CPU-days of a single processor at the Lyon computer centre.
 A factor of 3 improvement in
simulation speed is expected as a result of using the KM3 program 
instead of DADA for the
detector simulation, and a further factor of 2 gain is possible if the
fiducial volume is reduced. 

\begin{figure}
\begin{center}
\mbox{\epsfig{file=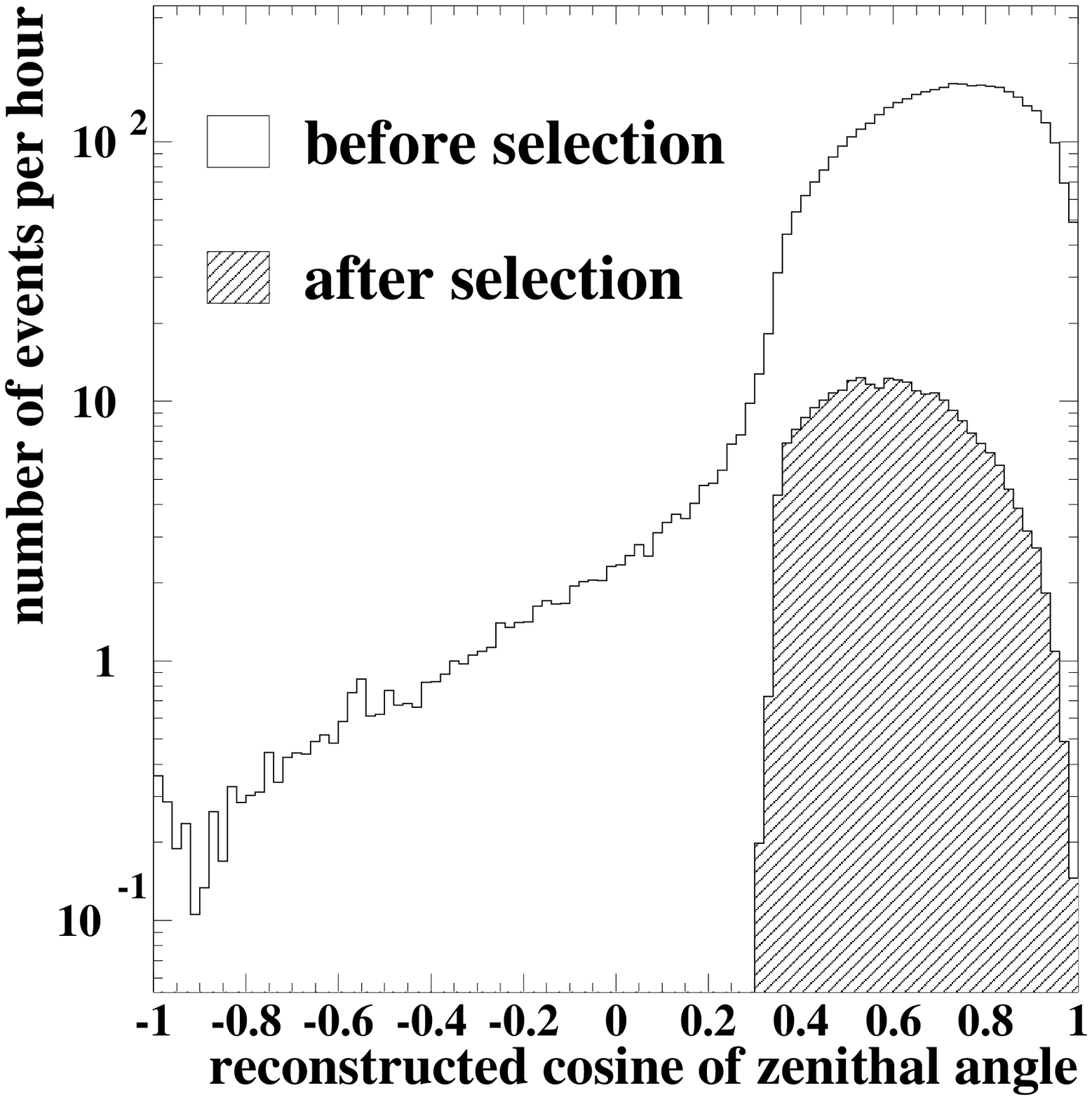,width=0.5\linewidth}}
\end{center}
\caption{\small Number of downward-going muons as a function of the
  reconstructed zenith angle.} 
\label{fg:fig_4-30}
\end{figure}

From all available simulations, no cosmic ray muon or multi-muon 
is accepted as an upward-going muon after reconstruction
and quality tests (see figure~\ref{fg:fig_4-30}).
Since the most dangerous event class will be the multi-muon events initiated
at primary energies $\geq$ 200~TeV, this implies, with the current
statistics, a rejection factor 
$N_{fake}/N_{atm. \nu} \leq 1/3$ at 90\% confidence level.
AMANDA and BAIKAL do not experience problems due to this background.
   
\subsection{Systematics on detector performance}

The performance of the detector depends on the values of various
simulation parameters relating to the properties of the photomultipliers,
the transparency of the water and the number of strings.  In the actual
detector these values are likely to be different from the preliminary
numbers used in the simulations, so it is essential to study the effect of
such variations on the performance of the detector.

\begin{figure}
\begin{center}
\mbox{\epsfig{file=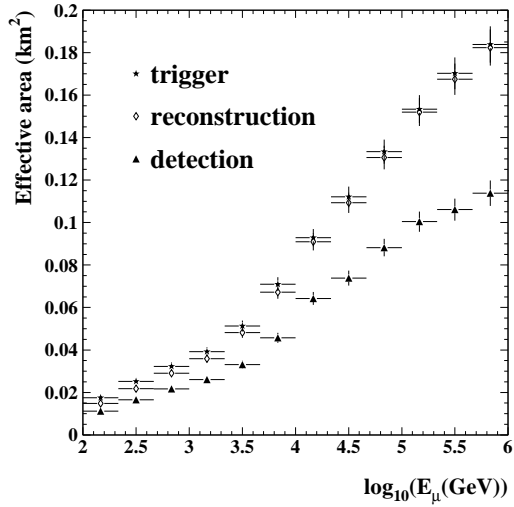,width=0.5\linewidth}}
\end{center}
\caption{\small Effective area for the optimised detector with the measured
  parameters.}
\label{fg:fig_5-5}
\end{figure}

As an example of the effects to be expected, 
the effective area of the detector
as a function of the muon energy, previously shown in figure 
\ref{fg:fig_4-9}, has been recalculated using the same detector layout with
parameters based on more recent work: the most recent optical
module angular dependence measurement, an absorption length of 47 rather 
than 55 m, an angular resolution of 0.3$^\circ$ and an overall timing
error of 2.5 ns.  The result is shown in figure \ref{fg:fig_5-5}.  

A more quantitative result is obtained by studying each parameter 
separately, beginning with photomultiplier parameters.  The final values
of these are not yet known since the choice of phototubes has not
yet been made.  Ten-inch tubes are currently used
in the simulations since they are functional and readily
available. However, photomultipliers with a larger 
photocathode are expected in the near future.

 Going from a 10-inch to a 12-inch PMT,
the potential rate for a $1/E^2_\nu$ spectrum increases by 40\%, and
$\sigma_t$ increases by 15\%. Therefore, 
the overall gain on $S/\sqrt{B}$ would be 10\%.

\begin{figure}[tp]
\begin{center}
\mbox{\epsfig{file=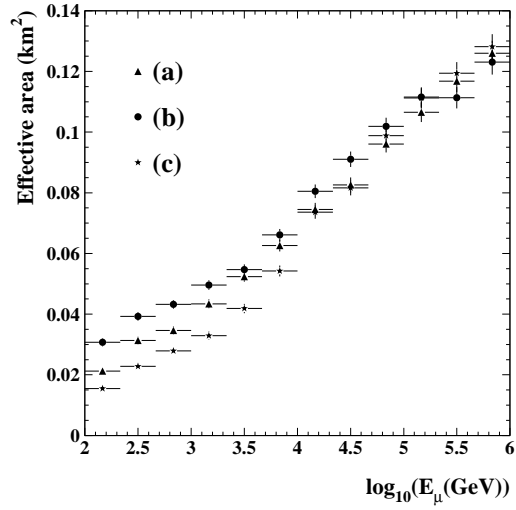,width=0.5\linewidth}}
\end{center}
\caption{\small Effective area sensitivity to PMT parameters: (a) standard
  parameters; (b) increase quantum efficiency by $20\%$; (c) change of the
  angular response.} 
\label{fg:fig_5-1}
\end{figure}

Figure~\ref{fg:fig_5-1} shows how an increase of
the absolute efficiency of 20\% and a change in the angular response
of the optical module
 affect the effective detector area. These parameters need to be measured
 accurately for a good understanding of the systematic
 errors.

\begin{figure}[tp]
\begin{center}
\mbox{\epsfig{file=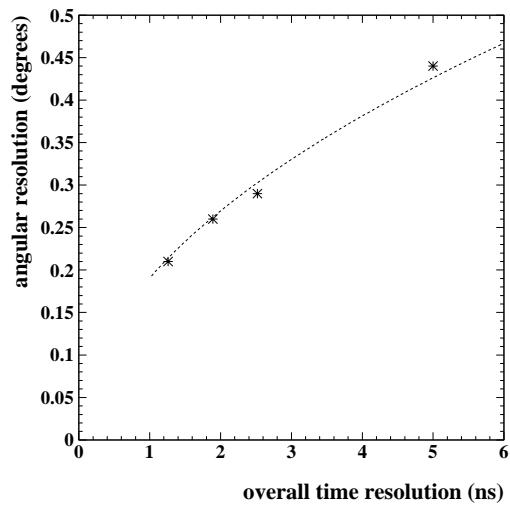,width=0.5\linewidth}}
\end{center}
\caption{\small The angular resolution as a function of the overall time
  resolution.} 
\label{fg:fig_5-2}
\end{figure}

The choice of phototube also affects the timing error $\sigma_t$, and 
consequently the angular resolution $\sigma_\theta$: $\sigma_\theta$
varies roughly as $\sqrt{\sigma_t}$, as shown in figure \ref{fg:fig_5-2}.
The 8-inch tubes used in OM tests to date have $\sigma_{pmt} = 1.1$ ns,
which combines with an alignment error of about 0.5 ns to give an
overall timing error of 1.3~ns and a corresponding angular resolution of
0.2$^\circ$. 
For the 10-inch PMT, the contribution of the PMT is larger 
($\sigma_{pmt}$ = 1.5~ns); if, in addition, 
the alignment error were as bad as 2~ns, 
the angular resolution would be 0.3$^{\circ}$.

\begin{figure}
\begin{center}
\mbox{\epsfig{file=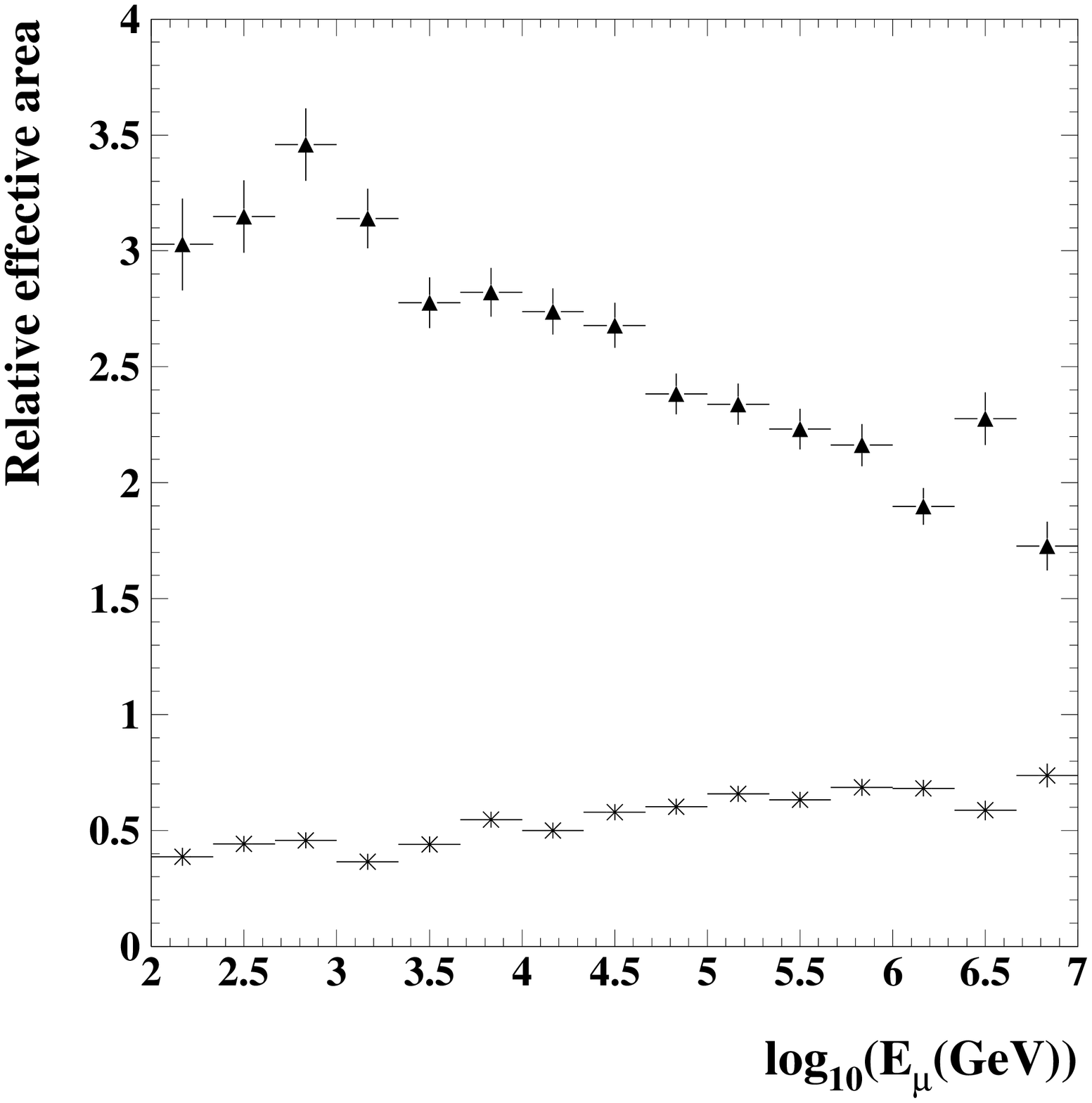,width=0.5\linewidth}}
\end{center}
\caption{\small Relative effective area versus the logarithm 
of the muon energy for 45 strings
(triangles) and 8 strings (crosses), normalised to 15 strings.}
\label{fg:fig_5-6}
\end{figure}

For a given string spacing, the effective area of the detector 
clearly depends on the number
of strings deployed.
Figure~\ref{fg:fig_5-6} shows the relative effective area of 8 and 45 string
detectors, normalised to the standard 15-string detector, as a function of
$E_\mu$. The effective area grows linearly
with the number of optical modules at low energy, but the effect decreases
above 10~TeV.

\begin{figure}[p]
\begin{center}
\mbox{\epsfig{file=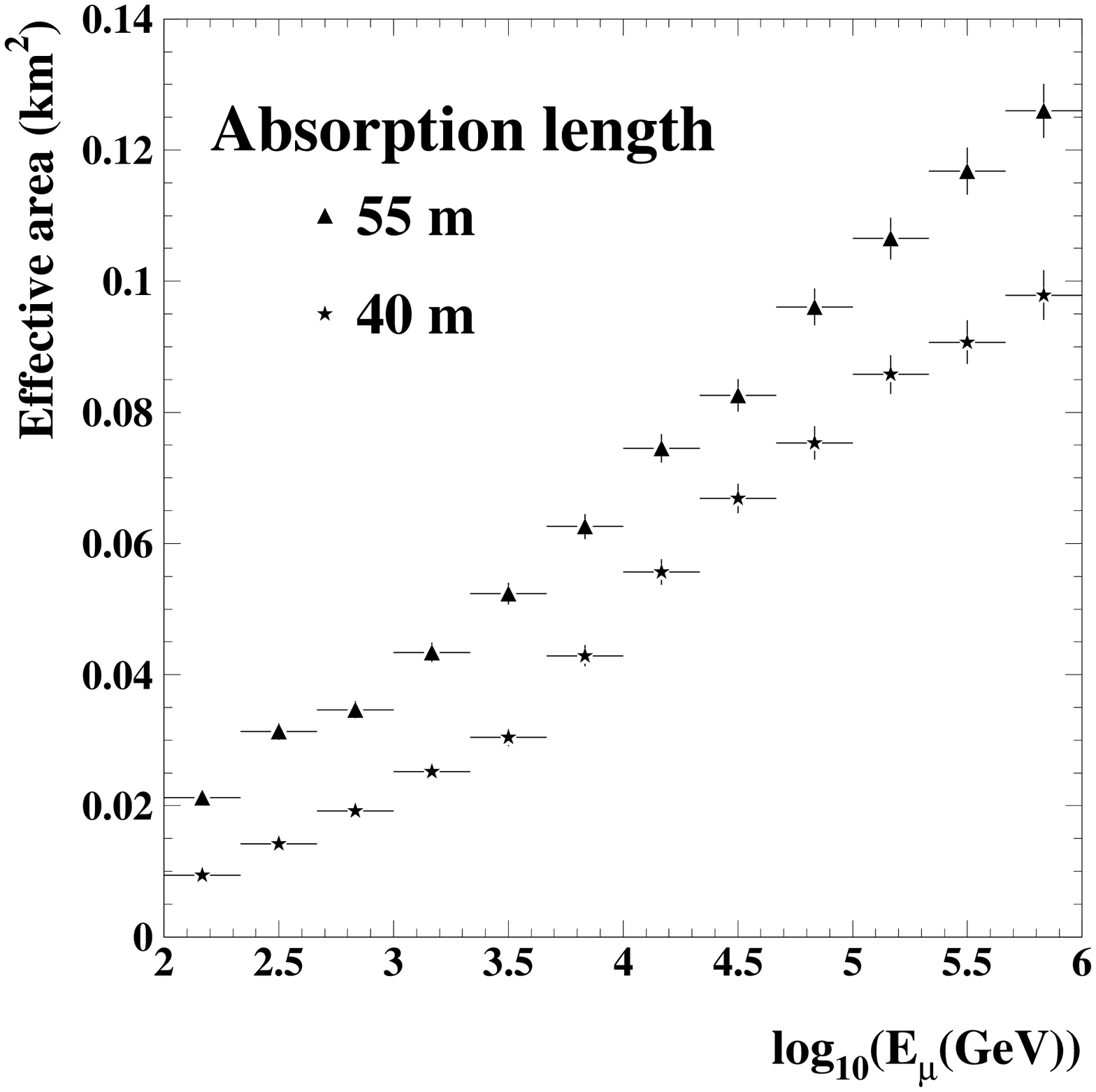,width=0.5\linewidth}}
\end{center}
\caption{\small Effective areas for absorption lengths of 40 and 55~m.}
\label{fg:fig_5-3}
\end{figure}   
   
The angular resolution improves when the number of strings 
increases: for eight strings $\sigma_\theta$ is 20\% greater than the
15-string value, and for 45 strings it is 15\% less.
Better angular resolution leads to improved sensitivity to 
point-like sources.  In practice, a larger detector would not simply be an
extension of the present array: operating experience with the 0.1 km$^2$
detector and additional optimisation studies would be required to develop
an appropriate design.

The effective area is also a function of the water transparency.
In the sea water at the detector site, effective attenuation lengths of
40--60 m have been measured, as discussed in section \ref{sect:trans_data}. 
The effect of varying the absorption
length in the simulation studies from 40 m to 55 m is shown in 
figure~\ref{fg:fig_5-3}.
It should be noted that
the reconstruction and selection criteria were optimised for 55~m, and may
not be optimal for 40 m.

The scattering effect gives fewer delayed photons (5\% at 25~m, 10\% at 45~m)
than those produced in catastrophic muon energy loss processes, so the angular 
resolution should not be affected.

\section{Neutrino oscillations}


Neutrino interactions below a few hundred GeV can be studied in
ANTARES because the neutrino energy can be estimated from
the range of contained muons observed in the detector
volume. These events can be used to study additional physics
channels such as neutrino oscillations and neutralino
annihilations. This section describes the simulations
that have been carried out to understand the
sensitivity to oscillations of atmospheric neutrinos.

The principal source of neutrinos for neutrino oscillation
studies is the decay of charged pions produced 
by cosmic rays interacting in the Earth's atmosphere.
Charged-current interactions of $\nu_\mu$ 
producing upward-going muons are selected 
so that the background from muons produced 
in the atmosphere is absorbed by
the Earth.

The simulation programs used for the astronomical neutrino
studies are also used for the oscillation studies.
The reconstruction algorithms, on the other hand, are
somewhat different. This chapter will describe the current
status relative to the simulation and reconstruction of the
signal and background for the neutrino oscillations studies,  
and the determination of the oscillation parameters.

\subsection{Track reconstruction}
\label{subsect:tracks_for_oscillations}

For the study of neutrino oscillations, events are reconstructed in two classes
depending on the number of strings with coincidences. If at least
two strings have coincidences, the event is classified as `multi-string',
while if only one string has coincidences, the event is
classified as 
`single-string'.  However, only 7\% of the multi-string
events have only two
coincidences and none of the well reconstructed single-string events has less
than two coincidences.

For the multi-string  neutrino oscillation events, 
a prefit similar to the one described earlier is used, but with
somewhat tighter cuts. 
For the single-string events, 
a method is used which estimates 
the angle of the track to the string using the
time differences between adjacent storeys;  
then three  hypotheses are tried
for the distance of the track from the string and the one with the
best fit likelihood is used as the start for the final fit. 
 
\begin{figure}
\begin{center}
\mbox{\epsfig{file=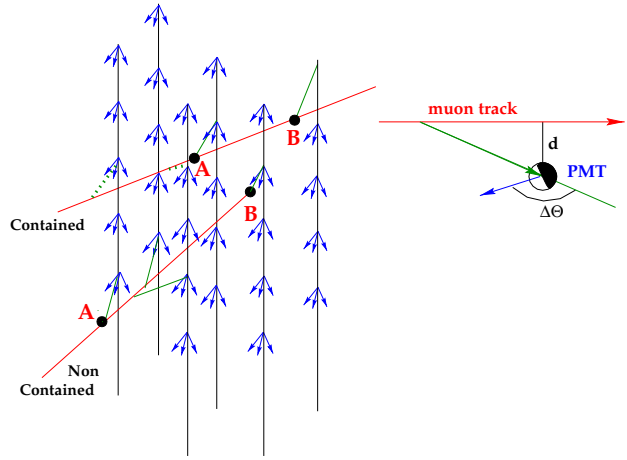,width=0.6\linewidth}}
\end{center}
    \caption{\small Illustration of the event containment 
    estimator $a_{npe}$.}
    \label{fig:containment}
\end{figure}

\begin{figure}
\begin{center}
\mbox{\epsfig{file=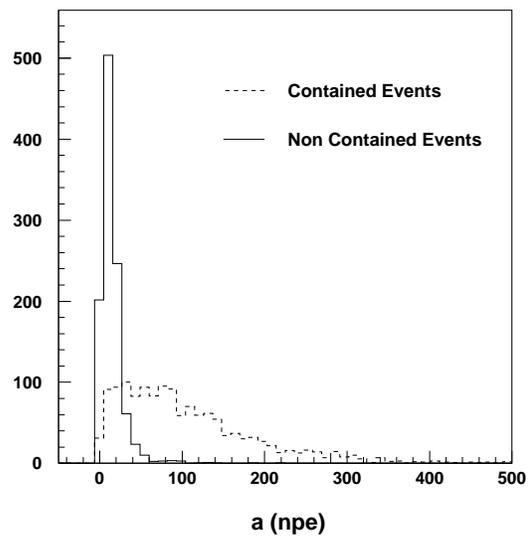,width=0.6\linewidth}}
\end{center}
    \caption{\small Distribution of $a_{npe}$ 
             for contained and
             non-contained events.}
    \label{fig:cont_examp}
\end{figure}

For the neutrino oscillation studies, 
the muon range is used to measure the muon energy from which
the neutrino energy is estimated. 
Events must be selected which are 
at least partially contained within the detector volume.
For the single-string events the containment is ensured
by requiring the reconstructed muon track to have
a zenith angle within $20^\circ$ of the vertical
and requiring the two highest and two lowest storeys of the string not to have
fired in the event.
A partial-containment condition is used for the multi-string
events, where tracks are used with zenith angles as large as 80$^\circ$.
Figure~\ref{fig:containment} illustrates the algorithm used.
Once the track is reconstructed, the points, where the first ($A$) and 
the last ($B$) detected Cherenkov photons were associated to
the track,
are located.
Then the track is extrapolated backwards and a test is made to see if any
photomultipliers could have seen that part of the track before the point $A$.
If none could, then the track is clearly not contained.

The estimator of the track containment is the sum of the amplitudes
of the hits that should have been seen on these optical modules
but were not seen. 
A simple
model is used to calculate these amplitudes. 
A minimum-ionizing particle seen at normal incidence by an optical module 
at a distance of 1~m yields an average of 
55 photoelectrons~\cite{bib:chap5_gamelle}. 
The sum of the amplitudes for a minimum ionizing track emitting
Cherenkov light at an angle $\Theta_i$ with respect to optical module
$i$ at a minimum distance $d_i$ from the track 
(figure~\ref{fig:containment}) is:

\bec
a_{npe} = \sum_{i}{\f{55}{d_i}f(\Delta \Theta_i)}
\eec

\noindent
Here $f(\Delta \Theta)$ accounts for the angular
response of the optical module.  Figure~\ref{fig:cont_examp} shows 
the distributions of this
quantity for contained and non-contained events;
a cut on $a_{npe}$ at a value of 50 
provides an efficient separation between
the two sets of events.

\subsection{Optimisation for oscillations}

The observation of atmospheric neutrinos for energies below
a few hundred GeV can be used to study neutrino oscillations for
values of the neutrino mass difference 
$\Delta m^2$ between $10^{-3}$ and $10^{-2}$~eV$^2$. 
The detector configuration proposed 
above
for neutrino astronomy 
has been evaluated for neutrino oscillations.
Other studies have been carried out to see 
if the optimal geometry could
be different for neutrino oscillations and astronomy.
Simulations have been made with
five different geometries, varying the number of optical modules
per storey and their orientation (pointing horizontally, 
down at 45$^\circ$, or vertically down).
The conclusion is that three optical modules per storey, 
pointing down at 45$^\circ$, 
is also optimal for neutrino oscillations. 
A denser vertical spacing is optimal for the oscillation
studies; therefore, the geometry proposed here 
has four strings with 8~m
vertical spacing and nine strings with 16~m spacing. The studies
performed so far are not exhaustive, and a better
compromise may be found using the same total number of strings and optical
modules. A horizontal spacing between strings of 60~m is optimal
both for astronomy  below 100~TeV and for oscillations.

\subsection{Event selection}

After reconstruction of an event, the quality of the
reconstruction is estimated using the fitted likelihood and
the error estimates provided by the likelihood fit. 
Another variable
used to discriminate against badly fitted tracks
is the angular distance between the track found in 
the prefit and the track found in the final fit.
All tracks are required to have at least 10 hit optical modules.
The mean number of hits associated with the tracks is 30
for multi-string tracks and 16 for single-string tracks. 
The mean number of hit coincidences 
(two out of three hits in a storey)
is 7.2 for multi-string events and 5.3 for single-string events.
Selected events are required to satisfy the containment and 
zenith angle conditions described in 
section \ref{subsect:tracks_for_oscillations}.

After reconstruction and containment cuts, the number of
charged-current \numu events expected, in the absence of neutrino
oscillations, is 2400 for three years of data taking with the
proposed detector. About 90\% of these events 
are partially-contained multi-string events; 
the remaining 10\% are fully-contained single-string events.

\subsection{Angular and energy resolution}

\begin{figure}
\begin{center}
\mbox{\epsfig{file=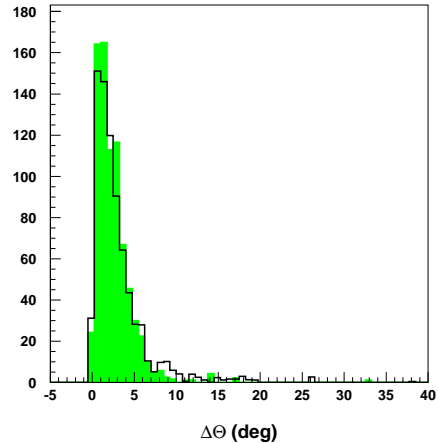,width=0.5\linewidth}}
\end{center}
   \caption{\small Angular resolution. 
   The shaded histogram is the difference in
angle between the \numu direction and the true $\mu$ direction and the
open histogram is the complete reconstruction error between the \numu
direction and the reconstructed $\mu$~direction.}
   \label{fig:AngRes} 
\end{figure}

\begin{figure}
 \begin{center}
\mbox{\epsfig{file=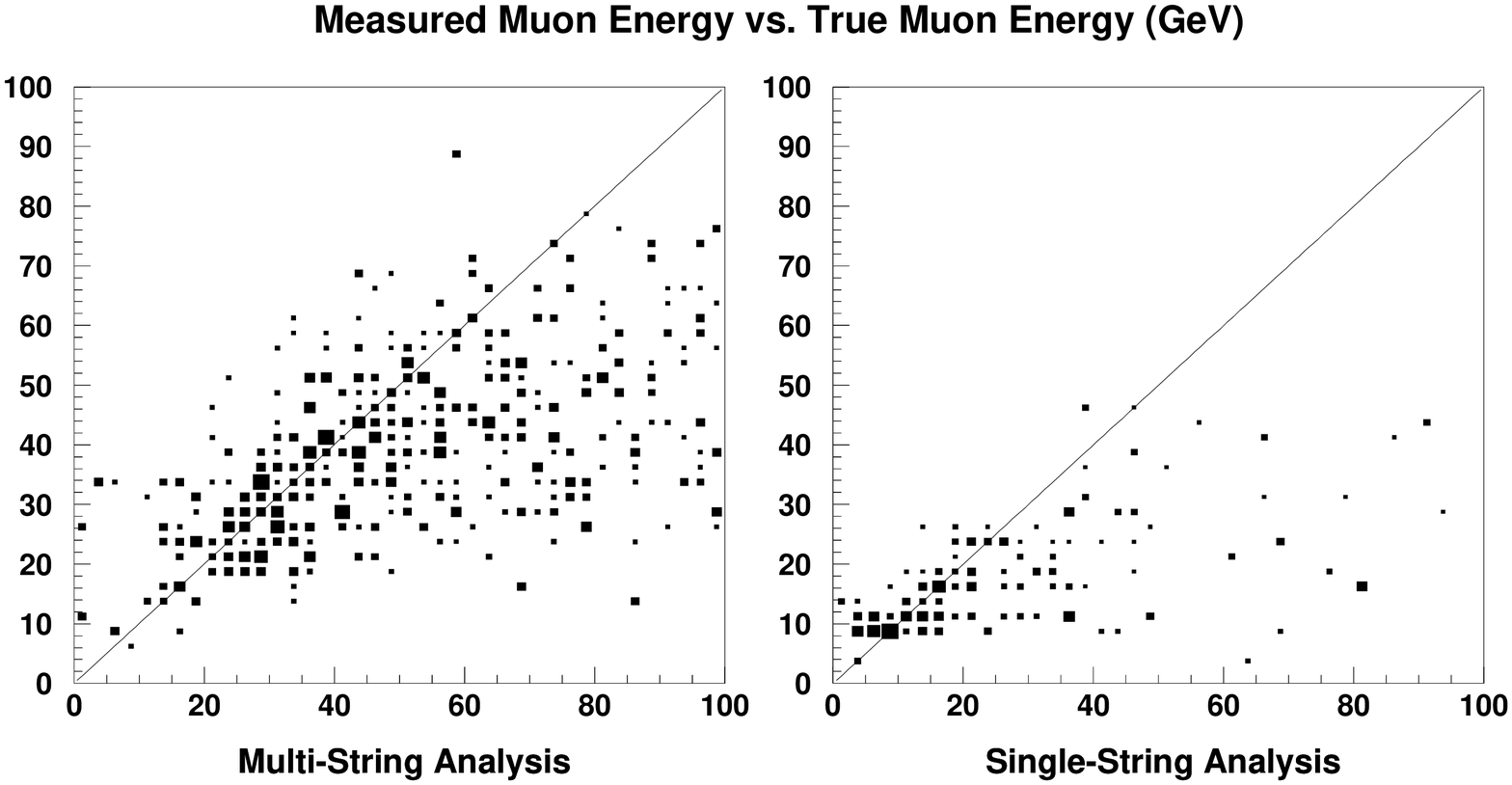,width=13cms}}
\end{center}
   \caption{\small Correlation between reconstructed and true muon energy
for tracks reconstructed as multi-string and single-string.} 
   \label{fig:EnergyCorrelation}
\end{figure}

The neutrino angular resolution for the selected events
is shown in figure~\ref{fig:AngRes}.
The experimental angular resolution is small
compared to the intrinsic difference between the neutrino and muon
directions. The median angular resolution from measurement errors 
is 0.4$^\circ$ for the
multi-string events and 1.8$^\circ$ for the single-string events.

The correlation between the
reconstructed muon  energy and its true energy is presented 
in figure~\ref{fig:EnergyCorrelation} for the two
types of events. 
The single-string events are mainly concentrated at low
energy. 
The true muon energy of the multi-string events 
extends above 100~GeV because the muon end-point is not required
to be inside the detector volume.
The intrinsic muon energy resolution 
of the ANTARES detector is 
adequate for the oscillation measurements. 
Nonetheless, attempts will be made to improve 
the resolution on the neutrino energy by estimating the
energy of the individual hadronic showers.

\begin{figure}
\begin{center}
\mbox{\epsfig{file=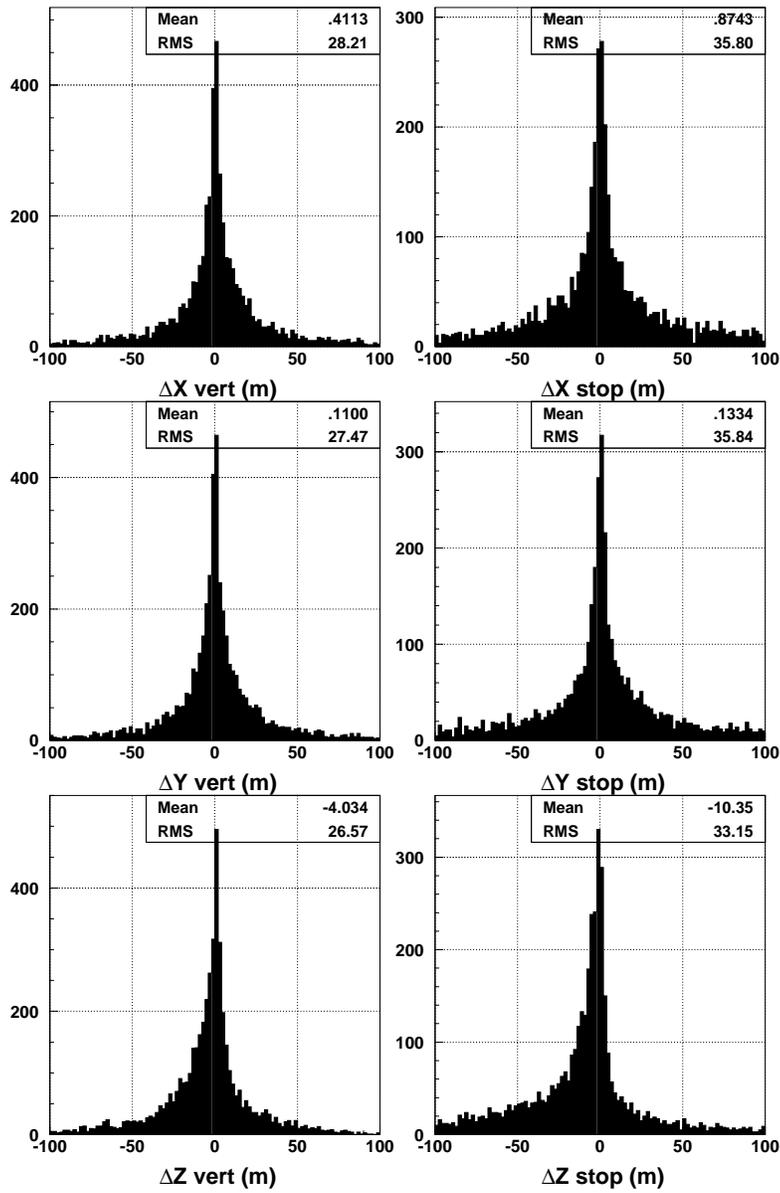,width=0.8\linewidth}}
\end{center}
   \caption{\small Resolution on the track starting point and stopping point
   for multi-string events.}
   \label{fig:VertStop}
\end{figure}

The spatial resolution for the starting point (vertex) 
and stopping point for 
the multi-string events is shown in figure~\ref{fig:VertStop}. 
The multi-string events can be reconstructed in all 
their spatial dimensions with
reasonable accuracy, but only the vertical coordinate is well
measured for the single-string events.
The distribution of the reconstructed
zenith angle and the energy for the muons 
from the partially-contained \numu events described above 
are shown in figures~\ref{fig:SigBak} (a) and (b).
         
\subsection{Backgrounds}

\begin{figure}
\begin{center}
\mbox{\epsfig{file=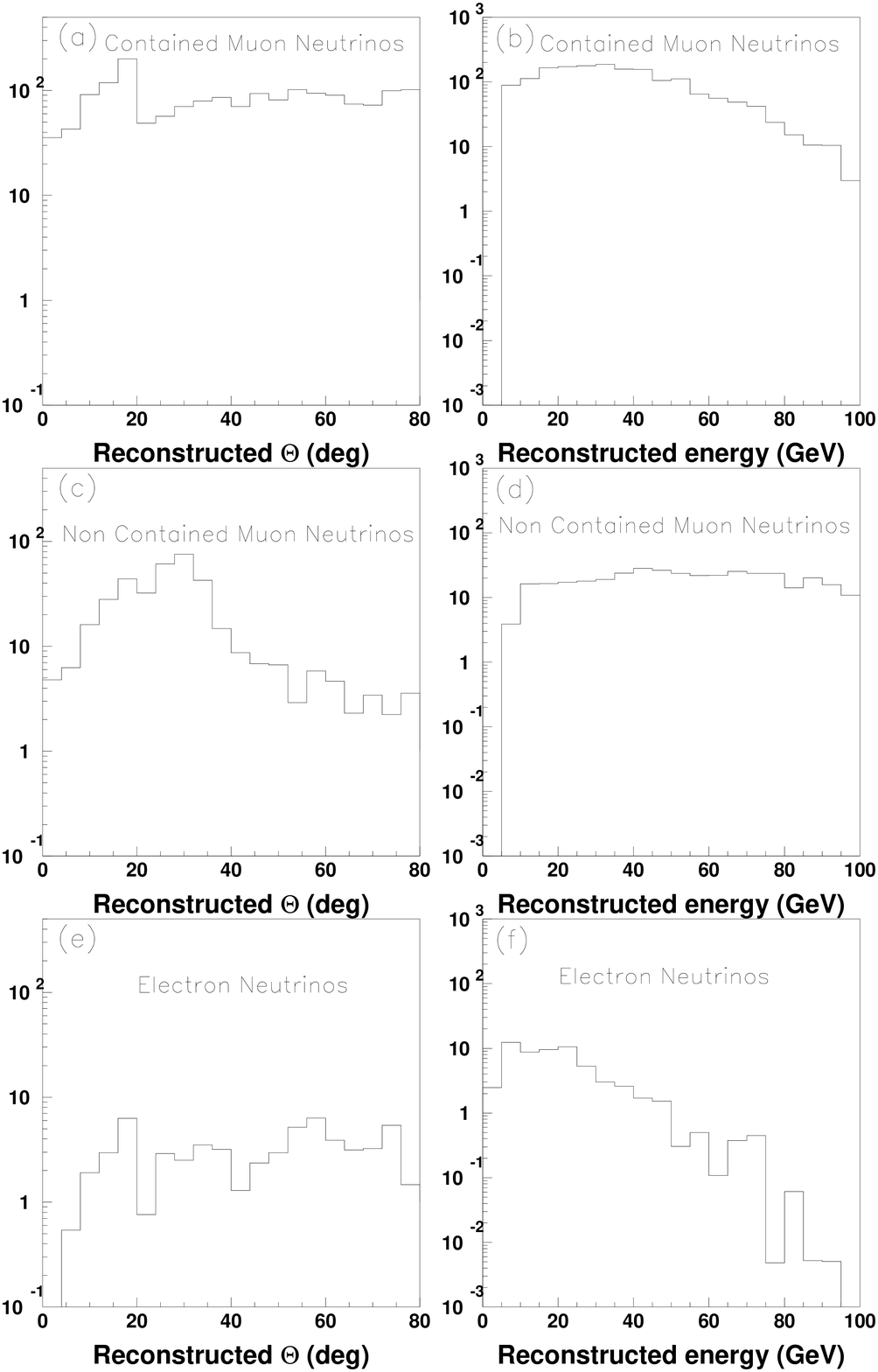,width=0.8\linewidth}}
\end{center}
\caption{\small Distributions of reconstructed muon angle and energy:
(a),(b) partially-contained signal \numu~events; 
(c),(d) non-contained \numu~events and (e),(f) background \nue~events.}
\label{fig:SigBak}
\end{figure}

The reconstructed zenith angle and the muon energy
for events from \nue interactions which pass
all the cuts as partially-contained \numu
events are shown in figure \ref{fig:SigBak} (e) and (f).
The \nue events amount to 4$\%$ of the signal \numu events;
their angular distribution is similar to the signal but they are
at lower mean energies.

Figure \ref{fig:SigBak} (c) and (d) show the distributions
for the \numu events which are truly from interactions
outside the detector, but which pass the cuts as contained events.
These events are 20$\%$ of the signal \numu events. Their
reconstructed energy has a mean of 55 GeV, but their real energy
is higher -- a mean of 140 GeV -- since they come from interactions
beyond the detector which pass the containment cuts due to
inefficiencies in the light detection.  

Another potential source of background is 
the upward-going muons from the decay of pions and
kaons from neutral-current interactions of neutrinos of 
all flavours. This background is small, however, 
less than 4\% of the \numu charged-current interactions. 
Moreover, since these events have lower energies 
due to the missing final-state neutrinos, 
the acceptance will be low, 
and the number of fake events coming from this background
can be neglected.

Background from muons formed in the atmosphere are more 
time consuming to generate, due to the immense rate, 
than the background from neutrinos. 
The rate of muons arriving at the detector
at angles truly below the horizon is completely dominated by
\numu interactions (figure~\ref{fig:fluxmu_angle}).
But high statistics samples are needed to measure the rate
of badly reconstructed down-going muons. 
With the simulations performed to date, 
no events from atmospheric showers pass the selection cuts,
but the statistics simulated correspond to only two days of data.
This sets a limit on the possible number of atmospheric muons 
mis-identified as \numu interactions, but higher statistics
are needed for a full evaluation.

\subsection{Sensitivity to oscillations}
         

An analysis has been carried out using simulated
interactions to explore the
sensitivity of the proposed experiment to the neutrino oscillation
parameters $\Delta m^2$ and sin$^2 2 \theta$. 
The analysis is based on partially-contained muons,
and the muon energy is taken as the visible energy in the
detector.
The number of signal events was calculated 
using the presently known values 
for the flux of atmospheric neutrinos 
and the expected experimental acceptance.
The neutrino flight distance
$L_\nu$ is related
to the neutrino zenith angle $\theta_z$ 
by $L_\nu \sim L_0\cos\theta_z$,
where $L_0$ is the diameter of the Earth 
(taken as a perfect sphere of diameter 12740~km).

\begin{figure}
\begin{center}
\mbox{\epsfig{file=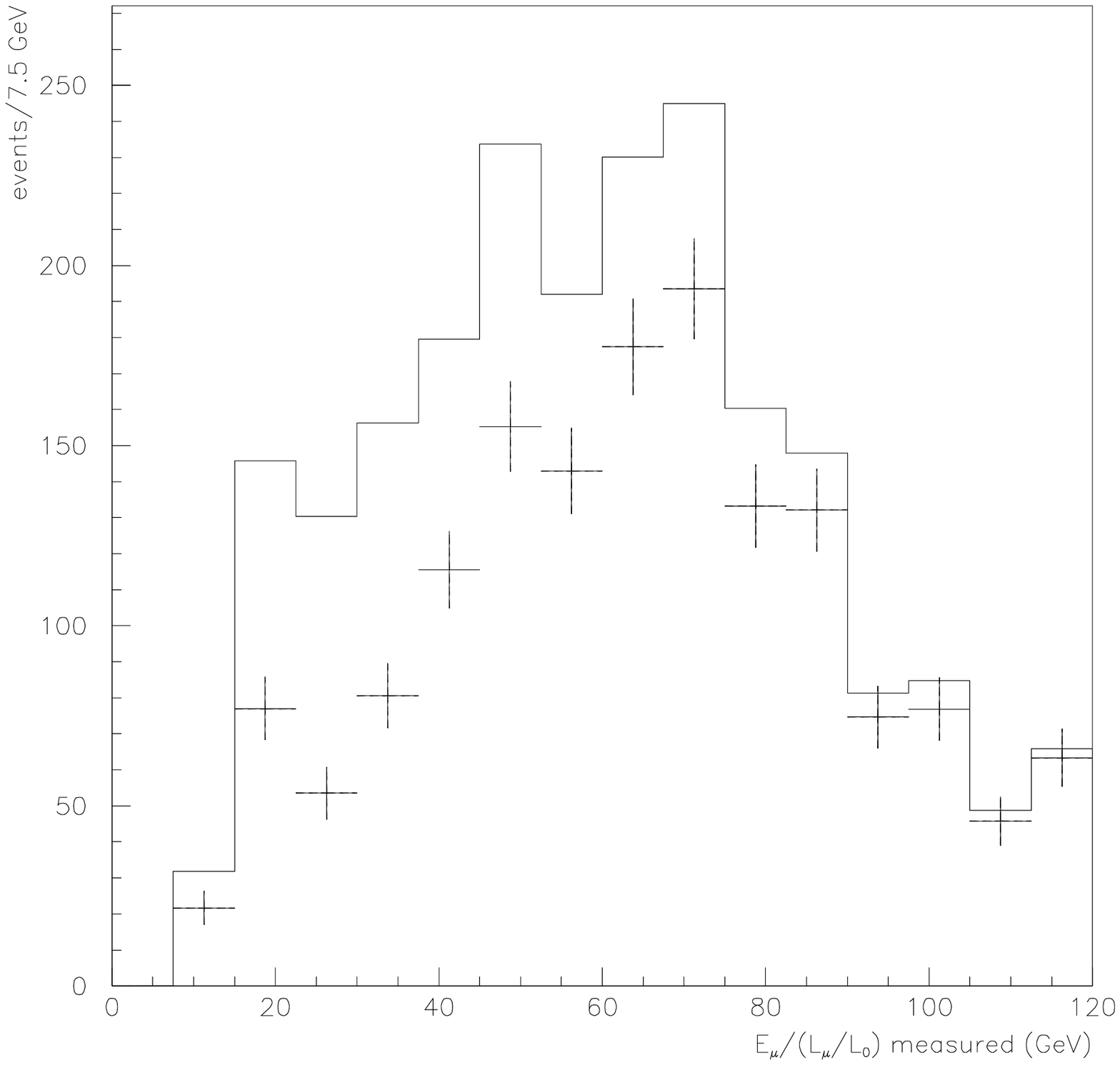,width=0.5\linewidth}}
\end{center}
\caption{\small Simulated number of events in 3 years with no
oscillations (histogram) and with oscillations (points) 
for sin$^2~2\theta =1.0$ and
$\Delta m^2 =0.0035$~eV$^2$.}
\label{fig:sim_oscil}
\end{figure}

The histogram in figure~\ref{fig:sim_oscil} shows the 
distribution of accepted
events as a function of $E_\mu/(L_\mu/L_0$), where $E_\mu$
is the energy and $L_\mu$ is the neutrino flight distance 
calculated using the zenith angle of the reconstructed muon
track, in place of the zenith angle of the neutrino.
The histogram represents a simulation of three years of
operation without neutrino oscillations.
The `data' points in figure~\ref{fig:sim_oscil} 
represent the same simulated events, but with each event
weighted by the probability:
$$  P = 1 - \sin^2 2\theta ~\sin^2 (1.27 \Delta m^2 L_\nu/E_\nu),  $$
using the true neutrino values of $E_\nu$ and $L_\nu$.

\begin{figure}
\begin{center}
\mbox{\epsfig{file=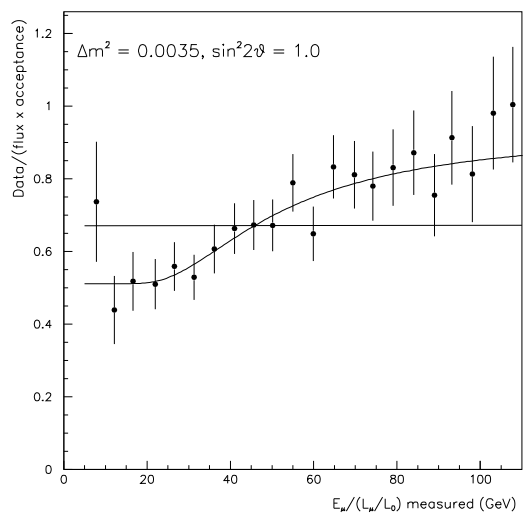,width=0.5\linewidth}}
\end{center}
\caption{\small Simulated spectrum of a typical experiment for 3 years
of data taking showing the ratio of data corresponding to 
oscillations with 
sin$^2 2\theta =1.0$ and $\Delta m^2 =0.0035$~eV$^2$,
to Monte Carlo expectation with no oscillations.
The horizontal line is the average level used to
test the statistical significance of the oscillations 
and the solid line is the fit used to 
extract the oscillation parameters.}
\label{fig:sim_example}
\end{figure}

To explore the $\Delta m^2$ and sin$^2 2 \theta$ parameter space,
a series of three-year simulations were carried out, 
each one producing a distribution of `data'
points like the points shown in figure~\ref{fig:sim_oscil}.
A large number of such simulations were carried out 
for each point in parameter space. 
Each of these distributions was compared to a
high-statistics simulation without oscillations.
An example of the ratio
oscillations/no-oscillations 
for one of these three-year simulations 
is shown in figure \ref{fig:sim_example}.

\begin{figure}
\begin{center}
\mbox{\epsfig{file=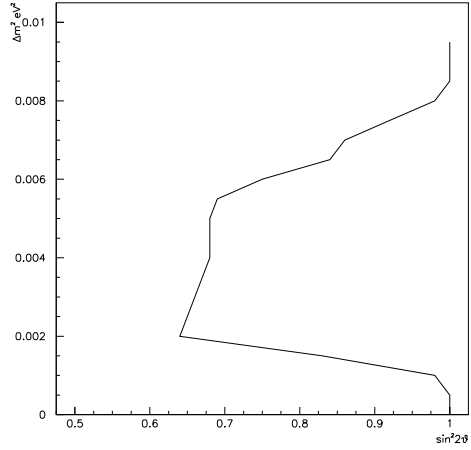,width=0.5\linewidth}}
\end{center}
\caption{\small For three years of data taking, neutrino
oscillations could be excluded at 90$\%$ confidence level 
in 80$\%$ of simulated experiments 
in the region to the right of the curve.}
\label{fig:sim_exclud}
\end{figure}

These simulations have been used to find the regions of
parameter space that could be excluded if no oscillations
are found in the data. 
To remove the effect of the absolute normalization, the
no-oscillations distribution is multiplied by a factor
which is varied as a free parameter for each of the
three-year simulations. 
Then the $\chi^2$ 
for the hypothesis that all of the
`data' points in the spectrum are compatible with 
no oscillations is
calculated. If the $\chi^2$ probability is 
greater than 10\%, then the experiment is said to be compatible
with no neutrino oscillations for the given parameters. 
Conversely, in the region of parameter space 
where the probability is less than 10\%, 
evidence for oscillations could be claimed with
90\% confidence.
Due to statistical fluctuations there is a spread in probabilities 
between different experiments for the same parameters. 
The exclusion plot shown in figure~\ref{fig:sim_exclud} gives the region
where 80$\%$ of experiments could give an exclusion at 
the 90\% confidence level.

\begin{figure}
\begin{center}
\mbox{\epsfig{file=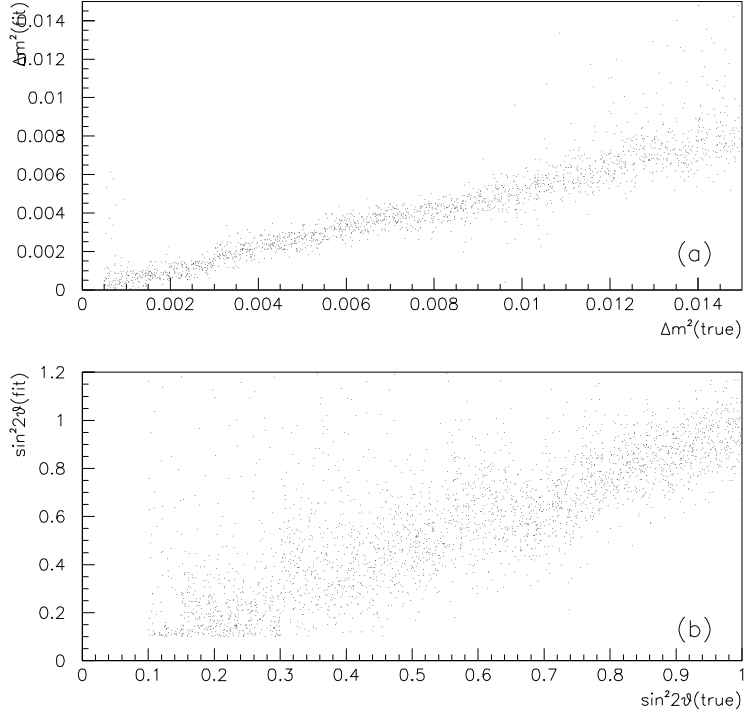,width=0.8\linewidth}}
\end{center}
\caption{\small Results of the fits to neutrino oscillation
parameters:
(a)~fitted and true values of $\Delta m^2$ for fixed 
sin$^2~2 \theta$~(true) = 1,
and (b)~fitted and true values of sin$^2 2 \theta$ 
for fixed $\Delta m^2$~(true) = 0.0035~eV$^2$.}
\label{fig:param}
\end{figure}

The simulations are also used to estimate the accuracy with
which the oscillation parameters can be extracted from the data. 
Distributions such as those shown in figure~\ref{fig:sim_example}
are fitted with a damped oscillation function:
$$ P_{fit} = 1 - \sin^2 2\theta
 [ 1/2 - ( 1/2 - \sin^2y ) e^{-2 (fy)^2}] $$
where $ y = (1.27 \Delta m^2 L_\mu /E_\mu)/\sqrt{1+f^2} $ 
and $f$ is a damping parameter.
In the analysis this damping parameter is obtained from a fit to a 
high-statistics Monte Carlo spectrum 
for each set of oscillation parameters. Then
the `data' distributions from the three-year simulations 
are fitted with $f$ fixed and $\Delta m^2$ 
and sin$^22\theta$ free.
Figure~\ref{fig:param} displays
the fitted oscillation parameters compared to 
the true generated parameters. A good correlation
between the fitted parameters and the true values 
is  seen over a large range
of the parameters 
but there is a bias in the fitted value of $\Delta m^2$
because the muon energy is smaller than the neutrino energy.
In the experiment, this bias would be corrected using the Monte Carlo.

\begin{figure}
\begin{center}
\mbox{\epsfig{file=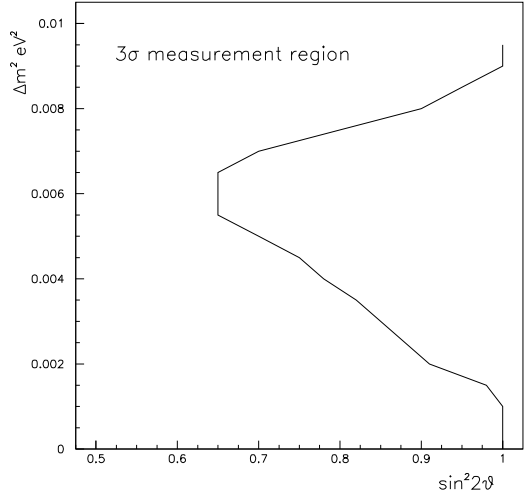,width=0.5\linewidth}}
\end{center}
\caption{\small For three years of data taking, 
the oscillation parameters $\Delta m^2$, sin$^2 2 \theta$ 
can be measured with fractional error less than 33$\%$ in
more than 80$\%$ of experiments in the region to the right of the curve.}
\label{fig:sim_good}
\end{figure}

Figure~\ref{fig:sim_good} shows the region
of parameter space where both $\Delta m^2$ and $\rm{sin}^22\theta$
can be measured with an error less than 33$\%$ in more than 80$\%$
of experiments.
It can be seen that the experiment can make accurate measurements
for $0.002<\Delta m^2<0.008$ and $\rm{sin}^22\theta>0.8$. The measurement
region is shifted up in $\Delta m^2$ compared to the exclusion region
shown in figure~\ref{fig:sim_exclud} 
because of the low energy cut-off in the
acceptance. 
A range of $E/L$ values around the minimum 
of the survival probability is needed to find the precise 
position of the minimum and make an accurate determination of 
$\Delta m^2$, 
whereas a measurement of the number of events at the calculated
position of the minimum is sufficient for exclusion.

\subsection{Systematics}

The present studies have shown that the proposed ANTARES detector 
can make useful measurements with the statistics expected in three
years. It is important to note, however, that the present analysis 
is exploratory and can be refined in many ways. 
The \numu quasi-elastic events 
(estimated to be about 15\% of the charged-current \numu events)
were not included in these simulations. These events
have a better correlation between the muon energy 
and the incident neutrino energy
than the deep-inelastic events used here.
Inclusion of these events will improve the experimental sensitivity to low
$\Delta m^2$. 
On the other hand, 
the sensitivity will suffer in the real experiment due to systematics 
resulting from
variations in the detector acceptance and assumptions about the neutrino flux
and its energy and angular dependence and about backgrounds.

The event reconstruction used at present has separate procedures
for multi-string and single-string events. In
the future, a uniform event reconstruction will be developed to
avoid the systematic errors caused at the edge of the acceptances.

The backgrounds have so far not been taken into account
in the analysis.
However, making a known background subtraction with the magnitude of
backgrounds presently estimated would have little impact on the results. 
The backgrounds from \nue and from non-contained \numu are related
to the atmospheric flux and are well simulated, so a
background subtraction can be made.
The present simulations indicate the backgrounds from muons
originating in the atmosphere are small. These simulations must
be refined by including light scattering in the water, and extended
to higher statistics.
The detector will have a significant acceptance for events above
the horizontal, and these events can be used to verify the 
muon background simulations for events below the horizontal.

In the analysis presented above, it was assumed that the
shape of the neutrino flux was known as a function of energy 
and zenith angle, but that the overall rate was unknown.
A different analysis can be performed which uses
the non-contained \numu events to normalise the contained \numu
events. This method would have similar statistical errors
to the method presented here, but very different systematics.

The knowledge of the flux of atmospheric neutrinos in the
energy range from 10~GeV to 1~TeV is of crucial importance for the
study of neutrino oscillations, because observed deviations
from the expected flux could be interpreted as a possible 
oscillation signal. Several calculations of this flux 
exist~\cite{volkova,mitsui,butkevich,bartol}. The comparison of the total
flux shows variations of the order of 20\% between the different calculations.
The energy dependence of the flux is also slightly different
in the various approaches. The difficulties in obtaining  consistent
results come from assumptions about the primary cosmic ray flux as well as
from the lack of knowledge concerning 
the development of atmospheric hadron showers.

\begin{figure}
\begin{center}
\mbox{\epsfig{file=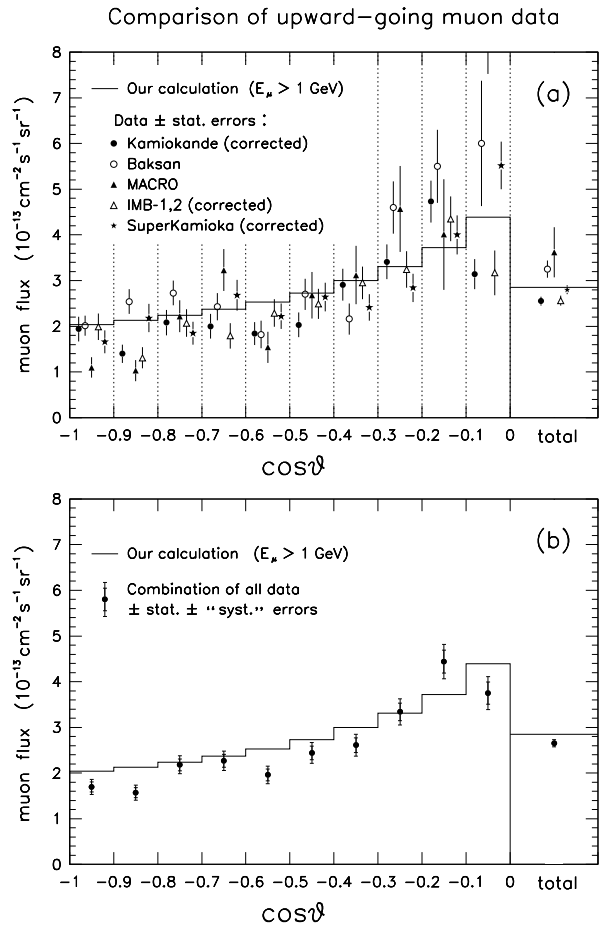,width=14cm}}
\vspace{-4cm}
\caption{\small Comparison of all upward-going muon spectra, 
rescaled to a common
threshold of 1~GeV, for (a) data from individual experiments, 
and (b) combined data. The figures are taken from Fogli et al.~\cite{fogli}.}
\label{upward}
\end{center}
\end{figure}

A compilation of present experimental data
on upward-going muons 
is shown in figure~\ref{upward}, taken from Fogli et al.~\cite{fogli}.
Only the zenith-angle dependence is available.
The experimental data show large 
fluctuations, but they are generally compatible. 
Nonetheless, even the combined data
shown in figure~\ref{upward}~(b) have 
at least a 20\%~bin-by-bin uncertainty.  This is likely to be the
limiting systematic on neutrino oscillation measurements by ANTARES,
and studies of theoretical calculations and archive data are under way
to reduce it as much as possible. 

%% file: chap7.tex

A number of related projects for the construction of neutrino
telescopes in either ice or water are currently underway. The status
of these projects is reviewed in this chapter.  The case of AMANDA is
given particular consideration, as the project is presently making
rapid progress.

\section{Deep ice detector: AMANDA }

AMANDA~\cite{Amanda} was installed in several stages 
in the deep Antarctic ice at the South Pole. 
The detector has a rather simple design: 
the analogue signals of optical modules (OMs) are directly transmitted to
surface through 2~km cables.
 In 1993-94, 80 OMs of AMANDA-A were
deployed on four strings at depths of 810 to 1000\,m; however,
measurements of the ice transparency at those depths showed that the
light scattering was unacceptable for maintaining good directional
information.  Subsequently, in the Austral summer 1995--96,
 86 OMs on four strings (AMANDA-B4),
were deployed at depths of 1500--2000\,m. The AMANDA-B4 array was mainly
used for calibration, and testing track reconstruction and cuts on the
real data.  The scattering
length for Cherenkov light in AMANDA-B depths was found to be
 two orders of magnitude larger than in shallower depths. 
In the following year, six new strings have
completed AMANDA-B, which now consists of 302 PMTs on 10 strings.
First analysis of the data from 85 days of effective lifetime resulted 
16 up-going
neutrino candidates, where 21 up-going atmospheric neutrinos were expected
from Monte Carlo simulations. 

During the 1997-98 season, the construction of AMANDA-II, which will
have an effective area of 30,000~m$^2$ has begun. Three 1.2~km long
strings were installed with modules that range from 1150 to
2350~metres in depth. In the 1999/2000 season 6 additional strings will be
deployed to complete the AMANDA-II array. The new strings will test a
variety of new techniques, such as the usage of 10$''$ photomultiplier
tubes, and 25--50 digital optical modules.

AMANDA is also proposing a km scale array (ICECUBE) of 5000 OMs on 80
strings to be deployed by 2007--08 at 1.4 to 2.4 km depths in the
Antarctic ice.

\subsection{AMANDA compared to ANTARES}
Apart from their technological differences, the main difference
between ANTARES and AMANDA is the use of water versus ice as the
detector medium.  Ice has the advantages of great stability (no
currents, hence no movement of deployed strings), no living organisms
to cause optical fouling or bioluminescence and low noise environment
due to absence of $^{40}$K.  The advantages of water are a much
longer scattering length, as measured by ANTARES test deployments, and
flexibility in experimental design (strings can be retrieved and
re-deployed).

The difference in scattering length between ice and water is
substantial: AMANDA measures $24\pm2$ m for ice at
1700~m~\cite{halzena}, whereas ANTARES test deployments indicate a
scattering length in excess of 200~m.  This leads to a significant
improvement in the angular resolution of ANTARES. An accuracy of
$0.3^\circ$ is achievable.

\subsection{Astrophysical sources}

 From the point of view of sky coverage, a detector at the mid-northern
 latitudes is an ideal complement to the South Pole location of
 AMANDA.  Neutrino telescopes look downwards, and are therefore
 sensitive to sources which are {\it not\/} circumpolar at their
 location: AMANDA `sees' the northern celestial hemisphere (sources
 with positive declination $\delta$).  ANTARES, if located at latitude
 $42^\circ\,50^\prime$ N, will see all sources with declination
 $\delta < +47^\circ\,10^\prime$: when combined with AMANDA this
 provides complete sky coverage, with overlap for confirmation and
 cross-calibration in the region $0^\circ < \delta < 47^\circ$.  It is
 worth noting that the Galactic centre, a region of interest for
 neutrino astronomy because of its high level of activity, is in the
 southern hemisphere and is therefore visible to ANTARES but not to
 AMANDA.

The low noise environment of AMANDA 
makes it more suitable for the detection of low-energy (MeV)
neutrinos such as those associated with supernovae.  However, neither
AMANDA nor ANTARES is really designed to compete with  dedicated
low-energy neutrino detectors such as Super-Kamiokande~\cite{superk},
SNO~\cite{sno-supernova} or
the proposed OMNIS~\cite{omnis} for supernova observations.

\subsection{Neutrino oscillations and neutralino searches}

Measurement of neutrino oscillations by the method described above
(effectively, measuring the energy spectrum) requires a low energy
threshold, closely spaced optical modules and good background
rejection.  As detailed in chapter \ref{chap:detector_performance}, 
ANTARES should be capable of
exploring the region of the Super-Kamiokande signal~\cite{superk} in a
few years of running.  It may be more difficult to do the energy
measurement by range in a more diffusive medium, such as the South
Pole ice.

As discussed in chapter 2, neutralino searches by neutrino telescopes
rely on the detection of neutrinos produced when neutralinos
gravitationally captured in the centre of the Earth or the Sun
annihilate.  The angular distribution of muons produced by neutrinos
from neutralino annihilation in the Earth's core can be used to infer
the neutralino mass~\cite{edsjo}, therefore, a good angular 
resolution is needed
to make best use of any signal.  The calculated width of the
neutrino-induced muon distribution ranges from about 2.5$^\circ$ for
high $m_\chi$ to $\sim 7.5^\circ$ for $m_\chi\sim80$ GeV/$c^2$, so an
angular resolution of better than 1$^\circ$ is highly desirable.  A
pair of neutralinos does not annihilate directly into a pair of
neutrinos, so the neutrino energy spectrum peaks at around one-third
to one-half of the neutralino mass. Therefore, a fairly low energy
threshold is useful.

In ANTARES both the angular resolution and the low energy threshold
can be achieved by using the analysis technique proposed
for the neutrino oscillation search.  The AMANDA
search~\cite{amandaneutralino} uses a conventional multi-string
analysis and loses sensitivity below a neutralino mass of about 100
GeV/$c^2$.

\subsection{Complementarity}
Detection of a new phenomenon by a single experiment is never entirely
satisfactory.  The fact that ANTARES and AMANDA are broadly comparable
experiments with similar capabilities in many areas promises rapid,
independent confirmation of any new discovery, avoiding the long
period of uncertainty associated with, for example, the solar neutrino
deficit.  Moreover, as shown above, ANTARES and AMANDA are also in
many respects complementary: AMANDA has lower background noise and is
already operational, ANTARES has a longer scattering length, hence
better angular resolution, and can be optimised for specific physics
targets.  The two together provide excellent all-sky coverage for
astrophysical point sources.  Overall, the combination of ANTARES and
AMANDA addresses all the physics goals of high-energy neutrino
telescopes better than either one could do alone.

\section{Deep water detectors}

\subsection{Baikal}

The Baikal~\cite{baikal} Neutrino Telescope is situated in the Siberian
Lake Baikal, at a depth of 1.1~km. It consists of eight strings
supporting 192 OMs. This project started in April
1993 by deploying 36 OMs in three half-strings. The system was
recovered after 300 days of running time. The main problem was the
biofouling deposited on the glass spheres, mainly on the upwards
facing optical modules.  In April 1996, four full strings carrying 96 OMs 
were deployed (NT-96). Only two layers of OMs were facing up. In winter
1997, 144 OMs were deployed, and the detector was
completed in 1998 with 192 OMs. 
In the analysis~\cite{NT-96} of 70 days of data taken with NT-96 array
$2.0\times 10^7$ events were reconstructed, nine of which were up-going. 
The expected number of atmospheric neutrinos from MC 
studies was 8.7, and the angular distribution of the events  
was consistent with that of atmospheric neutrinos. The effective area
of the detector had a maximum value of 400~m$^2$ for this analysis.  
A different method was developed specifically to look for vertically
up-going events, for which an effective area of about 1000~m$^2$ was
reached. This analysis found 4 events with zenith angles $>150^\circ$
where 3.7 events were expected. From this result Baikal set a $90\%$ C.L.
 upper limit of $1.1\times 10^{-13}$ $cm^{-2} s^{-1}$ on the muon 
flux from the centre of the Earth for energies $>10$~GeV.

\subsection{NESTOR}

NESTOR~\cite{Nestor} is planned to be installed in a 3\,800\,m deep
sea site off Pylos (Greece). The project consists of a tower of 12
hexagonal floors of 16\,m in radius supporting 168 OMs in total.  The
deployment of aluminium and of titanium floors has been tested at
2\,500\,m depth. Digital transmission of the signals is being
developed.

\section{Long baseline experiments}

The anomalous results seen in various atmospheric neutrino experiments
could, in principle, be explained either by $\nu _{\mu}-\nu _{e}$
oscillations or $\nu _{\mu}-\nu _{\tau}$ oscillations, even though the
latest Super-Kamiokande results favour $\nu _{\mu}-\nu _{\tau}$.  The
$\nu _{\mu}-\nu _{e}$ explanation has essentially been ruled out by
the results of the CHOOZ~\cite{CHOOZ} experiment. 
 The CHOOZ collaboration observed
electron neutrinos at a distance of one~km from a reactor complex by
observing inverse $\beta$-decay in a large heavily-shielded
gadolinium-loaded liquid scintillator target.  The disappearance of
electron neutrinos as they oscillated to muon neutrinos would have
resulted in a distortion of the observed electron spectrum from
inverse beta decay.  No distortion was seen, and the combination of
the long baseline and low background meant that the entire parameter
space for $\nu_{\mu}-\nu_{e}$ oscillations allowed by the existing
atmospheric neutrino experiments was ruled out.  A second
long-baseline reactor experiment is underway at Palo Verde in the
United States to verify this result.

Explanations for the observed atmospheric neutrino anomaly based on
$\nu_{\mu}-\nu_{\tau}$ oscillations could be tested by means of
accelerator-based long-baseline experiments. In these experiments, an
almost pure muon neutrino beam is produced from the decay in flight of
a pion beam at an accelerator facility and directed towards a distant
underground laboratory.  These muon neutrinos produce muons in the
detector by charged-current (CC) interactions.  The first signature of
neutrino oscillations would be a change in the total muon rate or a
distortion in the observed energy spectrum caused by the
energy-dependent disappearance of muon neutrinos.  Of course any
neutrino, with or without oscillations, can produce neutral-current
(NC) interactions in the detector, so a second signature for neutrino
oscillations would be a measured NC/CC ratio different than that
expected for a pure muon neutrino beam.  A third signature would be
the appearance of $\tau$~leptons coming from CC interactions of
$\tau$~neutrinos resulting from oscillations.  There are currently
three projects: K2K~\cite{K2K}, MINOS~\cite{MINOS}, and CERN/LNGS. 
 The potential advantage
of ANTARES is in the large fiducial mass of the detector which will
allow the measurement of atmospheric neutrinos above 5~GeV.

\subsection{K2K}

The 12~GeV primary proton beam at KEK is used to produce a muon
neutrino beam directed at the Super-Kamiokande detector 250~km away. It
first passes through two near detectors on the KEK site. As the energy
of the resulting neutrinos is too low to produce $\tau$~leptons in the
final state, the experiment searches for muon neutrino disappearance.

\begin{figure}[h]
\begin{center}
\mbox{\epsfig{file=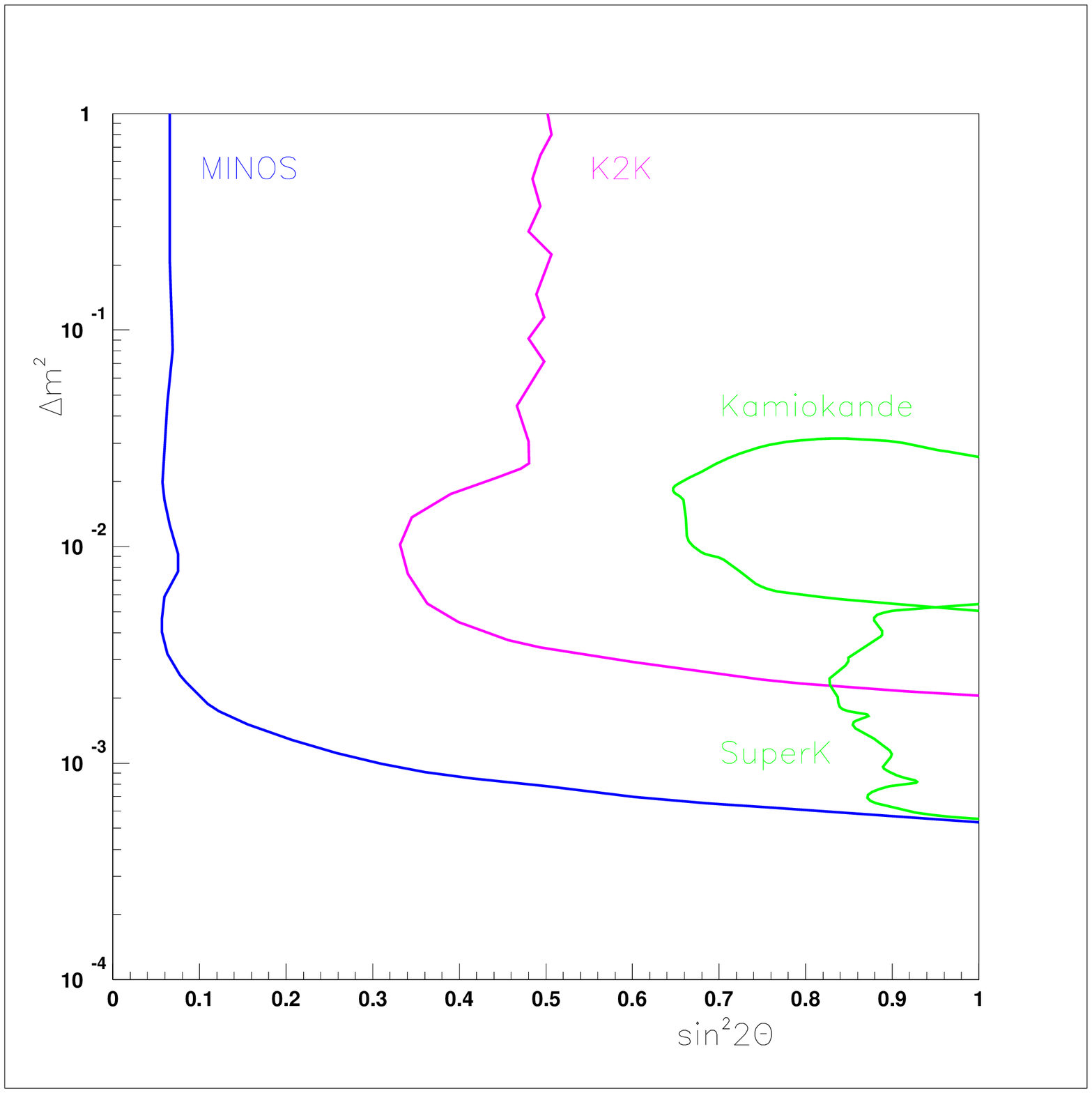,bbllx=0pt,bblly=150pt,bburx=540pt,bbury=660pt,width=0.9\linewidth}}
\caption{Expected sensitivity for K2K and MINOS compared with
Kamiokande and Super-Kamiokande signals.}
\label{fig:lblsens}
\end{center}
\end{figure}

The K2K experiment sent its first neutrinos towards the
Super-Kamiokande detector in March~1999.  The sensitivity of the
experiment is limited by the luminosity of the beam. However, the
experiment has the advantage of a considerable head start over the
other projects.  Figure~\ref{fig:lblsens} displays the sensitivity
after three years of running.

\subsection{MINOS}

The MINOS experiment will consist of a specially built beam-line
directing neutrinos  from the 120~GeV proton beam of the new
Main Injector at Fermi National Accelerator Laboratory (FNAL) to a
5.4~kton magnetised iron calorimeter to be built in the Soudan mine,
730~km away in northern Minnesota.  The existing Soudan II detector
will form an additional far detector with a smaller fiducial mass but
finer-grained calorimetry.  There will be a near detector on the FNAL
site.

The initial version of the detector is limited to muon neutrino
disappearance signatures.  The group plans to upgrade their detector
in the future, with a hybrid emulsion detector which would be capable
of imaging $\tau$~decays.  The predicted sensitivity of the
experiment is shown on figure~\ref{fig:lblsens}.

\subsection{CERN/LNGS}

There is much interest in constructing a third long-baseline facility
using a beam from CERN directed towards the Gran Sasso Laboratory.
This idea is proceeding to the point of formal proposals to be
submitted to the respective laboratories in autumn~1999.  
Because they emphasise $\tau$~appearance signatures, the
CERN/LNGS projects may be best
suited for precise measurements.
\vglue 0.5cm

\section{Gamma ray telescopes and Air Shower arrays}

Gamma ray telescopes and neutrino telescopes are obviously
complementary, though they may probe the same sources. Purely
electromagnetic sources would not be seen in neutrinos, but hadronic
sources will give both neutrinos and photons. Neutrinos are less
likely to be absorbed at the production site or during the transport
(photons above 10--100~TeV are absorbed by $\gamma \gamma$
interactions).

The Pierre Auger project~\cite{auger} is designed to
measure very high energy cosmic
rays (charged and neutral).  
Specifically, the experiment will also detect the showers caused by 
ultra-high-energy neutrinos (10$^{18}$~eV) travelling horizontally. 

%% file: chap8.tex


\hspace{3ex}The cost of a detector consisting of 4 strings
of 123 optical modules and 9 strings of 63 optical modules,
for an overall number of 13 strings and 1059 optical
modules,
has been estimated.
Table \ref{tab:cost_string} gives a summary of the cost of these
two types of strings.

\begin{table}[h]
\begin{center}
\begin{tabular}{||c||c|c||}
\hline
{\bf Cost of one string} & {\bf 63 OM} &{\bf 123 OM}\\ \hline
& {\bf kFF} & {\bf kFF} \\ \hline\hline
Acoustics: & 500 & 500 \\
4~rangemeters,  & & \\ 
concentrator, interface  & & \\ \hline
Optical modules & 1200 & 2340 \\ \hline
Top of string & 120 & 120 \\ \hline
Bottom of the string & 390 & 390 \\ \hline
Qualification, & 35 & 50 \\ 
transport, assembling &  & \\ \hline
elementary segments & 825 & 1610 \\ \hline
Electronics cost, & 150 & 175 \\ 
bottom of the string &  & \\ \hline
Container & 1050 & 2050 \\ \hline
              && \\
{\bf Total}& {\bf 4270} &  {\bf 7235} \\

   && \\ \hline\hline
\end{tabular}
\caption{Cost estimates for the two types of strings.}
\label{tab:cost_string}
\end{center}
\end{table}

Funding is already available for the construction of the first
string: the ``demonstrator''. The costs for
the further 12 strings are listed below:

\begin{itemize}
\item Strings:\\
      4 with 123~OM, 8 with 63~OM  \hglue 3.6cm    70~MFF\\
      including 10\% contingency
\item Electro-optical cables:\\
      2~cables / 12~fibres / 40~km \hglue 4.3cm 9~MFF
\item Deployment operations:\\
      80~days costed, 10-15~needed at sea \hglue 2.7cm 10~MFF
\item Sea floor interconnections and junction box:
      \hglue 1.2cm 4~MFF
\item Main power supply and transformer \hglue 2.7cm 2~MFF
\item Electronic cost for trigger, DAQ on the shore:\hglue
1.1cm 2~MFF
\item {\bf Total cost: \hglue 6.9cm 97~MFF}\\
\end{itemize}

These estimates are based on current experience.
Large scale production could lead to lower costs and enable extra strings to
be built.

Construction of a 1000 optical module detector would proceed in two
stages. The first stage would be a $\sim$ 500~OM detector,
costing 48~MFF, to be
built as soon as possible and enabling physics to start.
For this stage the electronics would be the critical path item.

The proposed schedule would be to build:
\begin{enumerate}
\item Seven strings (3 high density, 4 low density) 
in 1999-2000-2001, to be deployed in~2002;
\item The remaining strings to be ready to deploy in~2003.
\end{enumerate}

%% file: chap9.tex


In the summer of~1996, the ANTARES collaboration started to
study the feasibility of mounting a neutrino telescope
in the Mediterranean Sea near Toulon. The first phase of the project
was to measure parameters of the marine environment and to
acquire and develop the technology necessary to perform
an undersea experiment.

This phase is now completed. A location has been chosen
and explored using the submarine {\it Nautile}. The topography
has been mapped and the sea current measured during periods
covering two years. The rate and nature
of sedimentation at the site has been measured in different
seasons,
and the loss of transparency of the optical modules
was found to be less than 2$\%$ per year.
The transmission properties at the site
have been measured for blue light.
The absorption length has been determined to be of the order of 50 m, 
whereas the effect of scattering is found to be negligible. 
Studies of the optical background in the sea water
due to natural radioactive decays and biological organisms
indicate that these rates can be dealt with.
This background will lead to an
inefficiency of the detector of less than 5$\%$. 

The technology required for the experiment has been attained with
the indispensable help of IFREMER and INSU, as well as through
developments with industry. The prototype string has proven our
ability to deploy a detector string and the submarine cable
connections have been demonstrated. The design of mechanics
and electronics for a full detector is well advanced and most
essential elements have been illustrated.

This proposal has presented the extensive scientific programme possible
with an undersea  neutrino telescope.
The same apparatus could detect cosmic neutrinos
in the range from TeV to PeV with the ability to measure the diffuse
neutrino flux from active galactic nuclei, 
make searches for dark matter in
the form of relic neutralinos, and measure neutrino oscillation
parameters. Evidence for point sources of neutrinos could be obtained by
integrating the neutrino flux from the positions of known
candidate sources such as AGN and gamma-ray bursts. 
Simulations have been performed to show the physics
potential of the detector and optimize its geometry.

The next step is the proposed construction
of a detector consisting of 13~strings with about 1000~photomultipliers
in an array covering an area of 0.1~km$^2$. The detector will be
completely modular and can be deployed in the sea in stages.
The
present schedule calls for 7~strings to be deployed in~2002 with continual
tests and partial deployments at intermediate stages. Physics
measurements could start with this initial detector. 
The full 13~string detector should be operational in~2003. 
The cost of
the proposed detector is estimated at 15~MEuros (97~MFF).

The ANTARES project has now grown to a size of over 100~members
from five European countries.
While still actively searching and attracting new 
collaborators, the present institutes have the manpower
and resources required to build and operate a detector of the size
and complexity proposed.  

The collaboration between particle physicists and astrophysicists has a
long and fruitful history. It has given rise to entirely new ways of looking
at the Universe, such as X-ray and gamma-ray satellite-based
telescopes, and
recently ground-based gamma ray telescopes extending the 
field of astronomy above 10~TeV. There is every reason to believe that 
neutrino astronomy will open a new window on the Universe at least as 
exciting and intriguing.

Some of the goals outlined in the scientific programme
 cannot be achieved on a realistic 
time scale with a 0.1 km$^2$ detector.  For example, to observe most 
astrophysical point sources a detector on the square kilometre scale will
be required.  The experience gained in constructing and operating the 
detector described here will be essential for the success of a km-scale 
project.

%% file: proposal.bbl
\begin{thebibliography}{999}


\bibitem{rf:gzk} K.\,Greisen, Phys. Rev. Lett. {\bf 16} (1966) 748; \\
     G.\,T.\,Zatsepin and V.\,A.\,Kuz'min, JETP Lett. {\bf 4} (1966) 78; \\
     R.\,J.\,Gould and G.\,P.\,Schreder, Phys. Rev. Lett. {\bf 16} (1966) 252.
     

\bibitem{Berez90} V.S. Berezinsky et al., Astrophysics of Cosmic Rays,
North-Holland, 1990.
\bibitem{Bednarek97} W. Bednarek and R. J. Protheroe, 
Phys. Rev. Lett.  {\bf 79} (1997) 2616.
\bibitem{Esposito96} J.A. Esposito et al., Astrophys. J. {\bf 461} (1996) 820.
\bibitem{Gaisser95} T.\,K. Gaisser, F. Halzen and T. Stanev, Phys. Rep. {\bf
258} (1995) 173.
\bibitem{Egret94} C. E. Fichtel et al., Astrophys. J. Suppl. {\bf 94} 
(1994) 551.
\bibitem{TeV} From WHIPPLE Collaboration:\\
                    M. Punch et al., Nature {\bf 160} (1992)
		    477;\\
                    J. Quinn et al., IAU Circular 6169, Juin 1995.\\
                    From HEGRA Collaboration:\\
                    D. Petry et al., Astron. Astrophys. {\bf
		    311} (1996) L13.


\bibitem{Biermann89} P.L. Biermann, Proceedings of {\em ``Hotspots in
extragalactic radio sources"}, Astron. Astrophys. {\bf 272} (1993) 161.
\bibitem{Mannheim98} K.~Mannheim, Science {\bf 279} (1998) 684.
\bibitem{WaxBah99} E. Waxman and J. Bahcall, Phys. Rev. {\bf
D59} (1999) 023002.
\bibitem{SAX} IAU Circular 6649, 1997,
The BeppoSAX Science Data Center, http://www.sdc.asi.it.
\bibitem{Piran98} T. Piran, Physics Report (in press); astro-ph/9810256.

\bibitem{Galama99} T.J. Galama et al., astro-ph/9903021.

\bibitem{Baron} E. Baron, Nature {\bf 395} (1998) 635, 
and references therein.


















\bibitem{UHEdata}
D.J. Bird et al., Phys. Rev. Lett. {\bf 71} (1993) 3401;\\
M. Takeda et al., Phys. Rev. Lett. {\bf 81} (1998) 1163.

\bibitem{UHEreview} 
S. Yoshida and H. Dai, J. Phys. {\bf G24} (1998) 905.

\bibitem{UHEtopdef}
P. Bhattacharjee and G. Sigl, astro-ph/9811011.

\bibitem{UHEtopevol}
G. Vincent, N.D. Antunes and M. Hindmarsh, Phys. Rev. Lett.
{\bf 80} (1998) 2277.


\bibitem{UHEtopgamma}
P. Bhattacharjee, Q. Shafi and F.W. Stecker, Phys. Rev.
Lett. {\bf 80} (1998) 3698.

\bibitem{UHEbkv}
V. Berezinsky, M. Kachelrie{\ss} and A. Vilenkin, 
 Phys. Rev. Lett. {\bf 79} (1997) 4302.

\bibitem{UHEbs}
M. Birkel and S. Sarkar, Astropart. Phys. {\bf 9} (1998)
297.

\bibitem{UHEpartprod}
D.J.H. Chung, E.W. Kolb and A. Riotto, hep-ph/9802238;\\
V.A. Kuzmin and I. Tkachev, hep-ph/9802304.

\bibitem{UHEben}
K. Benakli, J. Ellis and D.V. Nanopoulos, hep-ph/9803333.

\bibitem{UHEaniso}
S. L. Dubovsky and P. G. Tinyakov, hep-ph/9802382.

\bibitem{UHEggs}
P. Gondolo, G. Gelmini and S. Sarkar, Nucl. Phys. {\bf
B392} (1993) 111.


\bibitem{kamioka} K.\,S.\,Hirata et al., Phys. Lett. {\bf
  B205}\,(1988)\,416; \\
  K.\,S.\,Hirata et al., Phys. Lett. {\bf B280}\,(1992)\,146; \\ 
  Y.\,Fukuda et al., Phys. Lett. {\bf B335}\,(1994)\,237.

\bibitem{imb} D.\,Casper et al., Phys. Rev. Lett. {\bf 66}\,(1991)\,2561; \\
  R.\,Becker-Szendy et al., Phys. Rev. {\bf D46}\,(1992)\,3720.

\bibitem{superk} Y.\,Fukuda et al., Phys. Rev. Lett. {\bf
81}\,(1998)\,1562.

\bibitem{soudan} W.\,W.\,M.\,Allison et al., Phys Lett. {\bf
B391}\,(1997)\,491.

\bibitem{frejus1} Ch.~Berger et al., Phys. Lett. {\bf B245}\,(1990)\,305.

\bibitem{frejus2} K.~Daum et al., Z. Phys. {\bf C66}\,(1995)\,417.

\bibitem{nusex} M.~Aglietta et al., Europhys. Lett. {\bf 8}\,(1989)\,611.

\bibitem{imb-stop} R.~Becker-Szendy et al., Phys. Rev.
Lett. {\bf 69}\,(1992)\,1010.

\bibitem{Halzentau} F. Halzen and D. Saltzberg,
Phys. Rev. Lett. {\bf 81} (1998) 4305.


\bibitem{cosmology} M.A. Strauss and J.A. Willick, 
Phys.~Rep.~{\bf 261} (1995) 271; 
\\A. Dekel, Ann.~Rev.~Astron.~Astrophys.~{\bf 32} (1994) 319.

\bibitem{susy} See, for example, H. E. Haber and G. L. Kane, 
Phys.~Rep.~{\bf 117} (1985) 75.

\bibitem{LSP} J.~Orloff, M.~de~J\'esus, C.~Tao,
``The Nature of the LSP", Rapport GDR Supersym\'etrie, 1998.

\bibitem{jungman} G. Jungman, M. Kamionkowski and K. Griest, Phys.~Rep.~{\bf 
267} (1996) 195.

\bibitem{bottino} A. Bottino, N. Farnengo, G. Mignola and L. Moscoso, 
Astropart.~Phys.~{\bf 3} (1995) 65; \\
V. Berezinsky et al., Astropart.~Phys.~{\bf 5} (1996) 333.

\bibitem{edsjo} J. Edsj\"o, ``Aspects of neutrino detection of neutralino 
dark matter", PhD thesis, University of Uppsala (1997).

\bibitem{gondolo} P.~Gondolo and J.~Silk, ``Dark Matter
Annihilation of the Galactic Center", MPI-PhT/99-10;
OUAST/99/9, March 1999.

\bibitem{ellis} J. Ellis, R. A. Flores and S. S. Masood, Phys.~Lett.~{\bf B294} 
(1992) 229.



\bibitem{Rubakov} V.A. Rubakov, JETP Lett. {\bf 33} (1981) 644. 

\bibitem{MACROMon} MACRO Collaboration, M. Ambrosio et al.,
Phys.~Lett.~{\bf B406} (1997) 249.

\bibitem{BAKSANMon}BAKSAN Collaboration, E.N. Alexeyev et
al, 21st ICRC, Adelaide, vol.10 (1990) 83. New limits have
been given by M. Boliev as private communications.

\bibitem{Parkerbound} E.N.Parker, Astrophys. J. {\bf 160} (1970) 383.\\
An improved bound is given by F.C. Adams et al., Phys. Rev.
Lett. {\bf 70} (1993) 2511.

\bibitem{BaikalMon}I. Sokalski (for the Baikal Collaboration),
Proceedings of the 1st International Conference on Non Accelerator New Physics
NANP-97, Russia (1997).

\bibitem{Weiler} T.W. Kephart and T.J. Weiler, Astropart.
Phys. {\bf 4} (1996) 271.

\bibitem{Qballs} A. Kusenko et al., Phys. Rev. Lett. {\bf 80} (1998)
3185.

\bibitem{BaikalQB} I.A. Belolaptikov et al., (Baikal Collaboration),
astro-ph/9802223.



\bibitem{ghandi}R. Ghandi et al., Astropart. Phys {\bf 5}
(1996) 81.

\bibitem{rn} D. Rein and L.M. Seghal,  Ann. of Phys.,
{\bf 133} (1981) 79.

\bibitem{H1ZEUS} T. Ahmed et al., Phys. Lett. {\bf B324} (1994) 241.\\
S. Aid et al., Z. Phys. {\bf C67} (1995) 565.\\
M. Derrick et al., Phys. Lett. {\bf B316} (1993) 412.

\bibitem{cteq} H.L. Lai et al., Phys. Rev. {\bf D51}
(1995) 4763.

%
\bibitem{mobley} Curtis D. Mobley, {\it Light and Water 
(Radiation Transfer in Natural Water)}, Academic Press (1994). 
%



\bibitem{Mazeas}F. Mazeas et al, R.INT.DITI/GO/MM 99-05
IFREMER-Brest.



\bibitem{bib:chap5_transmision_mes}
P. Galoumian, R. Berthier, P. Lamare, S. Loucatos,
``Glass and gel transparency tests"
ANTARES-Opmo/1998-004.

\bibitem{bib:chap5_mucage}
J. McNutt and A. Cade, 
``More $\mu$-metal Cage Studies", ANTARES note Nov. 1997.

\bibitem{bib:chap5_pmtspecif}
S. Basa and F. Montanet,
``Specifications for the Photomultipliers Used in the
ANTARES Project",  
ANTARES-Opmo/1998-003.

\bibitem{bib:chap5_pmtstudies}
C. Arpesella,  S. Basa and F. Montanet,
``Test results of a 14 stage 8'" Hamamatsu photomultiplier (R5912-02), 
ANTARES-Opmo/1997-005.

\bibitem{bib:chap5_gamelle} H. Lafoux,
``Late(st) Gamelle Results", ANTARES-Opmo/1998-002.

\bibitem{dn}
 V. Chaloupka et al., Technology development for a neutrino astrophysical
 observatory, L.O.I. to the D.O.E., LBL-38321, UC-412, February~1996.

\bibitem{bib:chap5_led}
J.S.~Kapustinsky et al.,
Nucl. Inst. Meth. {\bf A241} (1985), 612.



 


\bibitem{lepto} G. Ingelman et al., DESY 96-057 (1996).

\bibitem{okada} A. Okada, ``On the Atmospheric Muon Energy Spectrum in the Deep
 Ocean and its Parametrisation", ICRR-Report-319-94-14, ISSN 1340-3745 (1994).

\bibitem{propmu}P. Lipari and T. Stanev, 
    Phys. Rev. {\bf D44} (1991) 3543.


\bibitem{hemas} C. Forti et al, Phys. Rev. {\bf D42} (1990)
3668.

\bibitem{dada} C. Wiebusch, ``The Detection of Faint Light in Deep Underwater
  Neutrino Telescopes", Ph. D. Thesis, PITHA 95/37,
  RWTH-Aachen (1995).

\bibitem{km3}
D.M. Lowder, M. Moorhead, G. Shapiro, G. Smoot and S. Lowe,
``Stage I R\& D
for a km-Scale Neutrino Observatory: Physics and Detector
Simulations",
LBNL Report (1997).

\bibitem{DumandPrefit} V. Stenger,
``Track Fitting for the DUMAND Octogon", HDC-1-90, Univ.
Hawaii (1990).



\bibitem{sdss}
F.W. Stecker, C. Done, M.H. Salamon and P. Sommers, Phys. Rev. Lett.
{\bf 66} (1991) 2697; 
Phys. Rev. Lett. {\bf 69} (1992) 2738 (errata).

\bibitem{nmb}
L. Nellen, K. Mannheim and P.L. Biermann, Phys. Rev. {\bf
D47} (1993) 5270.

\bibitem{pro}
R.J. Protheroe, ``High Energy Neutrinos from Blazars", ADP-AT-96-7 and
astro-ph/9607165 (1996).

\bibitem{mannheim}
K. Mannheim, Astropart. Phys. {\bf 3} (1995) 295.

\bibitem{egret:cat} ``The Second EGRET Catalog of High-Energy Gamma-Ray
Sources", \\
http://cossc.gsfc.nasa.gov/cossc/egret/egret \\
catalog/cattex.html.

\bibitem{volkova} L.V. Volkova, Sov. J. Nucl. Phys. {\bf 31}
(1980) 1510.



\bibitem{mitsui} K. Mitsui, Y. Minorikawa and H. Komori, Nuovo Cimento
{\bf 9C} (1986) 995.

\bibitem{butkevich} A.V. Butkevich, L.G. Dedenko and I.M. Zheleznykh, 
Sov. J. Nucl. Phys. {\bf 50} (1989) 90.

\bibitem{bartol} V. Agrawal, T.K. Gaisser, P. Lipari and T. Stanev,
Phys. Rev. {\bf D53} (1996) 1314.

\bibitem{fogli} G.L. Fogli, E. Lisi and A. Marrone, Phys. Rev. {\bf D57}
(1998) 5893.



\bibitem{Amanda} F.\,Halzen, ``Results from Amanda and future'', talk given
at the {\it 18th International conference on Neutrino Physics and Astrophysics 
(Neutrino98)}, Takayama, Japan, June~4-9, 1998.

\bibitem{halzena} F. Halzen et al., ``The AMANDA Neutrino
Telescope", talk
presented at {\it Neutrino98}, Takayama, June 1998.

\bibitem{sno-supernova} J. F. Beacom, P. Vogel, Phys. Rev. {\bf D58} (1998)
 093012 (hep-ph/9806311).

\bibitem{omnis} P. F. Smith, talk given at {\it IDM98}, Buxton, September 1998.

\bibitem{amandaneutralino} F. Halzen et al., ``The AMANDA Neutrino Telescope 
and the Indirect Search for Dark Matter", talk given at {\it
DM98}, Santa Monica, 
February 1998.

\bibitem{baikal} G.\,V. Domogatsky, ``Results from Baikal'', talk given
at the {\it 18th International conference on Neutrino Physics and Astrophysics 
(Neutrino98)}, Takayama, Japan, June~4-9, 1998.

\bibitem{NT-96} Baikal Collaboration: V. A. Balkanov {\it et al.},
astro-ph/9903341, submitted to Astropart.Phys.

\bibitem{Nestor} B. Monteleoni, ``NESTOR a deep sea physics laboratory for 
the Mediterranean'' in {\em Proceeding of the 17th International conference 
on Neutrino Physics and Astrophysics (NEUTRINO'96)}, Helsinki, Finland, 
June~13-19, 1996, ed. K.\,Enqvist, K.\,Huitu, J.\,Maalampi (World Scientific, 
1997).
 
\bibitem{CHOOZ}CHOOZ Collaboration (M. Apollonio {\it et al.}),
 Phys.Lett. B420:397 (1998).

\bibitem{K2K} Yuichi Oyama for the K2K collaboration ), hep-ex/9803014.
    
\bibitem{MINOS} B. Barish, Nucl.Phys.Proc.Suppl.70:227 (1999). 

\bibitem{auger} K. S. Capelle, J. W. Cronin, G. Parente, E. Zas,
 Astropart.Phys. 8 (1998) 321-328 (astro-ph/9801313).




%










%









































\end{thebibliography}
